\documentclass[12pt]{article}
\usepackage{graphicx}
\usepackage{fullpage}
\usepackage[section]{placeins}
\usepackage{setspace}
\usepackage{chngcntr}
\usepackage{booktabs,caption,subcaption}
\usepackage{bm,array}
\usepackage{appendix}
\usepackage{titling}
\usepackage{adjustbox}
\usepackage{epsfig,latexsym,amsmath}
\usepackage{fancyhdr}
\usepackage{tabularx}
\usepackage{authblk}
\usepackage{float}
\usepackage{pdfpages}
\usepackage[labelfont=bf]{caption}
\usepackage[bottom]{footmisc}
\usepackage[round]{natbib}
\usepackage[english]{babel}
\usepackage[fixlanguage]{babelbib}
\usepackage{subcaption}
\usepackage{color,hyperref}
\definecolor{maroon}{RGB}{100,0,0}
\hypersetup{colorlinks,breaklinks,
            linkcolor=maroon,urlcolor=maroon,
            anchorcolor=maroon,citecolor=maroon}

\onehalfspace
\title{\huge Racial Disparities in Voting Wait Times: Evidence from Smartphone Data \\
\author{M. Keith Chen, \:Kareem Haggag,\: Devin G. Pope,\: and Ryne Rohla\thanks{Chen: Anderson School of Management, University of California at Los Angeles, keith.chen@anderson.ucla.edu, Haggag: Social and Decision Sciences, Carnegie Mellon University, kareem.haggag@cmu.edu. Pope: Booth School of Business, University of Chicago, devin.pope@chicagobooth.edu. Rohla: Anderson School of Management, University of California at Los Angeles, ryne.rohla@anderson.ucla.edu. The authors thank Stefano DellaVigna, Dean Karlan, Larry Katz, Charles Stewart III, and seminar audiences at Carnegie Mellon University, Stanford University, University of Chicago, and University of Illinois Urbana-Champaign for helpful comments and suggestions, Zachary Goldstein for research assistance, and A. Hoffman, R. Squire, and N. Yonack at SafeGraph for data access and technical assistance.}}}
\date{October 30, 2020}
\begin{document}
\maketitle

\begin{abstract}
\noindent Equal access to voting is a core feature of democratic government. Using data from hundreds of thousands of smartphone users, we quantify a racial disparity in voting wait times across a nationwide sample of polling places during the 2016 U.S. presidential election. Relative to entirely-white neighborhoods, residents of entirely-black neighborhoods waited 29\% longer to vote and were 74\% more likely to spend more than 30 minutes at their polling place. This disparity holds when comparing predominantly white and black polling places within the same states and counties, and survives numerous robustness and placebo tests. We shed light on the mechanism for these results and discuss how geospatial data can be an effective tool to both measure and monitor these disparities going forward.\\ 
\end{abstract}
\clearpage

\onehalfspacing
Providing convenient and equal access to voting is a central component of democratic government. Among other important factors (e.g. barriers to registration, purges from voter rolls, travel times to polling places), long wait times on Election Day are a frequently discussed concern of voters. Long wait times have large opportunity costs (\citealt{Stewart2015}), may lead to line abandonment by discouraged voters (\citealt{Stein2019}), and can undermine voters' confidence in the political process (\citealt{Alvarez2008, Atkeson2007, Bowler2015}). The topic of long wait times has reached the most prominent levels of media and policy attention, with President Obama discussing the issue in his 2012 election victory speech and appointing a presidential commission to investigate it. In their 2014 report, the Presidential Commission on Election Administration concluded that, ``as a general rule, no voter should have to wait more than half an hour in order to have an opportunity to vote.''

There have also been observations of worrying racial disparities in voter wait times. The Cooperative Congressional Election Study (CCES) finds that black voters report facing significantly longer lines than white voters (\citealt{Pettigrew2017, Alvarez2009, StewartIII2013}). While suggestive, the majority of prior work on racial disparities in wait times has been based on anecdotes and surveys which may face limits due to recall and reporting biases. 

In this paper, we use geospatial data generated by smartphones to measure wait times during the 2016 election. For each cellphone user, the data contain ``pings'' based on the location of the cellphone throughout the day. These rich data allow us to document voter wait times across the entire country and also estimate how these wait times differ based on neighborhood racial composition.

We begin by restricting the set of smartphones to a sample that passes a series of filters to isolate likely voters. This leaves us with a sample of just over 150,000 smartphone users who voted at one of more than 40,000 polling locations across 46 different states. Specifically, these individuals entered and spent at least one minute within a 60-meter radius of a polling location on Election Day and recorded at least one ping within the convex hull of the polling place building (based on building footprint shapefiles). We eliminate individuals who entered the same 60-meter radius in the week leading up to or the week after Election Day to avoid non-voters who happen to work at or otherwise visit a polling place on non-election days.

We estimate that the median and average times spent at polling locations are 14 and 19 minutes, respectively, and 18\% of individuals spent more than 30 minutes voting.\footnote{The time measure that we estimate in our paper is a combination of wait time in addition to the time it took to cast a ballot. We typically refer to this as just ``wait time'' in the paper. One may worry that the differences we find are not about wait times, but rather about differences in the amount of time spent casting a ballot. However, there is evidence to suggest this is not the case. For example, we find incredibly strong correlations between our wait time measures and survey responses that ask only about wait times as opposed to total voting time (``Approximately, how long did you have to wait in line to vote?'').} We provide descriptive data on how voting varies across the course of Election Day. As expected, voter volume is largest in the morning and in the evening, consistent with voting before and after the workday. We also find that average wait times are longest in the early morning hours of the day. Finally, as a validation of our approach, we show that people show up to the polls at times consistent with the opening and closing hours used in each state.

We next document geographic variation in average wait times using an empirical Bayes adjustment strategy. We find large differences across geographic units -- for example, average wait times across congressional districts can vary by a factor of as much as four. We further validate our approach by merging in data from the CCES, which elicits a coarse measure of wait time from respondents. Despite many reasons for why one might discount the CCES measures (e.g. reporting bias and limited sample size), we find a remarkably high correlation with our own measures -- a correlation of 0.86 in state-level averages and 0.73 in congressional-district-level averages. This concordance suggests that our wait time measures (and those elicited through the survey) have a high signal-to-noise ratio.

We next explore how wait times vary across areas with different racial compositions. We use Census data to characterize the racial composition of each polling place's corresponding Census block group (as a proxy for its catchment area). We find that the average wait time in a Census block group composed entirely of black residents is approximately 5 minutes longer than average wait time in a block group that has no black residents. We also find longer wait times for areas with a high concentration of Hispanic residents, though this disparity is not as large as the one found for black residents. These racial disparities persist after controlling for population, population density, poverty rates, and state fixed effects. We further decompose these effects into between- and within-county components, with the disparities remaining large even when including county fixed effects. We perform a myriad of robustness checks and placebo specifications and find that the racial disparity exists independent of the many assumptions and restrictions that we have put on the data.

In the Appendix, we consider the potential mechanisms behind the observed racial differences. We ultimately find that a host of plausible candidate explanations do little to explain the disparity in our cross-section, including differences in arrival times of voters, state laws (Voter ID and early voting), the partisan identity of the underlying population or the chief election official, county characteristics (income inequality, segregation, social mobility), and the number of registered voters assigned to a polling place; although we do find larger disparities at higher-volume polling locations. Overall, our results on mechanism suggest that the racial disparities that we find are widespread and unlikely to be isolated to one specific source or phenomenon. 

Our paper is related to work in political science that has examined determinants of wait times and also explored racial disparities. Some of the best work uses data from the CCES which provides a broad sample of survey responses on wait times (\citealt{Pettigrew2017, Alvarez2009, StewartIII2013}). For example, \citet{Pettigrew2017} finds that black voters report waiting in line for twice as long as white voters and are three times more likely to wait for over 30 minutes to vote. Additional studies based on field observations may avoid issues that can arise from self-reported measures, but typically only cover small samples of polling places such as a single city or county (\citealt{Highton2006,Spencer2010,Herron2016}). \citet{Stein2019} collect the largest sample to date, using observers with stopwatches across a convenience sample of 528 polling locations in 19 states. Using a sample of 5,858 voters, they provide results from a regression of the number of people observed in line on an indicator that the polling place is in a majority-minority area. They find no significant effect -- although they also control for arrival count in the regression. In a later regression, they find that being in a majority-minority polling location leads to a 12-second increase in the time it takes to check in to vote (although this regression includes a control for the number of poll workers per voter which may be a mechanism for racial disparities in voting times). Overall, we arrive at qualitatively similar results as the political science literature, but do so using much more comprehensive data that avoids the pitfalls of self-reports. Going forward, this approach could produce repeated measures across elections, which would facilitate a richer examination of the causal determinants of the disparities.

Our paper also relates to the broader literature on racial discrimination against black individuals and neighborhoods (for reviews, see \citealt{Altonji1999}, \citealt{Charles2011}, and \citealt{Bertrand2017}), including by government officials. For example, \citet{Butler2011} find that legislators were less likely to respond to email requests from a putatively black name, even when the email signaled shared partisanship in an attempt to rule out strategic motives. Similarly, \citet{White2015} find that election officials in the U.S. were less likely to respond and provided lower-quality responses to emails sent from constituents with putatively Latino names. Racial bias has also been documented for public officials that are not part of the election process. For example, \citet{Giulietti2019} find that emails sent to local school districts and libraries asking for information were less likely to receive a response when signed with a putatively black name relative to a putatively white name. As one final example, several studies have documented racial bias by judges in criminal sentencing (\citealt{Alesina2014,Glaeser2003,Abrams2012}). 

\section{Data}
The three primary datasets we use in this paper include: (1) SafeGraph cell phone location records, (2) Polling locations, and (3) Census demographics. 

We use anonymized location data for smartphones provided by SafeGraph, a firm that aggregates location data across a number of smartphone applications (\citealt{Chen2018}). These data cover the days between November 1st and 15th, 2016, and consist of ``pings'', which record a phone's location at a series of points in time. In general, GPS pings are typically accurate to within about a 5-meter radius under open sky, though this varies depending on factors such as weather, obstructions, and satellite positioning (\hyperlink{https://www.gps.gov/systems/gps/performance/accuracy/}{GPS.gov}). Pings are recorded anytime an application on a phone requests information about the phone's location. Depending on the application (e.g. a navigation or weather app), pings may be produced when the application is being used or at regular intervals when it is in the background. The median time between pings in our sample for a given device is 48 seconds (with a mode of 5 minutes).

The geolocation data used in this paper is detailed and expansive, allowing us to estimate wait times around the entire United States. This data, however, naturally raises concerns about representativeness. If we were trying to estimate individual choices, e.g. vote choice, the sample could only produce estimates that are at best representative of the approximately 77\% of U.S. adults who owned a smartphone in 2016. While \citet{Chen2019} show that the data are generally representative of the U.S. along several observable dimensions (with the exception of skewing more wealthy), they may differ on unobservables. However, our goal is to estimate a property of places rather than individuals. That is, we estimate an outcome of queues that have multiple individuals in them. While the restriction to smartphone users may limit the number of wait times we observe, as long as there is a queueing rule at polling places, we should still observe an unbiased estimate of the wait times faced by voters, both those with and without smartphones.\footnote{This is not to dismiss the potential issue of missing polling places or times of day. However, \textit{a priori}, these omissions do not point to systematic bias in a particular direction.}

Polling place addresses for the 2016 General Election were collected by contacting state and county election authorities. When not available, locations were sourced from local newspapers, public notices, and state voter registration look-up webpages. State election authorities provided statewide locations for 32 states, five of which required supplemental county-level information to complete. Four states were completely collected on a county-by-county basis. In twelve states, not all county election authorities responded to inquiries (e.g. Nassau County, New York).

When complete addresses were provided, the polling locations were geocoded to coordinates using the Google Maps API. When partial or informal addresses were provided, buildings were manually assigned coordinates by identifying buildings through Google Street View, imagery, or local tax assessor maps as available. Additionally, Google Maps API geocodes are less accurate or incomplete in rural locations or areas of very recent development, and approximately 8\% of Google geocodes were manually updated. 

Of the 116,990 national polling places reported in 2016 by the U.S. Election Assistance Commission, 93,658 polling places (80.1\%) were identified and geocoded and comprise the initial sample of polling places in this paper. Appendix Figure \ref{fig:locations} illustrates the location of the 93,658 polling places and separately identifies polling places for which we identify likely voters on Election Day and pass various filters that we discuss and impose below. 

Demographic characteristics were obtained by matching each polling place location to the census block group in the 2017 American Community Survey's five-year estimates. Census block groups were chosen as the level of aggregation because the number of block groups is the census geography that most closely aligns with the number of polling places and because it contains the information of interest (racial characteristics, fraction below poverty line, population, and population density).

\section{Methods}

In order to calculate voting wait times, we need to identify a set of individuals we are reasonably confident actually voted at a polling place in the 2016 election. To do so, we restrict the sample to phones that record a ping within a certain distance of a polling station on Election Day. This distance is governed by a trade-off -- we want the radius of the circle around each polling station to be large enough to capture voters waiting in lines that may spill out of the polling place, but want the circle to be not so large that we introduce a significant number of false positive voters (people who came near a polling place, but did not actually vote). 

\begin{figure}[H]
\begin{center}
\caption{Defining the Radius}
\label{fig:defineradius}
\vspace{-5pt}
\includegraphics[scale=.165]{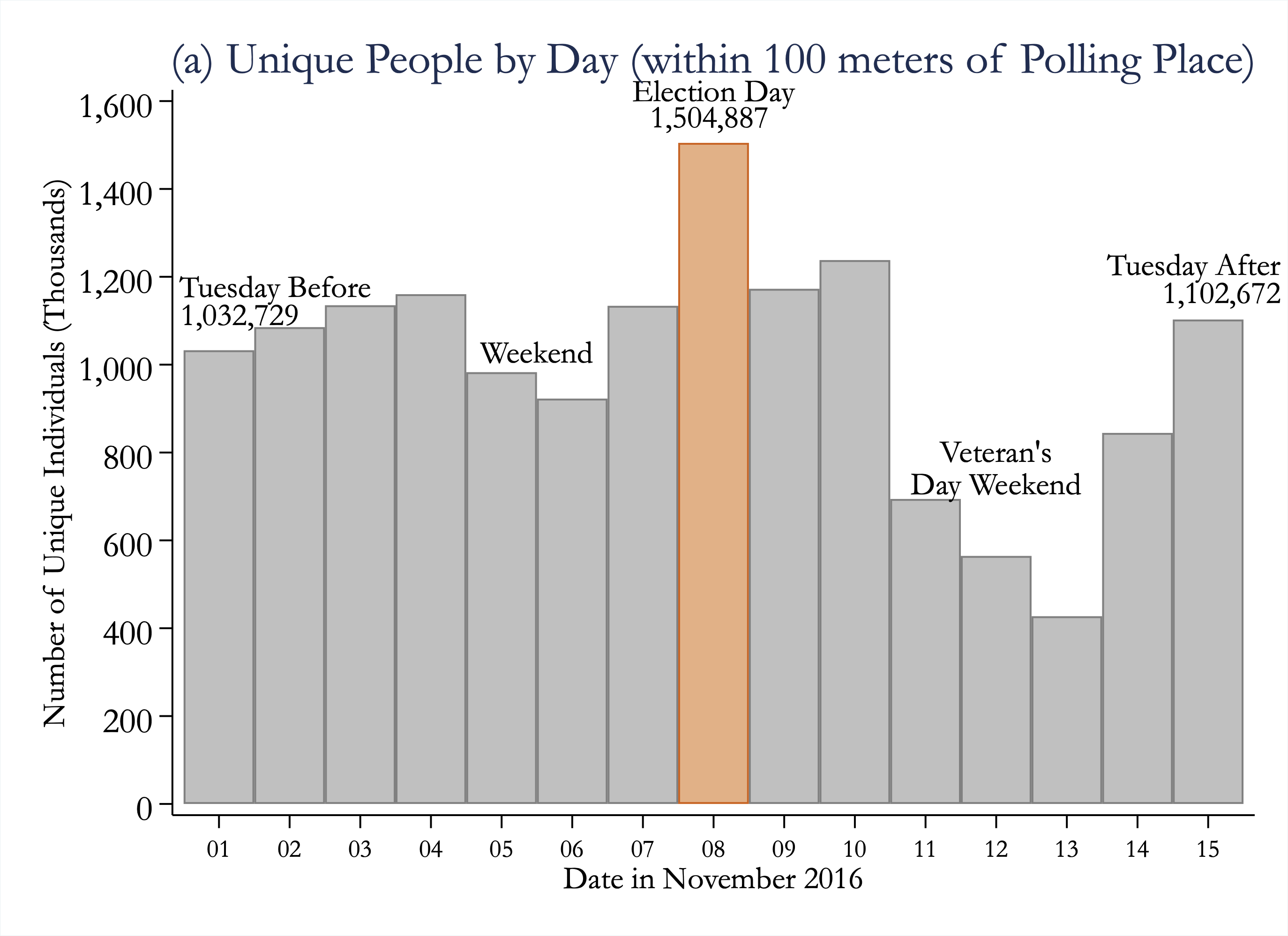}
\includegraphics[scale=.165]{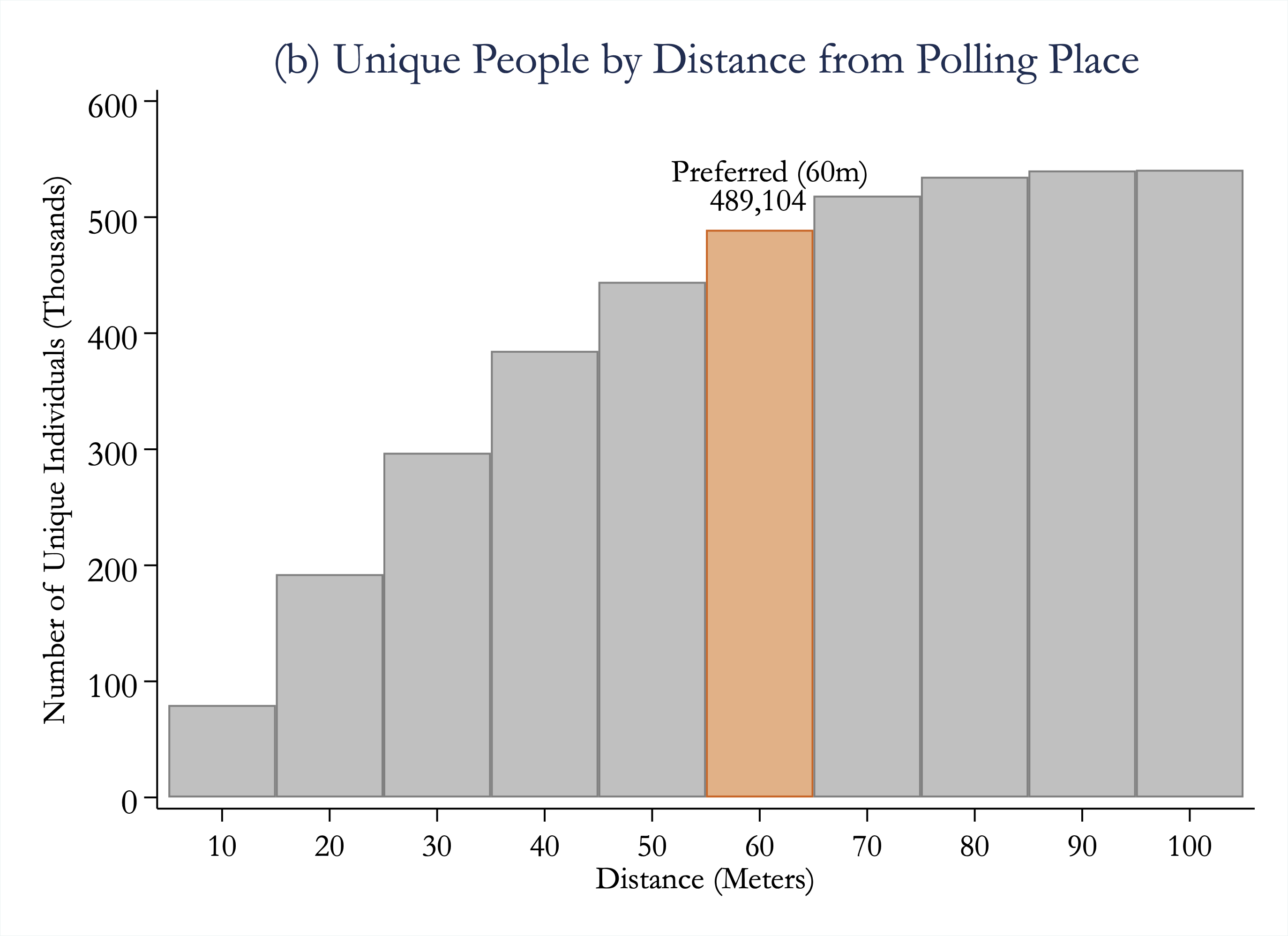}
\includegraphics[scale=.165]{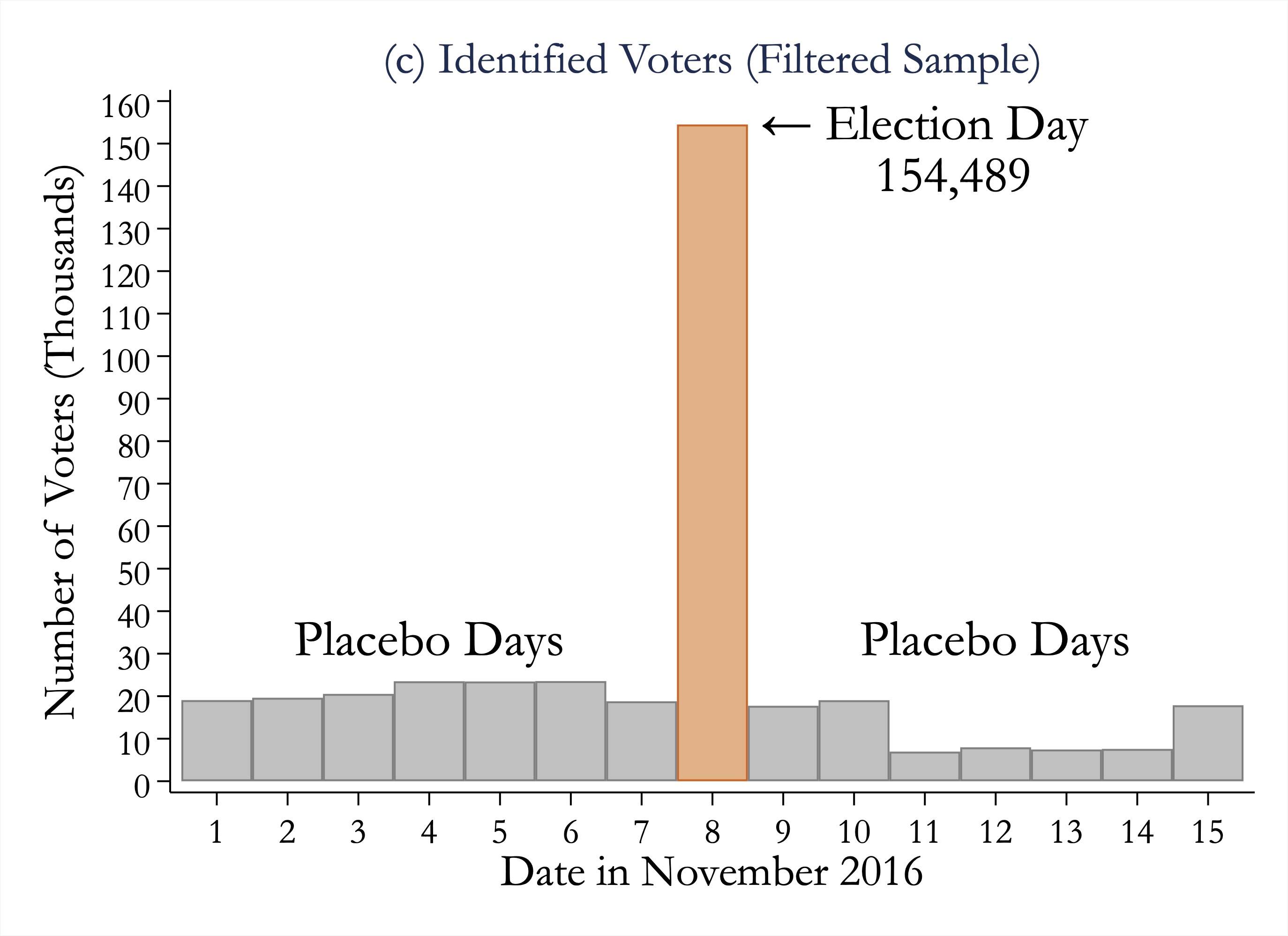}
\vspace{-10pt}
\caption*{\scriptsize \textbf{Notes}: Panel A plots the number of unique device IDs observed within 100 meters of polling place building centroids on each day from November 1 to November 15 -- Election Day (November 8) is highlighted in orange. Panel B plots the \textit{difference} in the number of unique devices that are within a particular radius of the polling place (10 meters to 100 meters) on Election Day \textit{relative to the average across all other days} -- our final radius of 60 meters is highlighted in orange. Panel C shows the sample of unique devices that are observed within 60 meters of a polling place building centroid after applying the full set of filters -- Election Day is highlighted in orange. Note that the Y-axes change across subfigures and that Veteran's Day was on Friday, November 11 in 2016. The initial sample of smartphones that recorded at least one ping on Election Day(November 8, 2016) consisted of 5.2 million unique devices. As Panel A shows, there are 1.5 million devices once we limit to those that recorded at least one ping within 100 meters of a polling place on that date. Limiting to those within 60 meters of a polling place (the final radius used in Panel C) drops this to 1.0 million devices. Further limiting to phones that recorded at least one ping in the convex hull of the polling place building drops this to 406k devices, and limiting to phones that recorded a consistent set of pings on Election Day (1 per hour for 12 hours) drops to 307k devices. Imposing the remaining filters discussed in the text drops to the final sample of 155k observed in the orange bar of Panel C. }
\end{center}
\end{figure}

We take a data-driven approach to determine the optimal size of the radius. In Panel A of Figure \ref{fig:defineradius}, we examine whether there are more unique individuals who show up near a polling place on Election Day relative to the week before and the week after the election (using a 100-meter radius around a polling location).\footnote{More precisely, we construct a 100-meter radius around the centroid of the building identified by Microsoft OpenStreetMap as the closest to the polling place coordinates.} As can be seen, there appear to be more than 400k additional people on Election Day who come within 100 meters of a polling place relative to the weekdays before and after. In Panel B of Figure \ref{fig:defineradius}, we plot the difference in the number of people who show up within a particular radius of the polling place (10 meters to 100 meters) on Election Day relative to the average across all other days. As we increase the size of the radius, we are able to identify more and more potential voters, but also start picking up more and more false positives. By around 60 meters, we are no longer identifying very many additional people on Election Day relative to non-election days, and yet are continuing to pick up false positives. Therefore, we choose 60 meters as the radius for our primary analysis. However, in Section 4.1, we demonstrate robustness of estimates to choosing alternative radii.

For each individual that comes within a 60-meter radius of a polling place, we would like to know the amount of time spent within that radius. Given that we do not receive location information for cell phones continuously (the modal time between pings is 5 minutes), we cannot obtain an exact length of time. Thus, we create upper and lower bounds for the amount of time spent voting by measuring the time between the last ping before entering and the first ping after exiting a polling-place circle (for an upper bound), and the first and last pings within the circle (for a lower bound). For example, pings may indicate a smartphone user was not at a polling location at 8:20am, but then was at the polling location at 8:23, 8:28, 8:29, and 8:37, followed by a ping outside of the polling area at 8:40am; translating to a lower bound of 14 minutes and an upper bound of 20 minutes. We use the midpoint of these bounds as our best guess of a voter's time at a polling place (e.g. 17 minutes in the aforementioned example). In the robustness section, we estimate our effects using values other than the midpoint.

Another important step in measuring voting times from pings is to isolate people who come within a 60-meter radius of a polling place that we think are likely voters and not simply passing by or people who live or work at a polling location. To avoid including passersby, we restrict the sample to individuals who had an upper bound measure of at least one minute within a polling place circle and for whom that is true at only one polling place on Election Day. To avoid including people who live or work at the polling location, we exclude individuals who we observe spending time (an upper bound greater than 1 minute) at that location in the week before or the week after Election Day. To further help identify actual voters and reduce both noise and false positives, we restrict the sample to individuals who had at least one ping within the convex hull of the polling place building on Election Day (using Microsoft OpenStreetMap building footprint shapefiles), logged a consistent set of pings on Election Day (posting at least 1 ping every hour for 12 hours), and spent no more than 2 hours at the polling location (to eliminate, for example, poll workers who spend all day at a polling place). In Section 4.1, we provide evidence of robustness to these various sample restrictions.

After these data restrictions, our final sample consists of 154,489 individuals whom we identify as likely voters across 43,413 polling locations. Panel C in Figure \ref{fig:defineradius} shows how many people pass our likely-voter filters on Election Day (154,489), and---as a placebo analysis---how many observations we would have on non-Election (``placebo'') days before and after the 2016 Election that would pass these same filters (modified to be centered around those placebo days). This analysis suggests that more than 87\% of our sample are likely voters who would not have been picked up on days other than Election Day. In Appendix Figure \ref{fig:app_placebodist}, we plot the distribution of wait times on each of these placebo non-election days. We find that the wait times of people who would show up in our analysis on non-election days are shorter on average than those that show up on Election Day. Thus, to the degree that we can not completely eliminate false positives in our voter sample, we expect our overall voter wait times to be biased upward. We also would expect the noise introduced by non-voters to bias us towards not finding systematic disparities in wait times by race. 

Appendix Table \ref{table:sumstats} provides summary statistics for our 154,489 likely voters. We find average voting wait times of just over 19 minutes when using our primary wait time measure (the midpoint between the lower and upper bound) and 18\% of our sample waited more than 30 minutes to vote. Weighted by the number of voters in our sample, the racial composition of the polling place block groups is, on average, 70\% white and 11\% black.

\section{Results: Overall Voter Wait Times}
\begin{figure}[H]
\caption{Overall Wait Times}
\begin{subfigure}{.5\textwidth}
  \caption{Wait Time Histogram}
  \includegraphics[width=.9\linewidth]{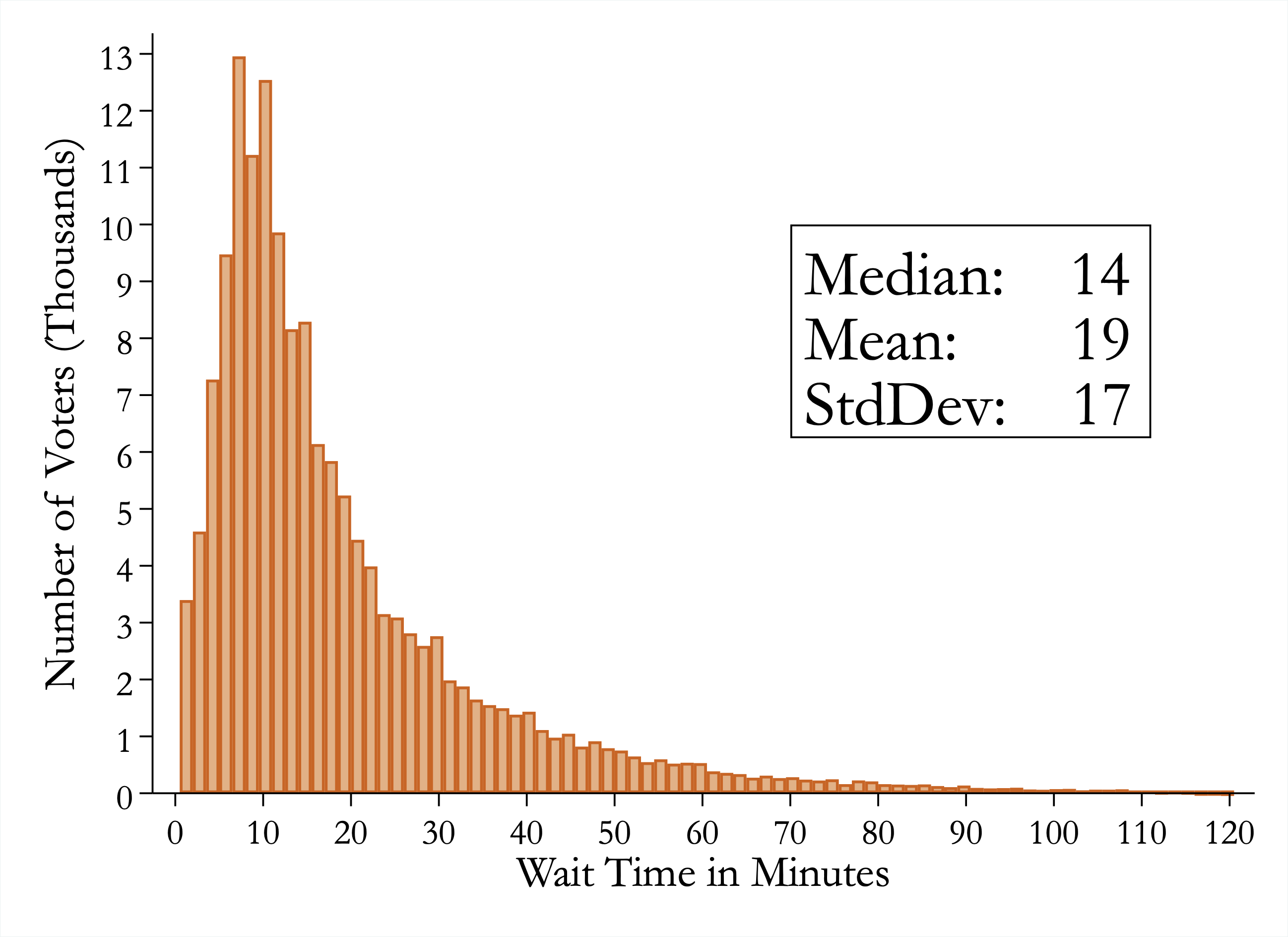}  
  \label{fig:sub-first}
\end{subfigure}
\begin{subfigure}{.5\textwidth}
  \caption{Geographic Variation}
  \includegraphics[width=1\linewidth]{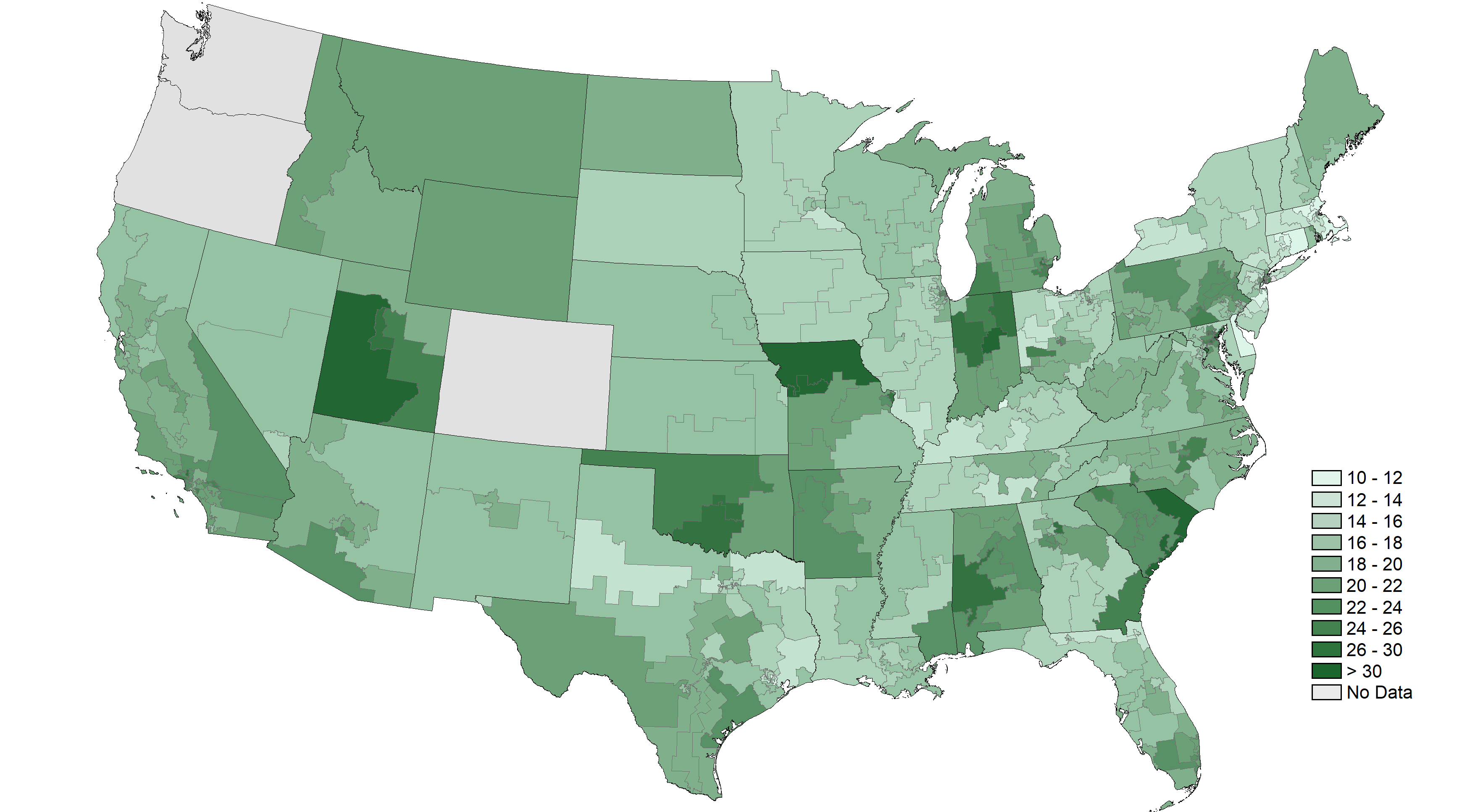}  
  \label{fig:sub-second}
\end{subfigure}
\begin{subfigure}{.5\textwidth}
  \caption{Number of Voters by Hour of Day}
  \includegraphics[width=.9\linewidth]{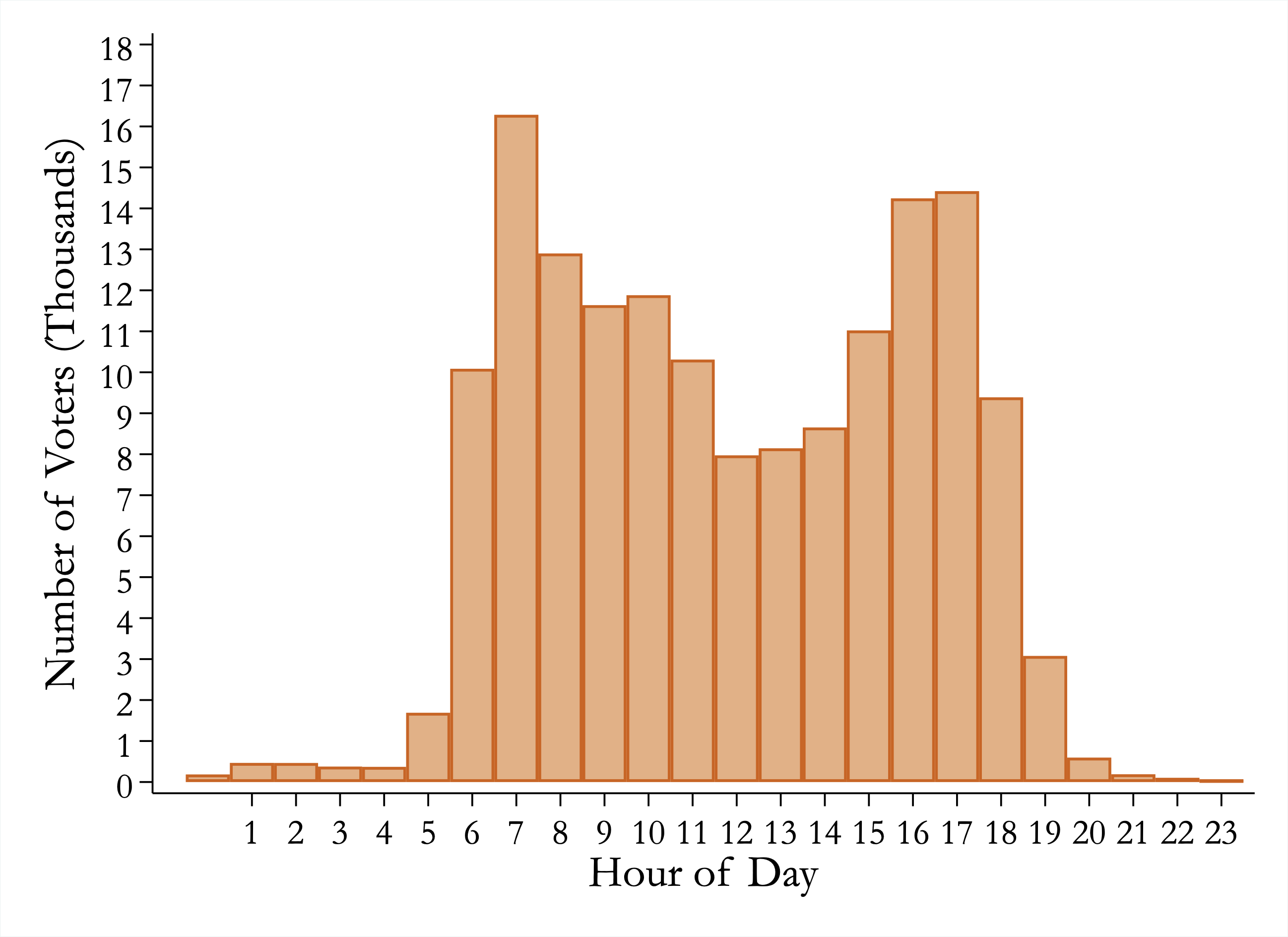}  
  \label{fig:sub-third}
\end{subfigure}
\begin{subfigure}{.5\textwidth}
  \caption{Average Wait Time by Hour of Day}
  \includegraphics[width=.9\linewidth]{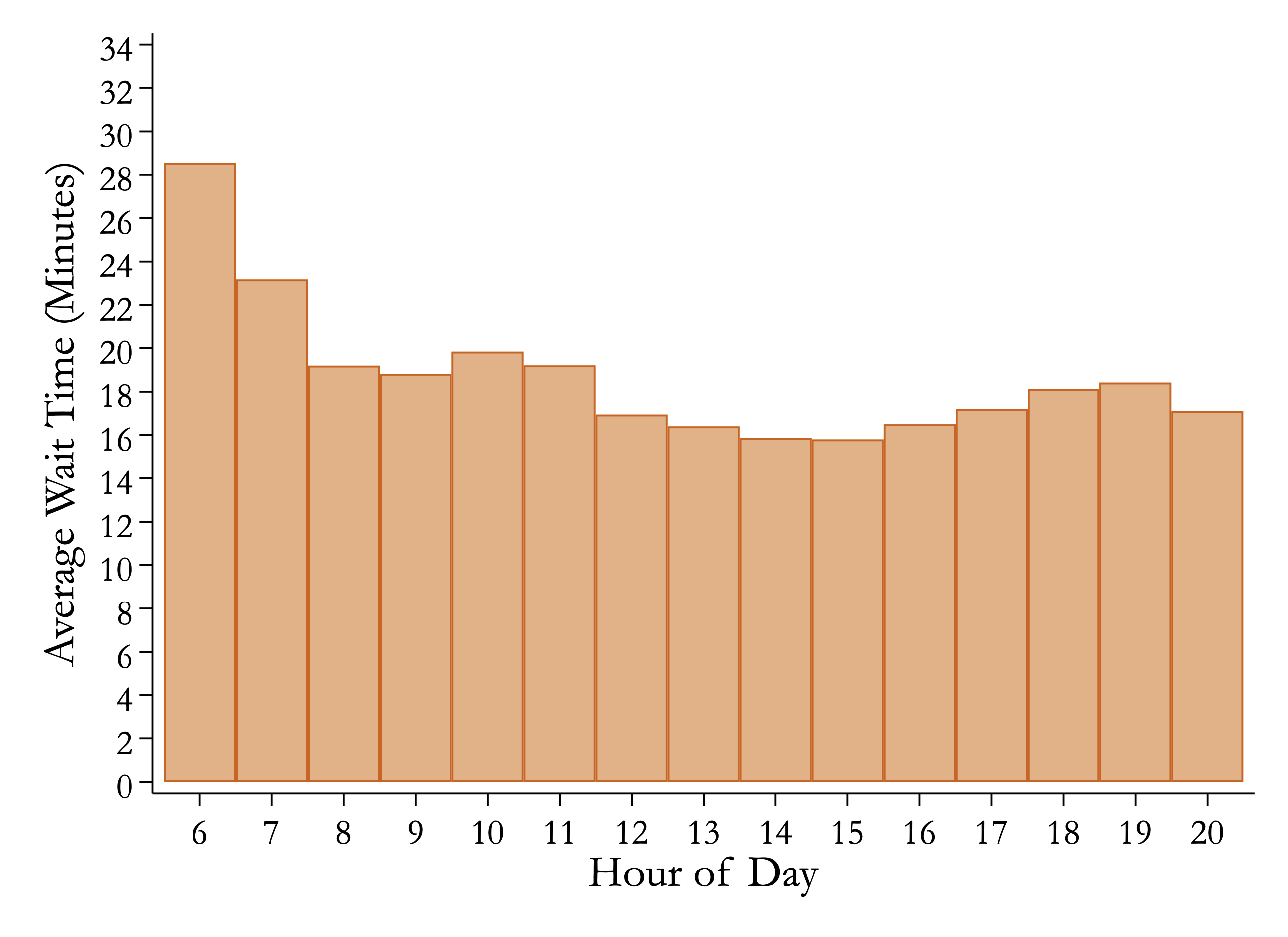}  
  \label{fig:sub-fourth}
\end{subfigure}
\label{fig:overallwait}
\caption*{\scriptsize \textbf{Notes}: Panel A plots a histogram corresponding to the 154,495 cell phones who pass the filters used to identify likely voters (using 1.5 minute bins). Panel B shows variation in (empirical-Bayes-adjusted) average wait times by congressional district (115th Congress). Panel C plots the total number of voters (volume) by hour of arrival. Panel D plots the average wait time for each hour of arrival.}
\end{figure}

We plot the distribution of wait times in Panel A of Figure \ref{fig:overallwait}. The median and average times spent at polling locations are 14 and 19 minutes, respectively, and 18\% of individuals spent more than 30 minutes voting. As the figure illustrates, there is a non-negligible number of individuals who spent 1-5 minutes in the polling location (less time than one might imagine is needed to cast a ballot). These observations might be voters who abandoned after discovering a long wait time. Alternatively, they may be individuals who pass our screening as likely voters, but were not actually voting. \\

We next display the number of people who arrive to vote at the polling locations by time of day. This descriptive analysis of when people vote may be of interest in and of itself, but it also serves as a validation of whether people in our sample are indeed likely voters (e.g. if our sample consists primarily of people showing up at the polling locations at 3am, then one should worry about whether our sample is primarily composed of voters). Panel C of Figure \ref{fig:overallwait} shows the distribution of arrival times where the ``hour of day'' is defined using the ``hour of arrival'' for a given wait time (i.e. the earliest ping within the polling place radius for a given wait time spell). As expected, people are most likely to vote early in the morning or later in the evening (e.g. before or after work) with nearly twice as many people voting between 7 and 8am as between noon and 1pm. As a consistency check, Appendix Figure \ref{fig:openclose} repeats this figure separately by state's opening and closing times -- the figures show that likely-voter arrivals match state-by-state poll opening and closing times. Finally, Panel D of Figure \ref{fig:overallwait} plots the average wait time by time of arrival, showing that the longest averages are early in the morning.

In addition to temporal variation in wait times, we can also explore how voting wait times vary geographically. Appendix Tables \ref{table:app_state_descriptives} - \ref{table:app_county_descriptives} report average wait times by state, congressional district, and the 100 most populous counties, along with accompanying standard deviations and observation counts, as well as an empirical-Bayes adjustment to account for measurement error.\footnote{Even if all states in the U.S. had the same voter wait time, we would find some dispersion in our measure due to sampling variation. Due to sample size, this measurement error in our estimates would result in the smallest states being the most likely to show evidence of having either very short or very long wait times. Thus, throughout the paper, whenever we discuss voter wait times or racial disparities that have been aggregated up to either the county, congressional district, or state level, we will report estimates that have been adjusted for measurement using a Bayesian shrinkage procedure. This iterative procedure (discussed in detail in \citet{Chandra2016}) shrinks estimates toward the average of the true underlying distribution. The amount of adjustment toward the mean is a function of how far the estimate for each state/county is from the mean and the estimate's precision. The resulting adjusted estimate is our ``best guess'' (using Bayesian logic) as to what the actual wait time or disparity is for each geographic unit.} Focusing on the empirical-Bayes adjusted estimates, the states with the longest average wait times are Utah and Indiana (28 and 27 minutes, respectively) and the states with the shortest average wait time are Delaware and Massachusetts (12 minutes each). In Panel B of Figure \ref{fig:overallwait} we map the empirical-Bayes-adjusted average voting wait time for each congressional district across the United States. Average wait times vary from as low as $\sim11$ minutes in Massachusetts's Sixth and Connecticut's First Congressional District to as high as $\sim40$ minutes in Missouri's Fifth Congressional District. These geographic differences are not simply a result of a noisy measure, but contain actual signal value regarding which areas have longer wait time than others. Evidence for this can be seen by our next analysis correlating our wait time measures with those from a survey.

We correlate our average wait time measures at both the state and congressional district level with the average wait times reported by respondents in the the 2016 wave of the Cooperative Congressional Election Study \citep{DVN/GDF6Z0_2017}. The 2016 CCES is a large national online survey of 64,600 people conducted before and after the U.S. general election. The sample is meant to be representative of the U.S. as a whole.\footnote{\url{https://dataverse.harvard.edu/dataset.xhtml?persistentId=doi\%3A10.7910/DVN/GDF6Z0}} There are several reasons one might be pessimistic that the wait time estimates that we generate using smartphone-data would correlate closely with the wait times reported from the CCES survey. First, given sample sizes at the state and congressional district level, both our wait times and survey wait times may have a fair bit of sampling noise. Second, our wait time measures are a combination of waiting in line and casting a ballot, whereas the survey only asks about wait times. Third, the question in the survey creates additional noise by eliciting wait times that correspond to one of five coarse response options (``not at all'', ``less than 10 minutes'', ``10 to 30 minutes'', ``31 minutes to an hour'', and ``more than an hour'').\footnote{There are 34,353 responses to the ``wait time'' question in the 2016 CCES. We restrict the sample of responses to just use individuals who voted in person on Election Day (24,378 individuals after dropping the 45 who report ``Don't Know''). Following \citet{Pettigrew2017}, we translate the responses to minute values by using the midpoints of response categories: 0 minutes (``not at all''), 5 minutes (``less than 10 minutes''), 20 minutes (``10 to 30 minutes'') or 45 minutes (``31 minutes to an hour''). For the 421 individuals who responded as ``more than an hour'' we code them as waiting 90 minutes (by contrast, \citet{Pettigrew2017} uses their open follow-up text responses.)} Lastly, the survey does not necessarily represent truthful reporting. For example, while turnout in the U.S. has hovered between 50 and 60 percent, more than 80\% of CCES respondents report voting. Given these reasons for why our wait time results may not correlate well with those from the survey, we find a remarkably strong correlation between the two. Using empirical-Bayes-adjusted estimates for both state-level wait time estimates from the cellphone data and those found in the CCES, we find correlation of 0.86 between the two. We find a similarly strong correlation at the congressional district level (correlation = 0.73). Our wait-time estimates are, on average, slightly longer than those in the survey, which is likely a reflection of the fact that our measure includes both wait time and ballot-casting time. Scatter plots of the state and congressional district estimates may be found in Appendix Figure \ref{fig:cces}. Overall, the strong correlations between the wait times we estimate and those from the CCES survey provide  validation for our wait time measure (and for the CCES responses themselves).

\section{Results: Racial Disparities in Wait Times}
\begin{figure}[H]
\begin{center}
\caption{Wait Time: Fraction Black 1st vs. 10th Decile}
\label{fig:disparity}
\vspace{-5pt}
\includegraphics[width=1\linewidth]{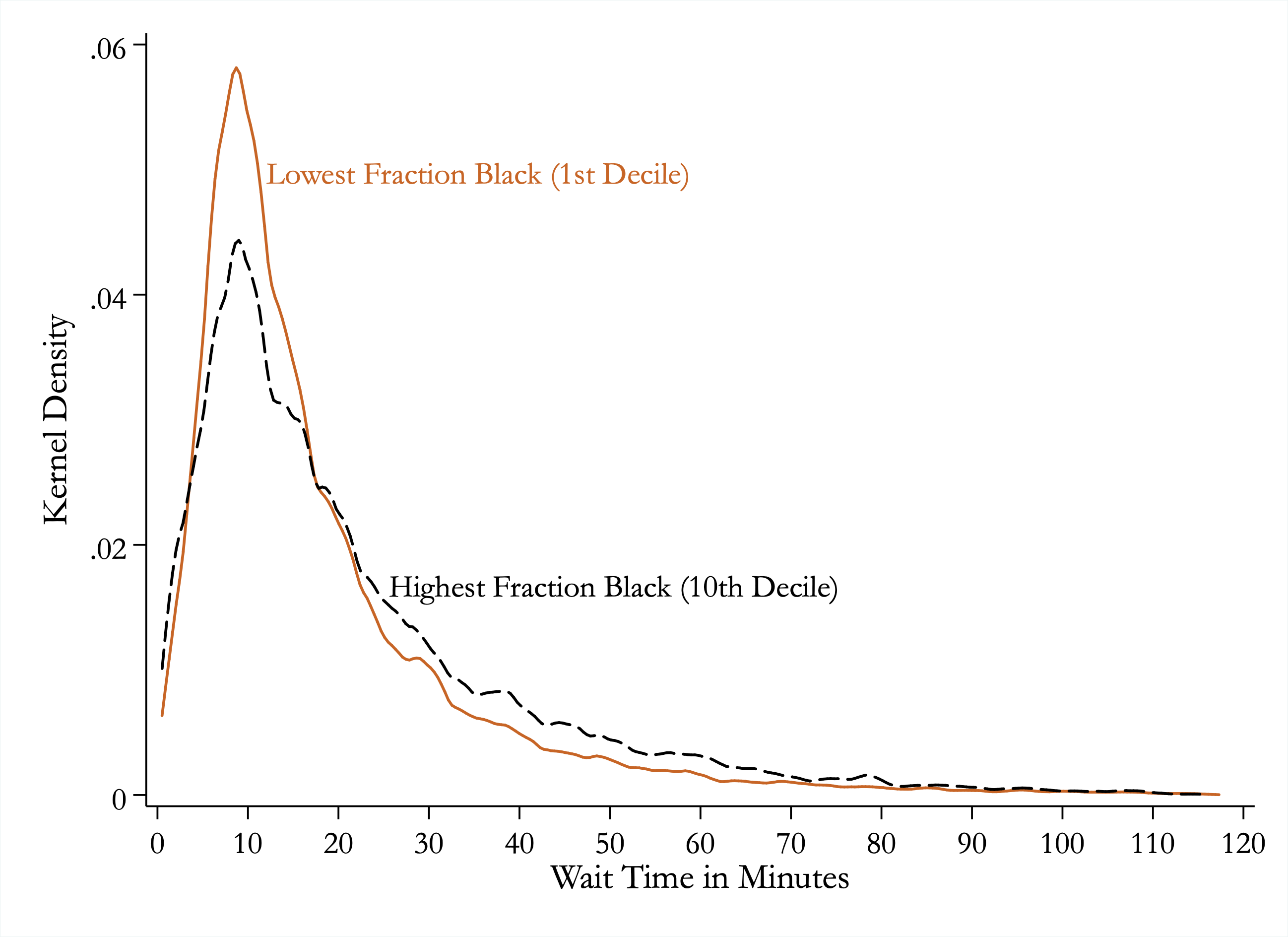}
\vspace{-10pt}
\caption*{\scriptsize \textbf{Notes}: Kernel densities are estimated using 1 minute half-widths. The 1st decile corresponds to the 34,420 voters across 10,319 polling places with the lowest percent of black residents (mean = 0\%). The 10th decile corresponds to the 15,439 voters across the 5,262 polling places with the highest percent of black residents (mean = 58\%).}
\end{center}
\end{figure}

In this section, we provide evidence that wait times are significantly longer for areas with more black residents relative to white residents. We begin with a simple visualization of wait times by race. Figure \ref{fig:disparity} plots the smoothed distribution of wait times separately for polling places in the top and bottom deciles of the fraction-black distribution. These deciles average 58\% and 0\% black, respectively. Voters from areas in the top decile spent 19\% more time at their polling locations than those in the bottom decile. Further, voters from the top decile were 49\% more likely to spend over 30 minutes at their polling locations. Appendix Figures \ref{fig:app_otherraces} and \ref{fig:app_poverty} provide similar density functions of wait-time comparisons for other demographic characteristics. \\

\begin{table}[H]
\begin{center}
\caption{Fraction Black and Voter Wait Time}
\label{table:main}
\vspace{-10pt}
\scalebox{0.85}{{
\def\sym#1{\ifmmode^{#1}\else\(^{#1}\)\fi}
\begin{tabular}{l*{5}{c}}
\toprule
&\multicolumn{1}{c}{(1)}&\multicolumn{1}{c}{(2)}   &\multicolumn{1}{c}{(3)}&\multicolumn{1}{c}{(4)}&\multicolumn{1}{c}{(5)}\\
\multicolumn{5}{l}{\textbf{Panel A: Ordinary Least Squares (Y = Wait Time)}} \\
\hline
\midrule
Fraction Black                &        5.23\sym{***}&        5.22\sym{***}&        4.96\sym{***}&        4.84\sym{***}&        3.27\sym{***}\\
                              &      (0.39)         &      (0.39)         &      (0.42)         &      (0.42)         &      (0.45)         \\
\addlinespace
Fraction Asian                &                     &       -0.79         &       -2.48\sym{***}&        1.30\sym{*}  &       -1.10         \\
                              &                     &      (0.72)         &      (0.74)         &      (0.76)         &      (0.81)         \\
\addlinespace
Fraction Hispanic             &                     &        1.15\sym{***}&        0.43         &        3.90\sym{***}&        1.50\sym{***}\\
                              &                     &      (0.37)         &      (0.40)         &      (0.46)         &      (0.50)         \\
\addlinespace
Fraction Other Non-White      &                     &       12.01\sym{***}&       11.76\sym{***}&        1.66         &        2.04         \\
                              &                     &      (1.94)         &      (1.95)         &      (1.89)         &      (1.93)         \\
\midrule
N                             &     154,411         &     154,411         &     154,260         &     154,260         &     154,260         \\
$R^2$                         &        0.00         &        0.00         &        0.01         &        0.06         &        0.13         \\
DepVarMean                    &       19.13         &       19.13         &       19.12         &       19.12         &       19.12         \\
Polling Area Controls? &No&No&Yes&Yes&Yes \\
State FE?                       &No&No&No&Yes&Yes \\
County FE?                       &No&No&No&No&Yes \\
\hline
\multicolumn{5}{l}{\textbf{Panel B: Linear Probability Model (Y = Wait Time $>$ 30min)}} \\
\hline
\midrule
Fraction Black                &        0.12\sym{***}&        0.12\sym{***}&        0.11\sym{***}&        0.10\sym{***}&        0.07\sym{***}\\
                              &      (0.01)         &      (0.01)         &      (0.01)         &      (0.01)         &      (0.01)         \\
\addlinespace
Fraction Asian                &                     &       -0.00         &       -0.04\sym{**} &        0.04\sym{**} &       -0.02         \\
                              &                     &      (0.02)         &      (0.02)         &      (0.02)         &      (0.02)         \\
\addlinespace
Fraction Hispanic             &                     &        0.03\sym{***}&        0.01         &        0.08\sym{***}&        0.03\sym{***}\\
                              &                     &      (0.01)         &      (0.01)         &      (0.01)         &      (0.01)         \\
\addlinespace
Fraction Other Non-White      &                     &        0.21\sym{***}&        0.21\sym{***}&        0.03         &        0.05         \\
                              &                     &      (0.04)         &      (0.04)         &      (0.04)         &      (0.04)         \\
\midrule
N                             &     154,411         &     154,411         &     154,260         &     154,260         &     154,260         \\
$R^2$                         &        0.00         &        0.00         &        0.01         &        0.04         &        0.10         \\
DepVarMean                    &        0.18         &        0.18         &        0.18         &        0.18         &        0.18         \\
Polling Area Controls? &No&No&Yes&Yes&Yes \\
State FE?                       &No&No&No&Yes&Yes \\
County FE?                       &No&No&No&No&Yes \\
\hline\hline
\bottomrule
\multicolumn{6}{l}{\footnotesize \sym{*} \(p<0.10\),    \sym{**} \(p<0.05\), \sym{***} \(p<0.01\)} \\
\end{tabular}
}
}
\end{center}
\vspace{-20pt}
\caption*{\scriptsize \textit{Notes}: Robust standard errors, clustered at the polling place level, are in parentheses. Unit of  observation is a cellphone identifier on Election Day. \textit{DepVarMean} is the mean of the dependent variable. The dependent variable in Panel B is a binary variable equal to 1 if the wait time is greater than 30 minutes. \textit{Polling Area Controls} includes the population, population per square mile, and fraction below poverty line for the block group of the polling station. ``Asian'' includes ``Pacific Islander.'' ``Other Non-White'' includes the ``Other,'' ``Native American,'' and ``Multiracial'' Census race categories.}
\end{table}

Of course, Figure \ref{fig:disparity} focuses just on polling places that are at the extremes of racial makeup. We provide a regression analysis in Table \ref{table:main} in order to use all of the variation across polling places' racial compositions and to provide exact estimates and standard errors. Panel A uses wait time as the dependent variable. In column 1, we estimate the bivariate regression which shows that moving from a census block group with no black residents to one that is entirely composed of black residents is associated with a 5.23 minute longer wait time. In column 2, we broaden our focus by adding additional racial categories,revealing longer wait times for block groups with higher fractions of Hispanic and other non-white groups (Native American, other, multiracial) relative to entirely white neighborhoods. Column 3 examines whether these associations are robust to controlling for population, population density, and fraction below poverty line of the block group (see Appendix Tables \ref{table:app_ols} and \ref{table:app_lpm} for the full set of omitted coefficients). The coefficient on fraction black is stable when adding in these additional covariates. Column 4 adds state fixed effects and the coefficient on fraction black only slightly decreases, suggesting that racial disparities in voting wait times are just as strong within state as they are between state.

In column 5, we present the results within county. We find that the disparity is mitigated, but it continues to be large and statistically significant. This suggests that there are racial disparities occurring both within and between county. Understanding the level at which discrimination occurs (state, county, within-county, etc.) is helpful when thinking about the mechanism. Further, the fact that we find evidence of racial disparities within county allows us to rule out what one may consider spurious explanations such as differences in ballot length between counties that could create backlogs at other points of service (\citealt{Pettigrew2017,Edelstein2010,Gross2013}). 

Panel B of Table \ref{table:main} is analogous to Panel A, but changes the outcome to a binary variable indicating a wait time longer than 30 minutes. We choose a threshold of 30 minutes as this was the standard used by the Presidential Commission on Election Administration in their 2014 report, which concluded that, ``as a general rule, no voter should have to wait more than half an hour in order to have an opportunity to vote'' (\citealt{Bauer2014}). We find that entirely black areas are 12 percentage points more likely to wait more than 30 minutes than entirely white areas, a 74\% increase in that likelihood. This remains at 10 percentage points with polling-area controls and 7 percentage points within county.  

\subsection{Robustness}
We have made several data restrictions and assumptions throughout the analysis. In this section, we document the robustness of the racial disparity estimate to using alternative restrictions and assumptions.

\begin{figure}[H]
\begin{center}
\caption{Robustness to Different Data Construction Choices}
\label{fig:robust}
\vspace{-10pt}
\includegraphics[width=.9\linewidth]{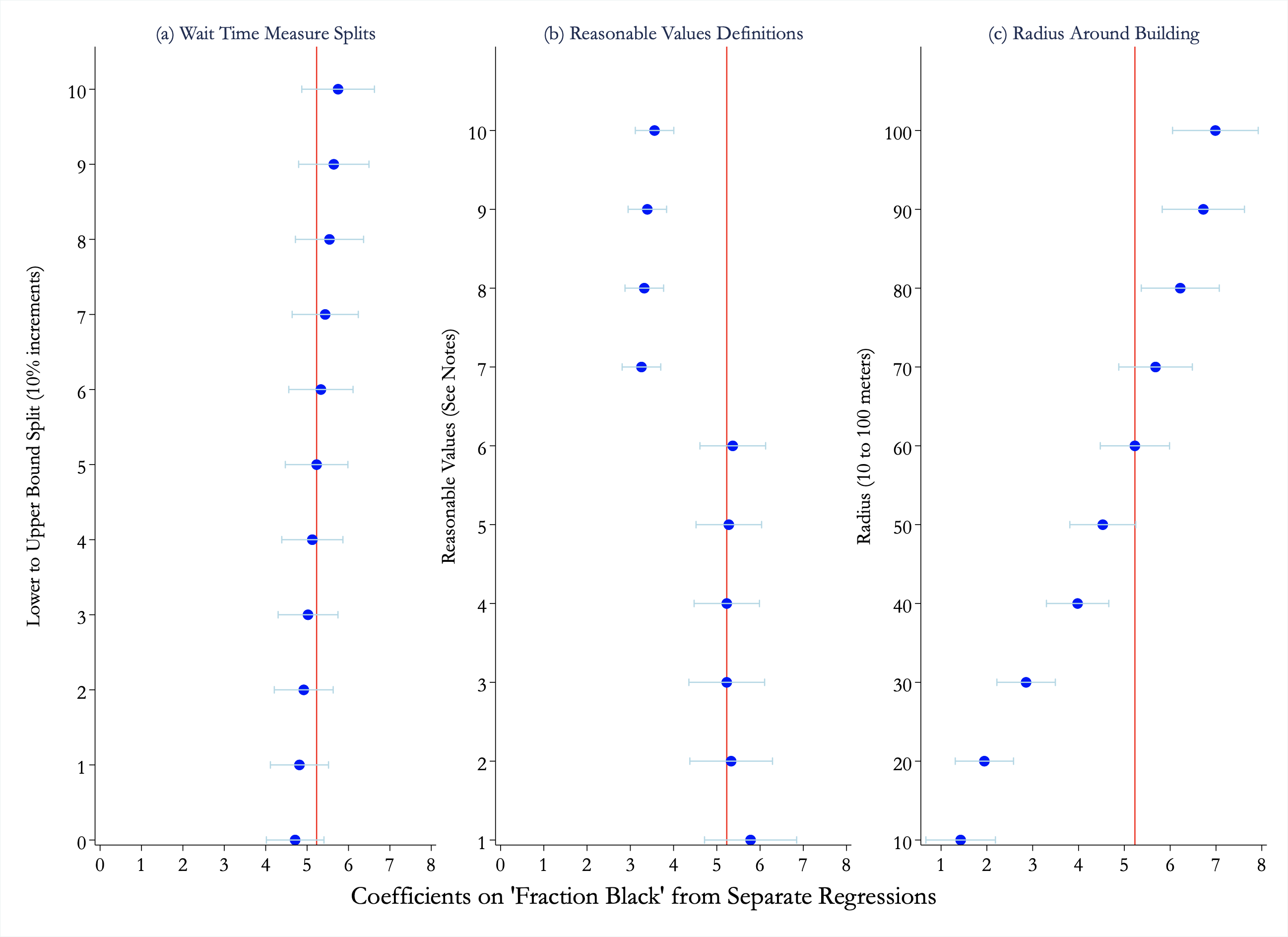}
\caption*{\scriptsize \textbf{Notes}: Points correspond to coefficients on ``Fraction Black'' from separate regressions (+/- 1.96 robust standard errors, clustered at the polling place level). Unit of  observation is a cellphone identifier on Election Day. All specifications are of the form used in Column 1 of Panel A, Table 1. Panel A varies the dependent variable across splits between the lower and upper bounds for our wait time measure (as described in Data and Methods); the first point (y = 0) corresponds to the lower bound, the last point (y = 10) corresponds to the upper bound measure, and all other points are intermediate deciles of the split (e.g. y = 5 corresponds to the midpoint of the two measures). Panel B varies the ``reasonable values'' (RV) filter, as follows: [RV1] Upper Bound under 5 hours (N = 159,046; Mean of Dependent Variable = 22.92) [RV2] Upper Bound under 4 hours (N = 158,167; Mean = 21.79) [RV3] Upper Bound under 3 hours (N = 156,937; Mean = 20.63) [RV4] Upper Bound under 2 hours (N = 154,411; Mean = 19.13) [RV5] Upper Bound under 2 hours and over 1.5 minutes (N = 154,014; Mean = 19.17) [RV6] Upper Bound under 2 hours and over 2 minutes (N = 153,433; Mean = 19.24) [RV7] Upper Bound under 1 hour and over 2 minutes (N = 141,170; Mean = 15.64) [RV8] Upper Bound under 1 hour and over 2.5 minutes (N = 140,470; Mean = 15.71) [RV9] Upper Bound under 1 hour and over 3 minutes (N = 139,788; Mean = 15.78) [RV10] Upper Bound under 1 hour and over 4 minutes (N = 138,452; Mean = 15.91). Panel C varies the bounding radius around the polling station centroid from 10 meters (N = 60,821; Mean = 12.09) up to 100 meters (N = 113,797; Mean = 21.81). The red line on each figure corresponds to the coefficient from the choice we use in our primary analysis, i.e. the midpoint wait time measure (Panel A), a filter of upper bounds under 2 hours (Panel B), and a radius of 60 meters (Panel C).}
\end{center}
\end{figure}

In our primary analysis we use the midpoint between the lower and upper bound of time spent near the polling location as the primary measure of wait time. In Panel A of Figure \ref{fig:robust}, we vary the wait time measure from the lower bound to the upper bound in 10 percent increments, finding that it has little impact on the significance or magnitude of our estimates. We further vary the wait time trimming thresholds in Panel B and the radius around a building centroid used to identify the polling location in Panel C. While these do move the average wait times around, and the corresponding differences, we find that the difference remains significant even across fairly implausible adjustments (e.g. a tight radius of 20 meters around a polling place centroid). We show the associated regression output for this figure in Appendix Table \ref{table:app_robust}. 

Another set of assumptions was in limiting the sample to individuals who (a) spent at least one minute at a polling place, (b) did so at only one polling place on Election Day, and (c) did not spend more than one minute at that polling location in the week before or the week after Election Day. As a robustness check, we make (c) stricter by dropping anyone who visited \textit{any other polling place} on any day in the week before or after Election Day, e.g. we would thus exclude a person who only visited a school polling place on Election Day, but who visited a church (that later serves a polling place) on the prior Sunday. This drops our primary analysis sample from 154,489 voters down to 68,812 voters, but arguably does a better job of eliminating false positives. In Appendix Table \ref{table:app_stricttable} and Appendix Figure \ref{fig:app_strictfigure} we replicate our primary analysis using this more restricted sample and find results that are very similar to our preferred estimates.

As a placebo check, we perform our primary regression analysis using the same sample construction methods on the non-Election days leading up to and after the actual Election Day. Specifically, we repeat the regression used in Table 2, Panel A, Column 1 for each of these days. Appendix Figure \ref{fig:app_placebocoef} shows the coefficients for each date. We find that none of these alternative dates produces a positive coefficient, suggesting that our approach likely identifies a lower bound on the racial gap in wait times. 

As a final robustness/validation, we correlate the racial disparities in wait times that we identify using the smartphone data with the racial disparities in wait times found using the CCES survey (discussed in the previous section). As we found when correlating our overall wait time measure with the CCES, there is a strong correlation at the state level (0.72). The correlation at the congressional district level is much more modest (0.07). \\

\section{Discussion and Conclusion}

Exploiting a large geospatial dataset, we provide new, nationwide estimates for the wait times of voters during the 2016 U.S. presidential election. In addition to describing wait times overall, we document a persistent racial disparity in voting wait times: areas with a higher proportion of black (and to a lesser extent Hispanic) residents are more likely to face long wait times than areas that are predominantly white. These effects survive a host of robustness and placebo tests and are also validated by being strongly correlated with survey data on voter wait times. 

While the primary contribution of our paper is to carefully document voting wait times and disparities at the national level, it is natural to ask why these disparities exist. In the Appendix, we explore the mechanism and do not find conclusive evidence in favor of arrival bunching, partisan bias, early voting, or strict ID laws. We find suggestive evidence that the effects could be driven by fewer resources that leads to congestion especially in high-volume polling places. We are left with the fact that these racial disparities are not limited to just a few states or areas with particular laws or party affiliations that might reflect strategic motivations. Rather, there is work to be done in a diverse set of areas to correct these inequities. A simple explanation is that government officials in general tend to focus more attention on areas with white constituents at the expense of those with black constituents. For example, this could be due to politicians being more responsive to white voters' complaints about voting administration than those from black voters (and relatedly, white voters lodging more complaints), in line with prior work demonstrating lower responsiveness to black constituents across a variety of policy dimensions (e.g. \citealt{Butler2011,Giulietti2019,White2015}).

Our results also demonstrate that smartphone data may be a relatively cheap and effective way to monitor and measure progress in both overall wait times and racial disparities in wait times across various geographic areas. The analysis that we conduct in this paper can be easily replicated after the 2020 election and thereby generate a panel dataset of wait times across areas. Creating a panel dataset across the country may be useful to help pin down the mechanism for disparities (e.g. using difference-in-differences designs to test if disparities in voter wait times change when different laws or election officials take over in a state). We hope that future work can build on the results in this paper to provide a deeper understanding of disparities in voting wait times and their causes.

\newpage
\nocite{*}
\bibliographystyle{ecta}
\bibliography{voterwaittimes.bib}
\newpage

\section*{Appendix A: Figures and Tables}
\appendix
\renewcommand\thefigure{A.\arabic{figure}}    
\setcounter{figure}{0} 
\renewcommand\thetable{A.\arabic{table}}    
\setcounter{table}{0}

\begin{figure}[H]
\begin{center}
\caption{Geographic Coverage}
\label{fig:locations}
\vspace{-5pt}
\includegraphics[scale=.52]{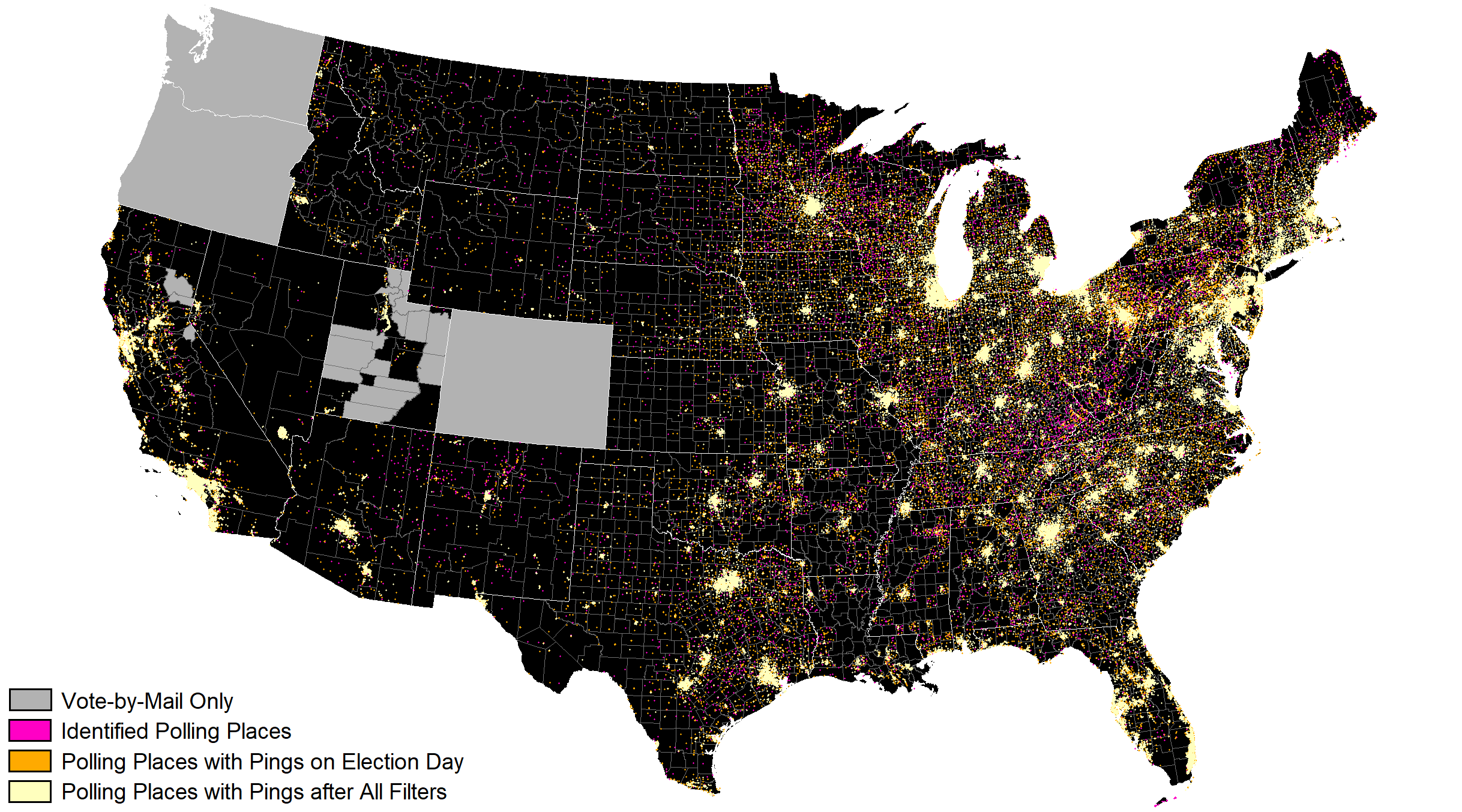}
\caption*{\scriptsize \textbf{Notes}: This figure shows polling place locations (overlaid on county shapes) colored by whether smartphone pings were observed.}
\end{center}
\end{figure}

\begin{figure}[H]
\begin{center}
\caption{Placebo Day Wait Time Histograms}
\label{fig:app_placebodist}
\vspace{-10pt}
\includegraphics[width=.31\linewidth]{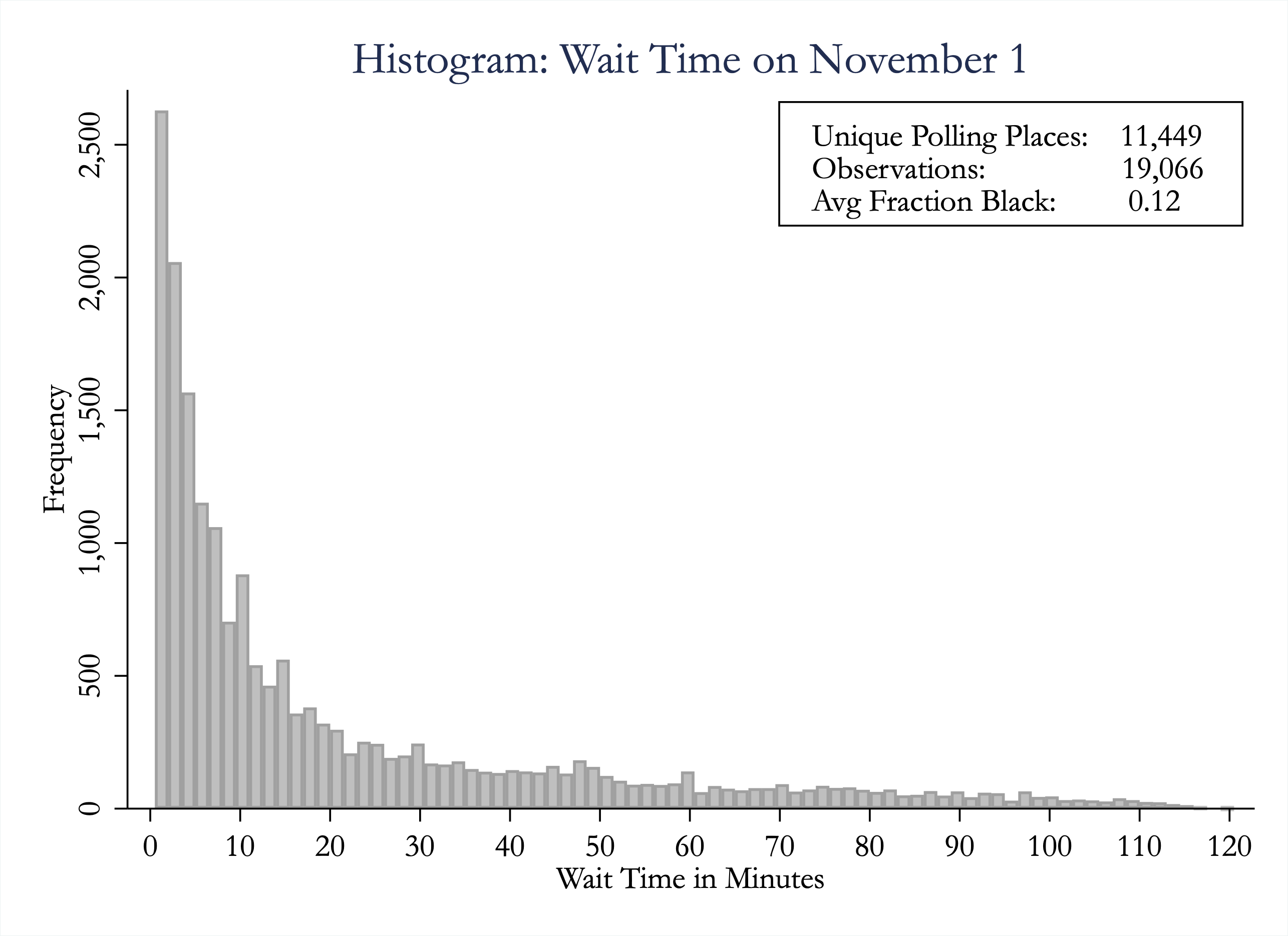}
\includegraphics[width=.31\linewidth]{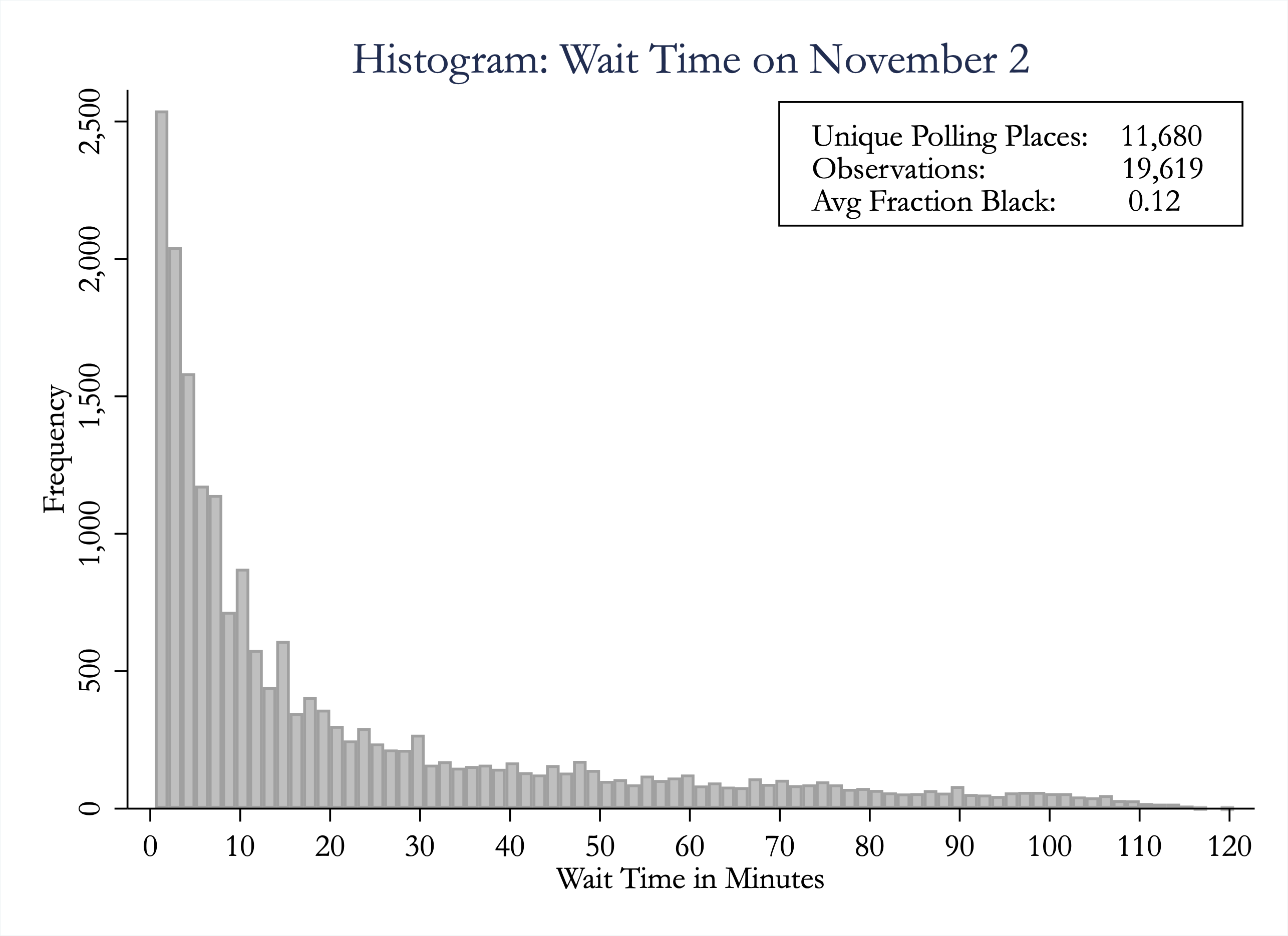} 
\includegraphics[width=.31\linewidth]{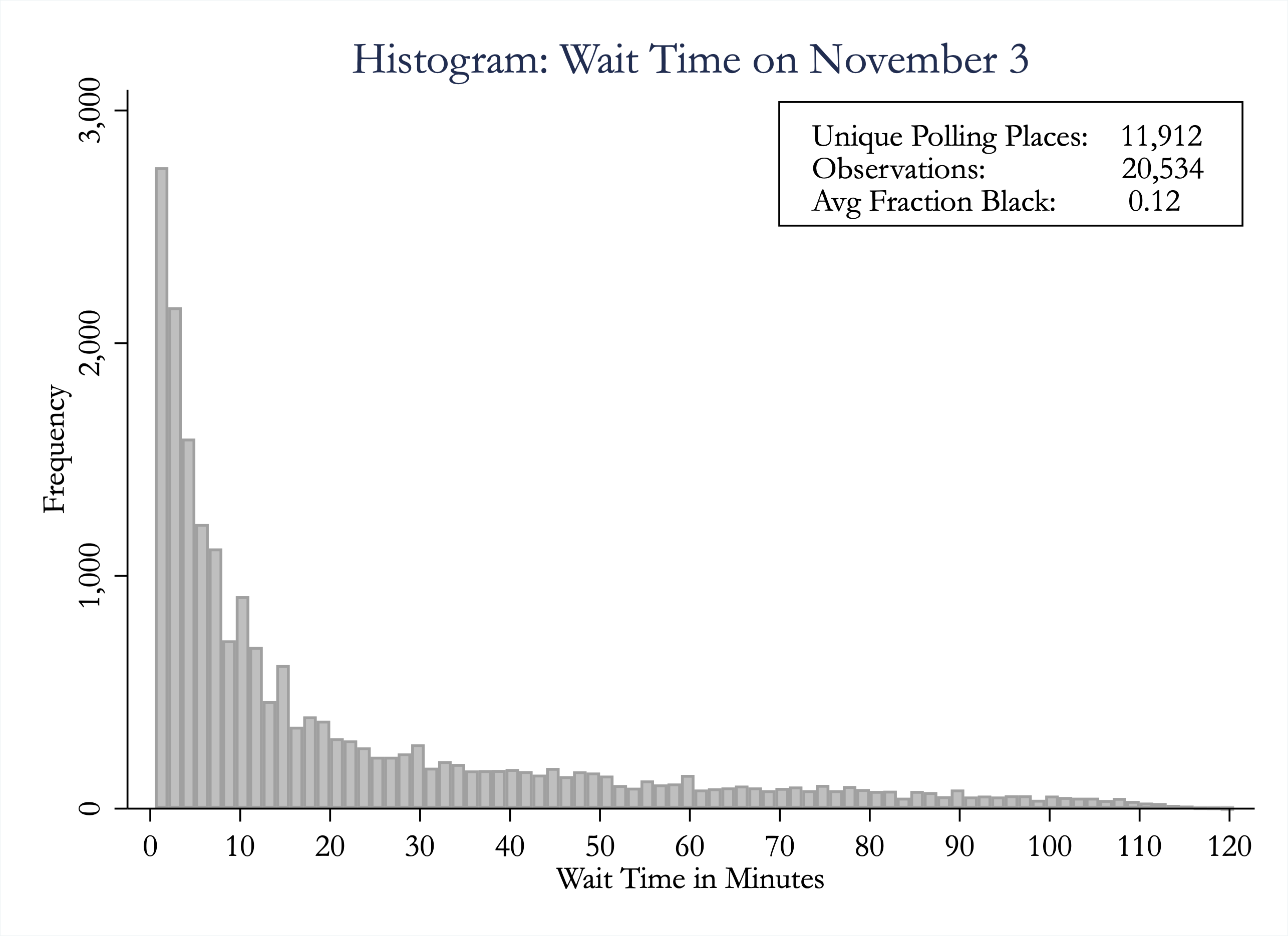} \\
\includegraphics[width=.31\linewidth]{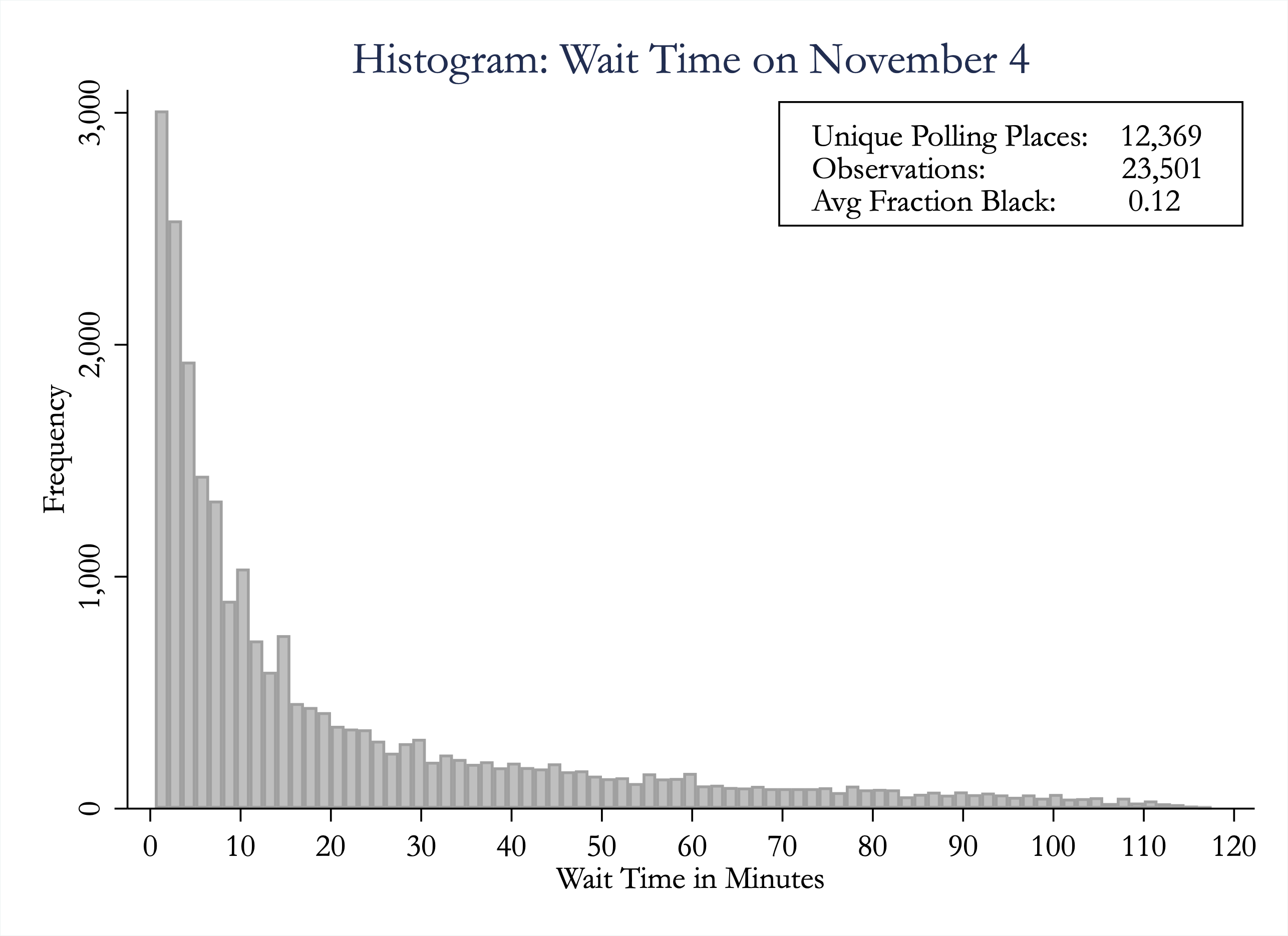}
\includegraphics[width=.31\linewidth]{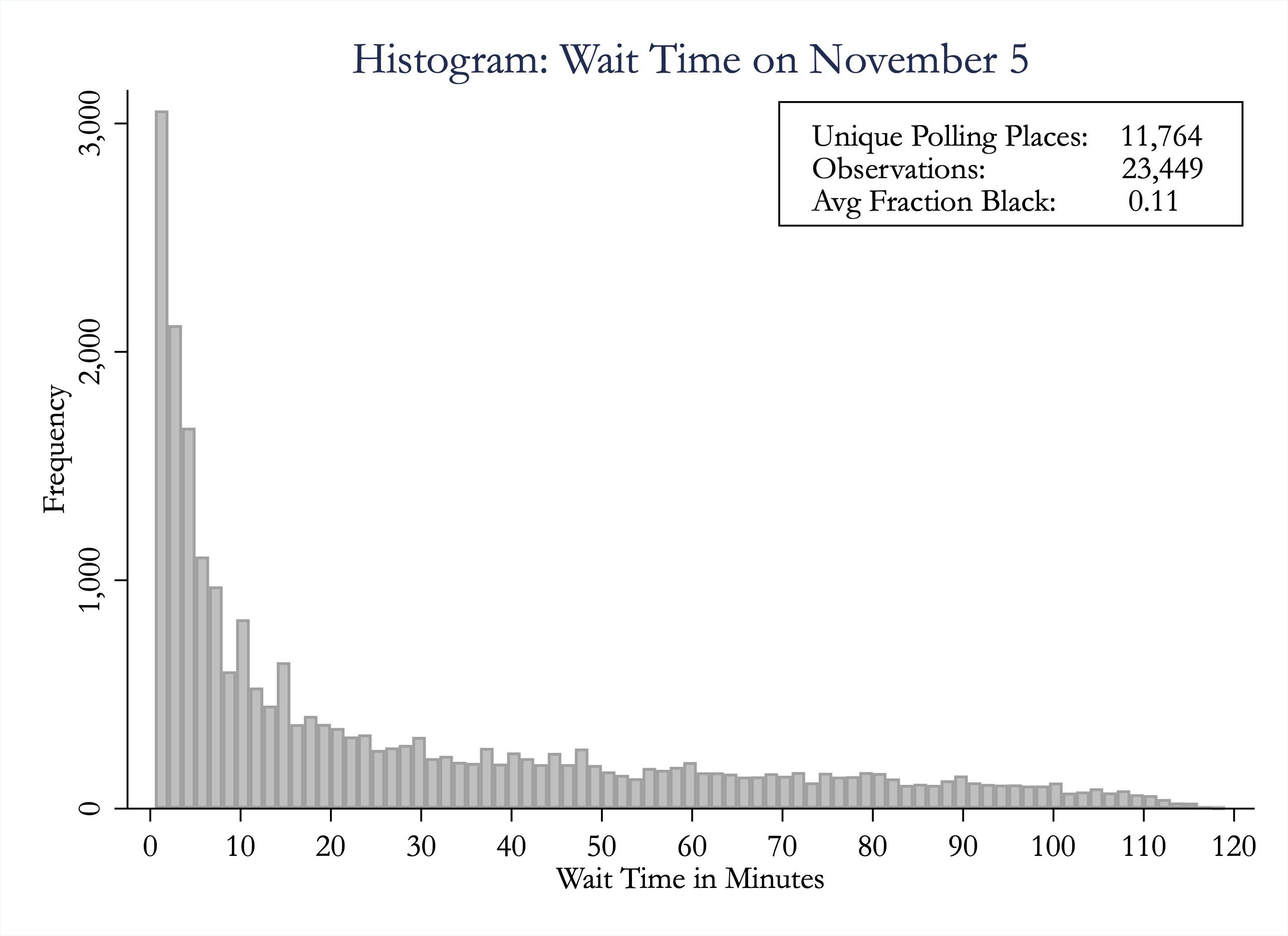} 
\includegraphics[width=.31\linewidth]{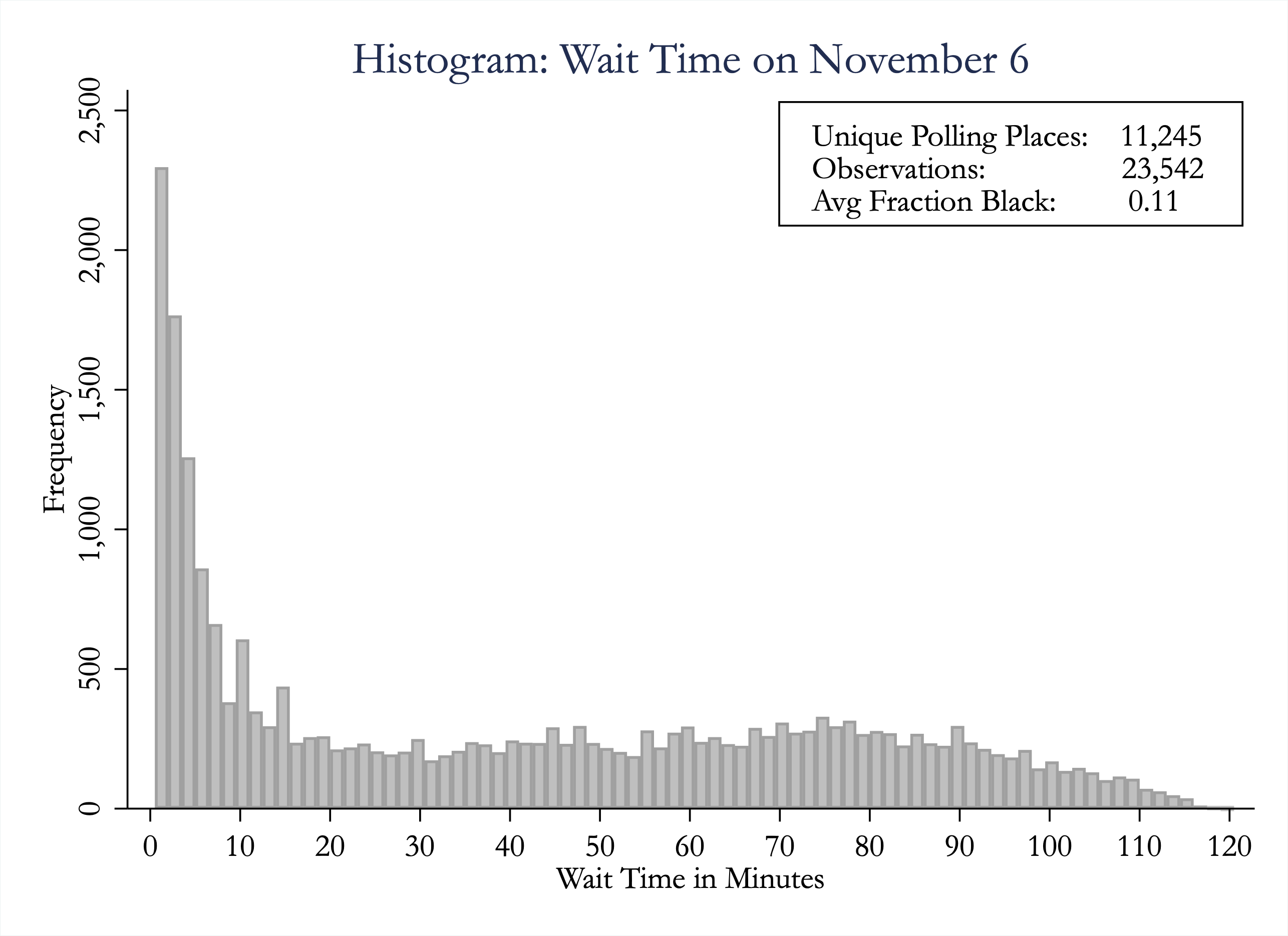} \\
\includegraphics[width=.31\linewidth]{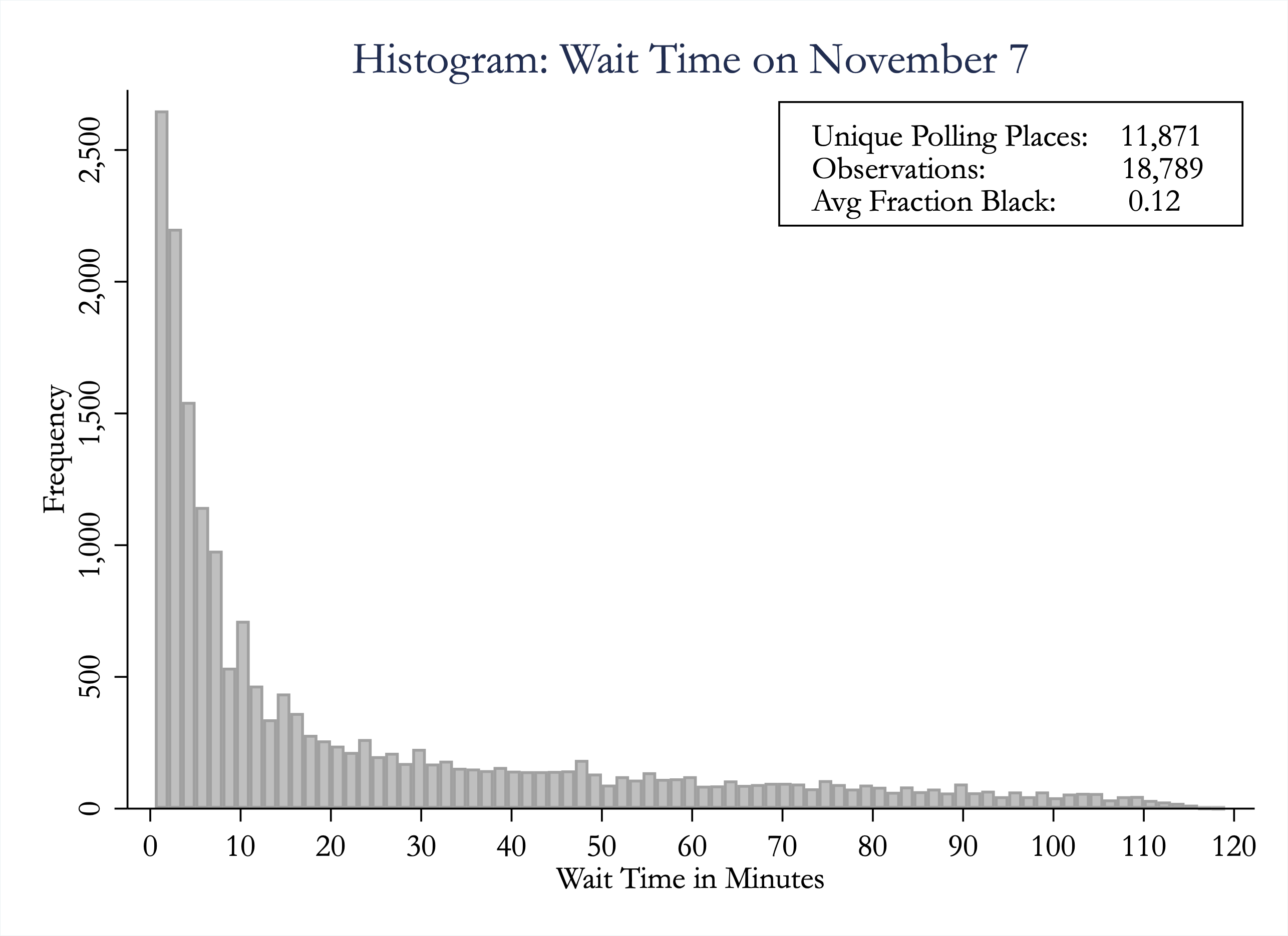}
\includegraphics[width=.31\linewidth]{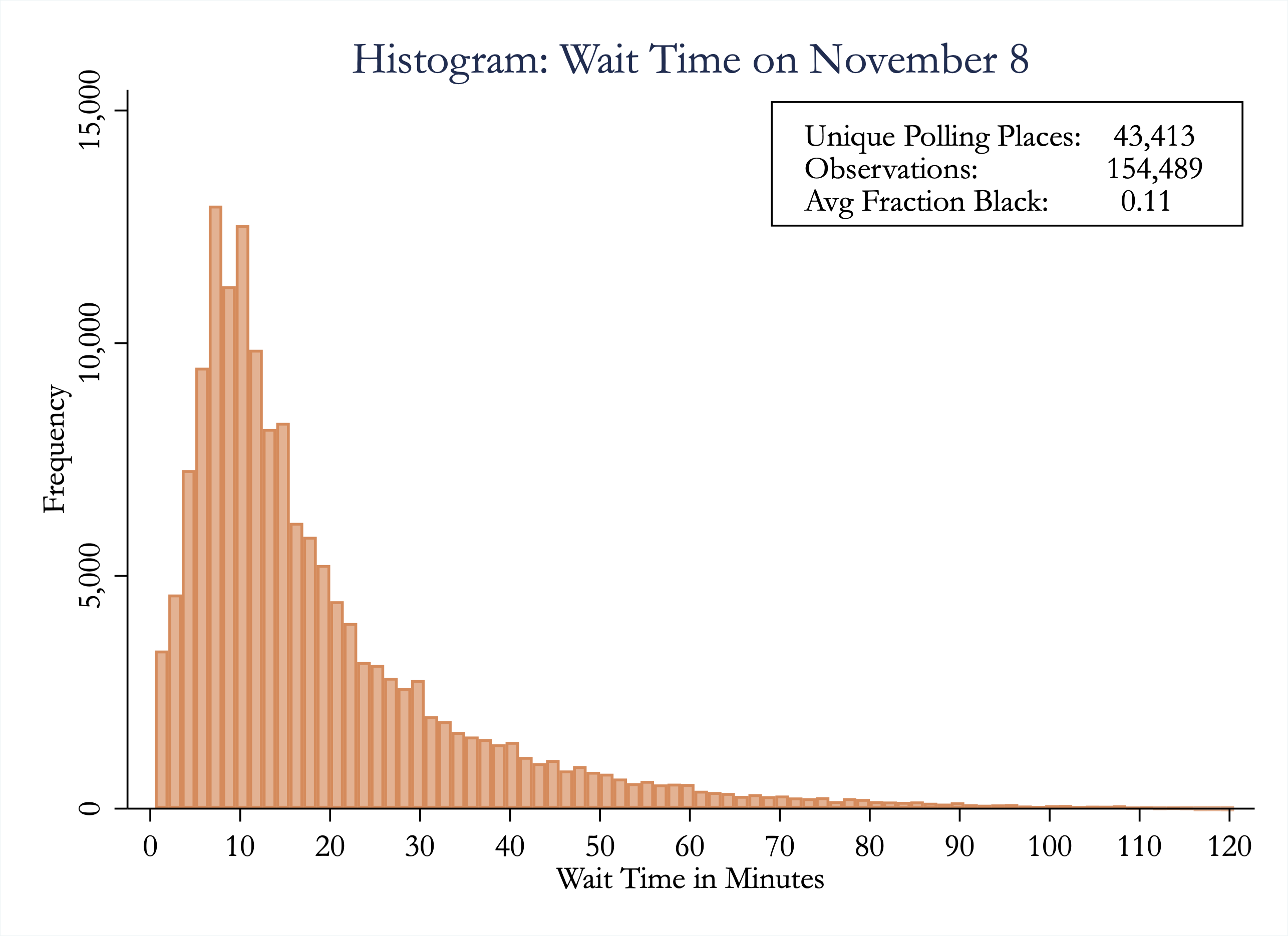} 
\includegraphics[width=.31\linewidth]{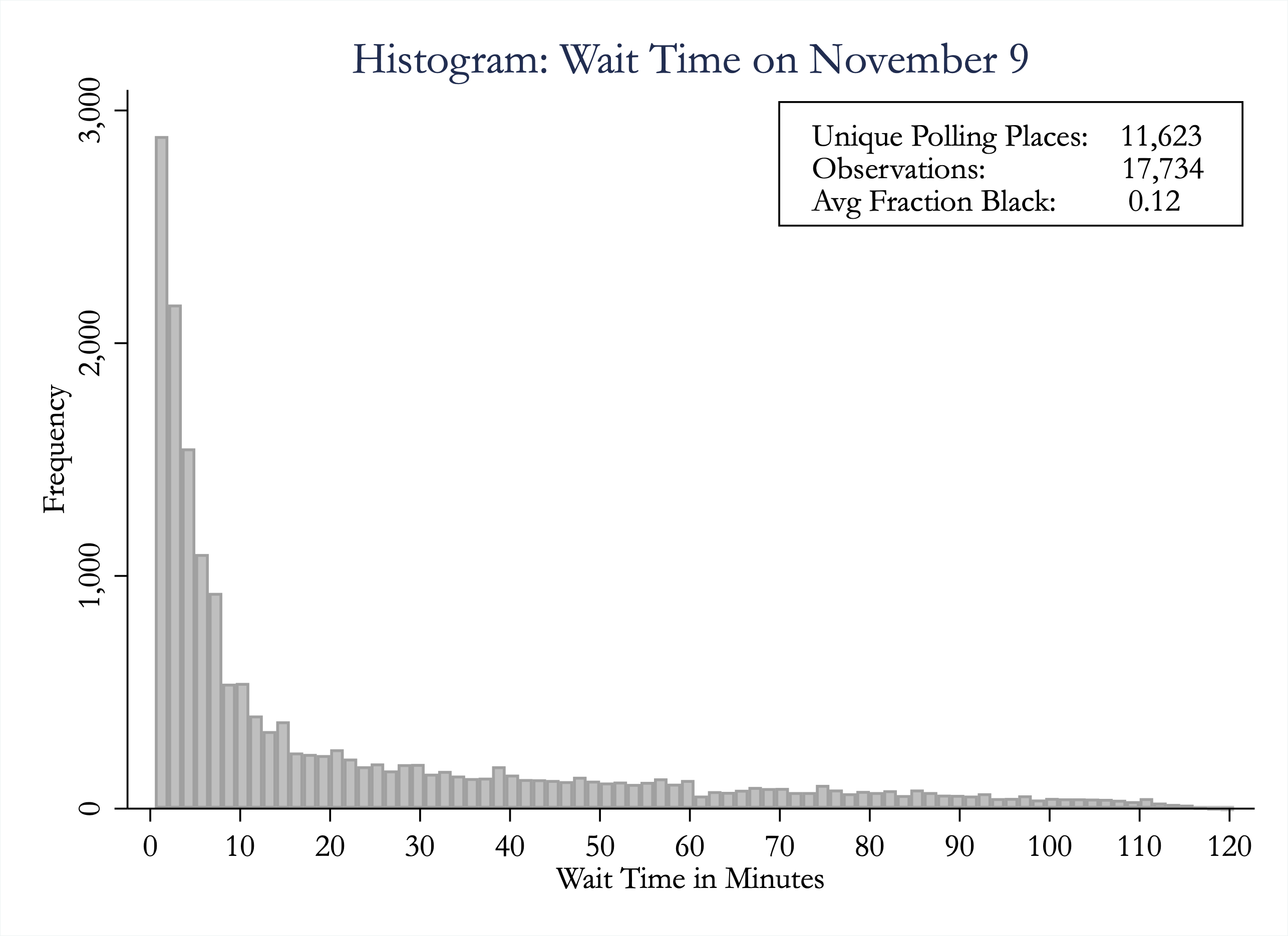} \\
\includegraphics[width=.31\linewidth]{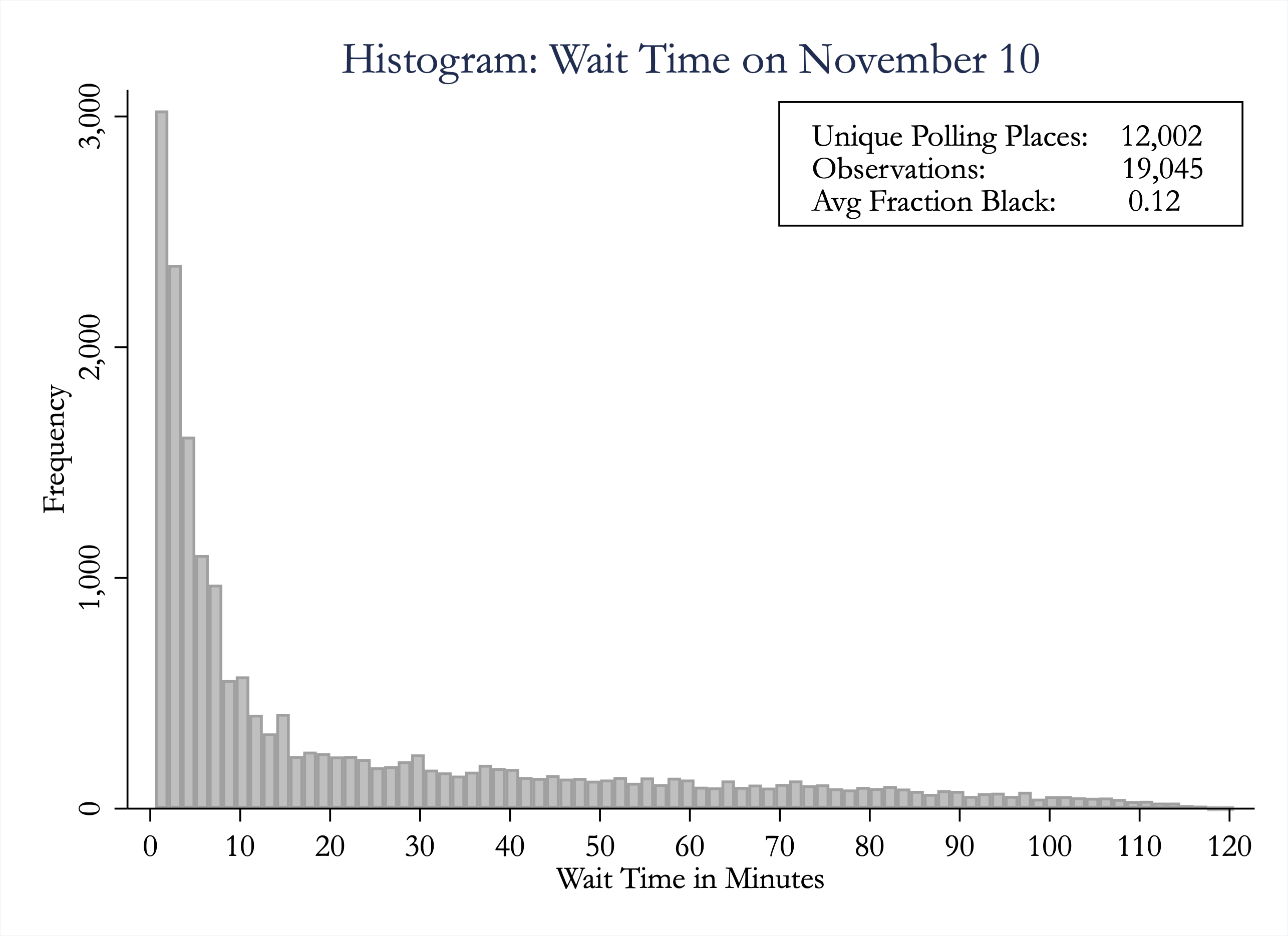}
\includegraphics[width=.31\linewidth]{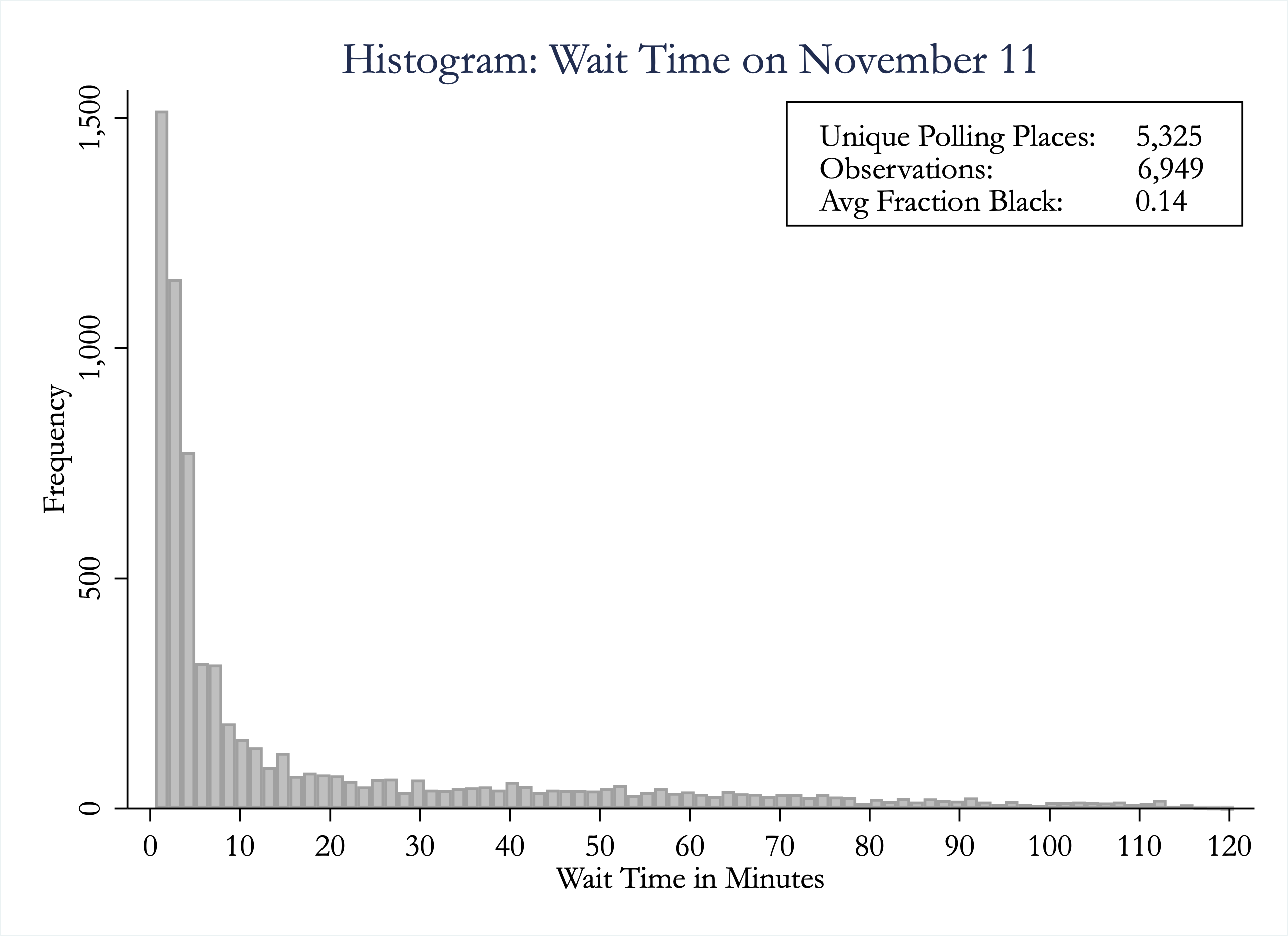} 
\includegraphics[width=.31\linewidth]{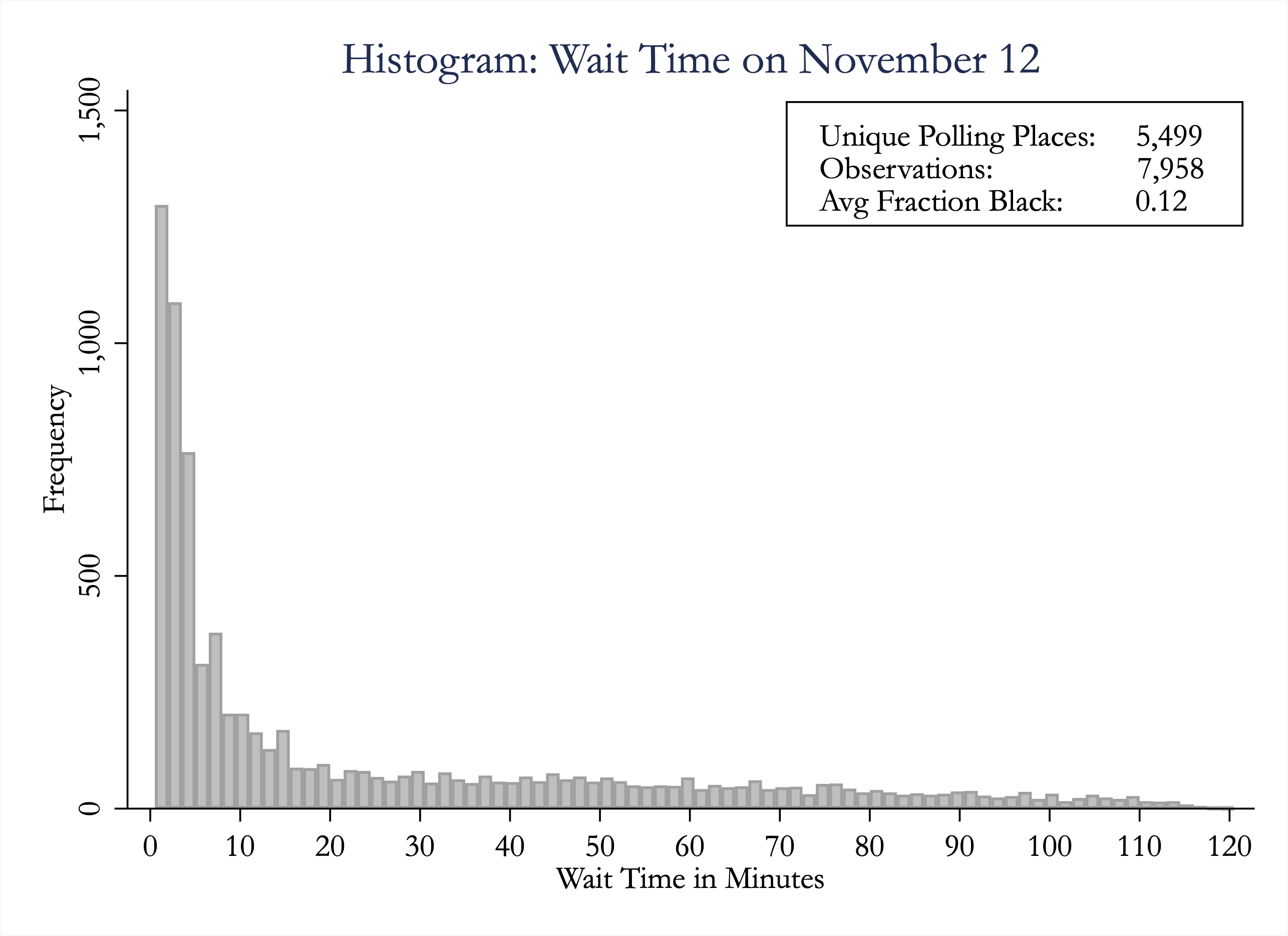} \\
\includegraphics[width=.31\linewidth]{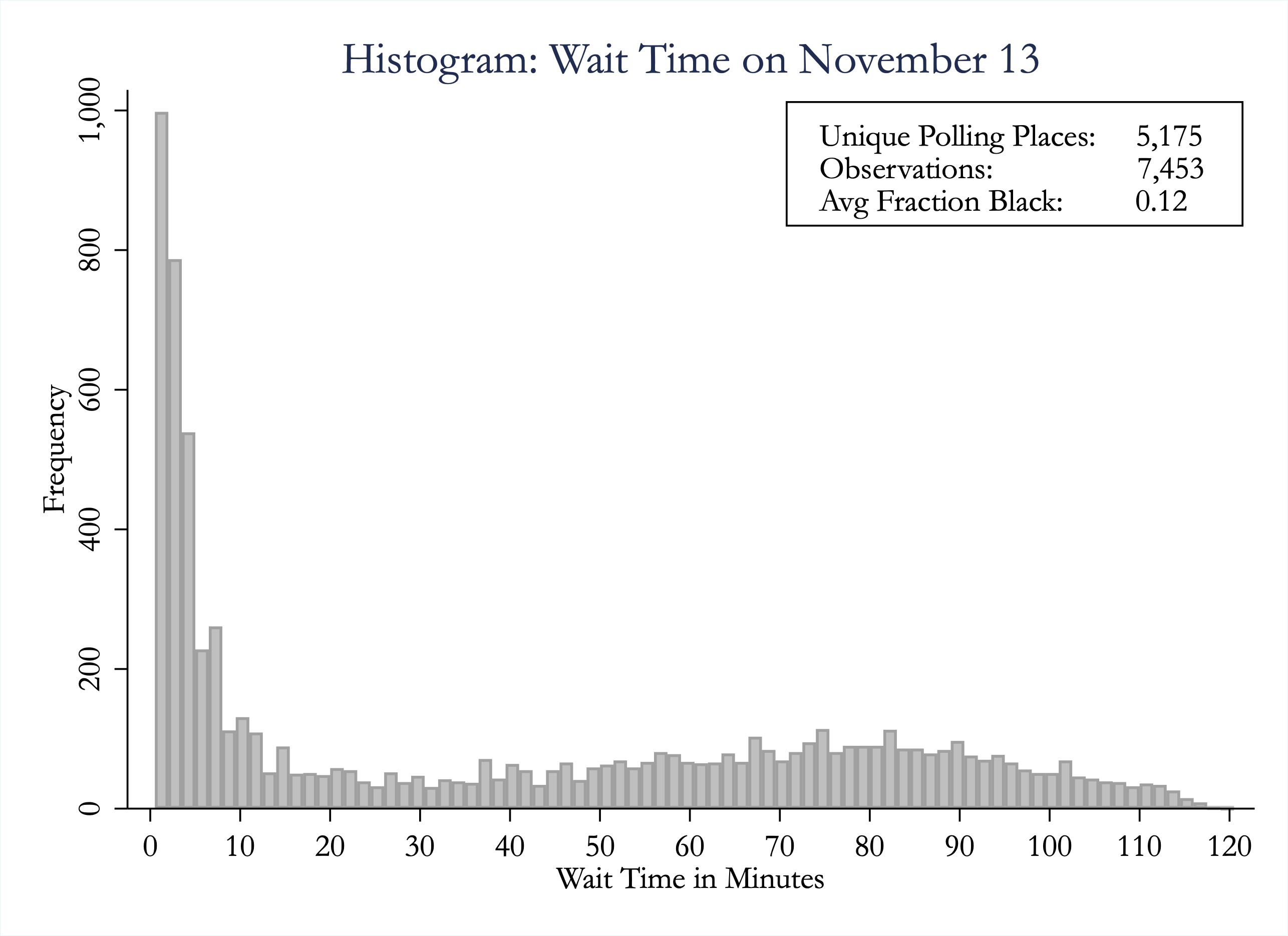}
\includegraphics[width=.31\linewidth]{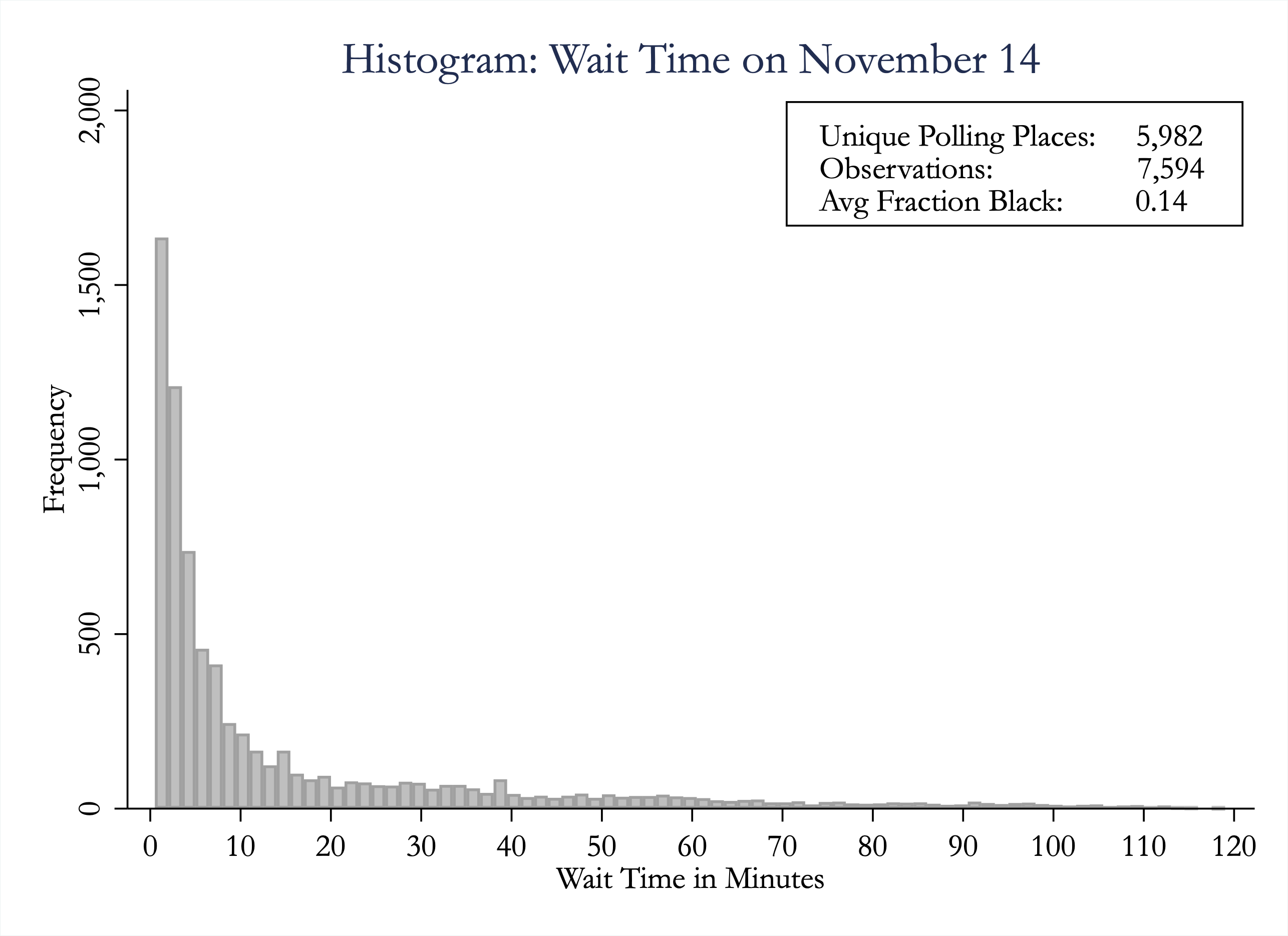} 
\includegraphics[width=.31\linewidth]{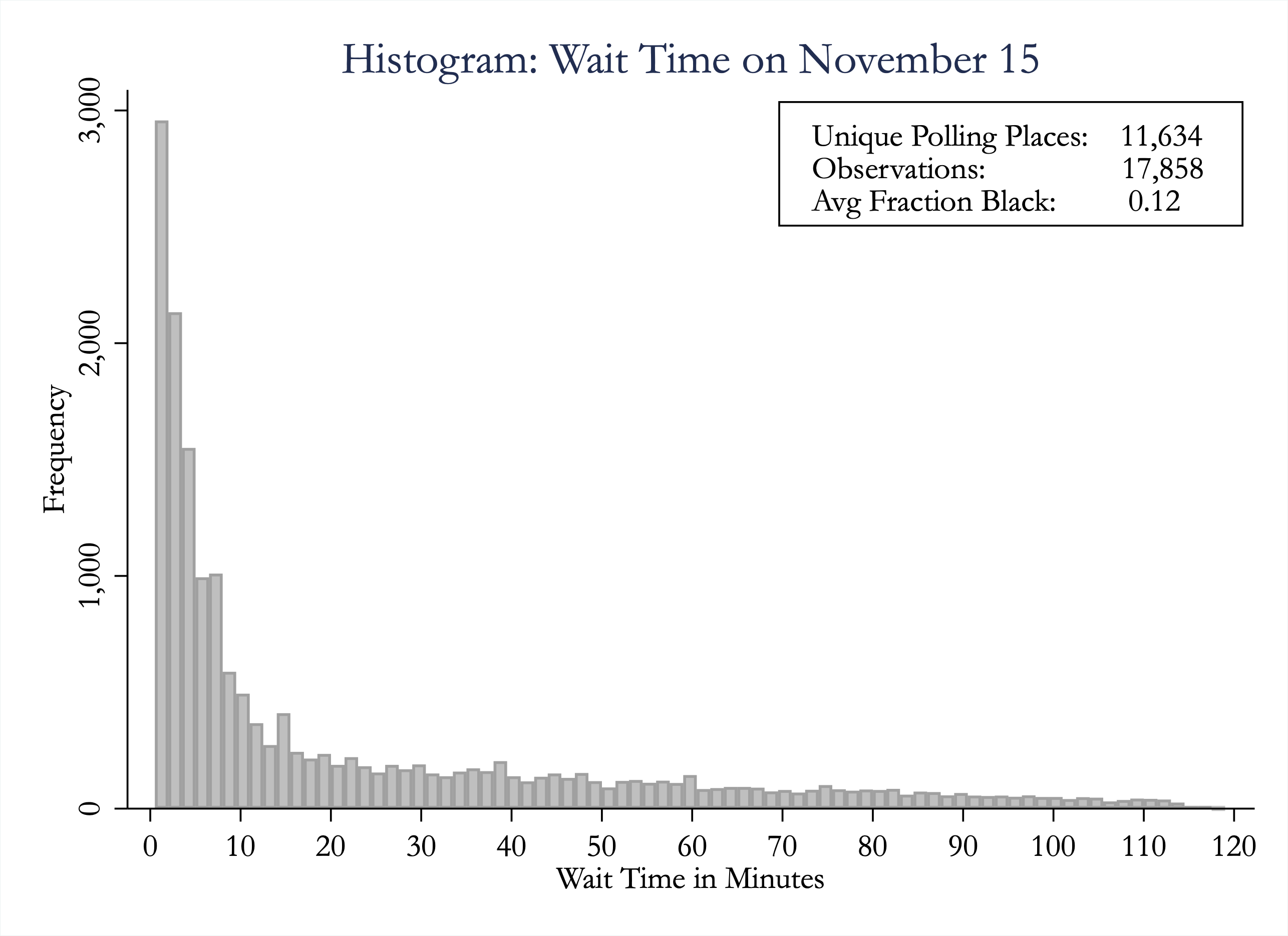} \\
\caption*{\scriptsize \textbf{Notes}: In this figure, we replicate our sample construction across 14 placebo days (i.e. we apply our filters to identifying a ``likely voter'' but replace the sample and the date used in each filter definition to the placebo date). The figure corresponding to Election Day (i.e. Figure \ref{fig:overallwait} of the paper) is also shown, highlighted in orange. The figure illustrates that our filters identify a plausible distribution of wait times on Election Day, but that applying the same set of filters (with dates shifted accordingly) produces a very different distribution shape on other dates. Note that the Y-axes change across sub-figures.}
\end{center}
\vspace{-10pt}
\end{figure}

\begin{table}[H]
\begin{center}
\caption{Summary Statistics for Voter Wait Time Measures}
\label{table:sumstats}
\vspace{-10pt}
\scalebox{.85}{{
\def\sym#1{\ifmmode^{#1}\else\(^{#1}\)\fi}
\begin{tabular}{l*{1}{cccccccc}}
\hline\hline
&\multicolumn{1}{c}{(1)}&\multicolumn{1}{c}{(2)} &\multicolumn{1}{c}{(3)}&\multicolumn{1}{c}{(4)} &\multicolumn{1}{c}{(5)}&\multicolumn{1}{c}{(6)} &\multicolumn{1}{c}{(7)}&\multicolumn{1}{c}{(8)} \\        &N&Mean&SD&Min&p10&Median&p90&Max \\
\hline
\textbf{Wait Time Measures}&            &            &            &            &            &            &            &            \\
Primary Wait Time Measure (Midpoint)&     154,489&       19.13&       16.89&        0.51&        5.02&       13.57&       40.83&      119.50\\
Lower Bound Wait Time Measure&     154,489&       11.26&       16.19&        0.00&        0.00&        5.52&       30.62&      119.08\\
Upper Bound Wait Time Measure&     154,489&       27.00&       20.33&        1.02&        9.28&       20.30&       54.52&      119.98\\
Wait Time Is Over 30min&     154,489&        0.18&        0.38&        0.00&        0.00&        0.00&        1.00&        1.00\\
\textbf{Race Fractions in Polling Area}&            &            &            &            &            &            &            &            \\
Fraction White      &     154,411&        0.70&        0.26&        0.00&        0.27&        0.79&        0.96&        1.00\\
Fraction Black      &     154,411&        0.11&        0.18&        0.00&        0.00&        0.03&        0.31&        1.00\\
Fraction Asian      &     154,411&        0.05&        0.09&        0.00&        0.00&        0.02&        0.14&        0.96\\
Fraction Hispanic   &     154,411&        0.11&        0.17&        0.00&        0.00&        0.05&        0.31&        1.00\\
Fraction Other Non-White&     154,411&        0.03&        0.04&        0.00&        0.00&        0.02&        0.07&        0.99\\
\textbf{Other Demographics}&            &            &            &            &            &            &            &            \\
Fraction Below Poverty Line&     154,260&        0.11&        0.12&        0.00&        0.01&        0.07&        0.26&        1.00\\
Population (1000s)  &     154,489&        2.12&        1.87&        0.00&        0.84&        1.71&        3.56&       51.87\\
Population Per Sq Mile (1000s)&     154,489&        3.81&        9.44&        0.00&        0.20&        1.99&        7.04&      338.94\\
\hline\hline
\end{tabular}
}
}
\end{center}
\vspace{-20pt}
\caption*{\scriptsize \textit{Notes}: Race fractions and other demographics are defined at the Census block group of the associated polling place. These demographics correspond to the 2017 American Community Survey's five-year estimates.}
\end{table}

\begin{figure}[H]
\begin{center}
\caption{Voter Volume by Hour of Day (Early vs. Late Open and Close States)}
\label{fig:openclose}
\vspace{-5pt}
\includegraphics[width=.47\linewidth]{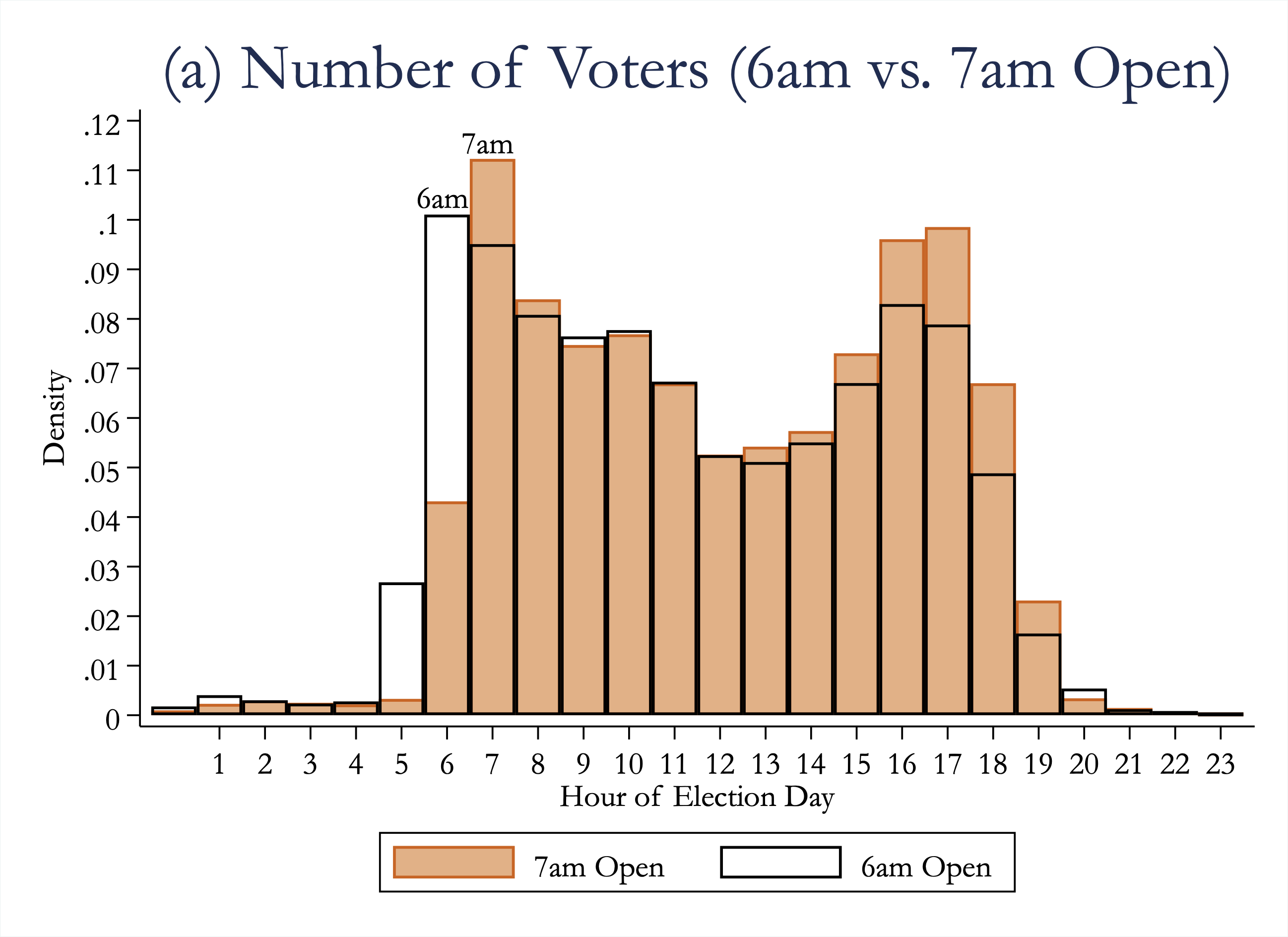}
\includegraphics[width=.47\linewidth]{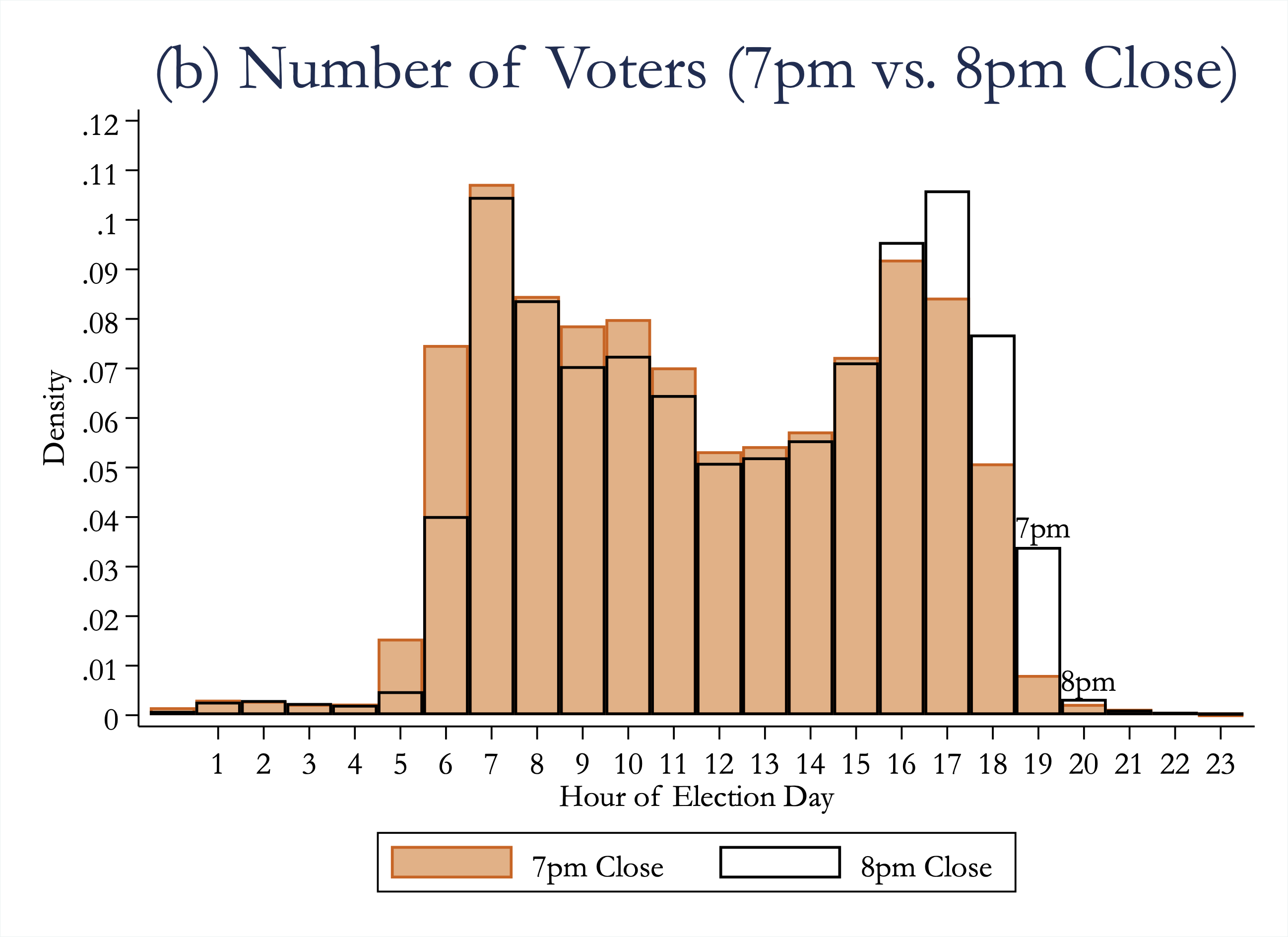}
\vspace{-10pt}
\caption*{\scriptsize \textbf{Notes}: In this figure, we use state poll opening and close times to further validate our filters as identifying likely voters. Panel A separately plots the histogram for the 10 states where polls open at 6am and the 22 that open at 7am; Panel B plots the histograms for the 17 states that close at 7pm versus the 18 states that close at 8pm. We see relative spikes at 7am for the states that open at 7am (orange histogram), and that the number of voters falls substantially at 7pm for states that close at 7pm (orange histogram). [State open and close times are taken from: \url{https://ballotpedia.org/State_Poll_Opening_and_Closing_Times_(2016)\#table}. We omit states which do not have standardized open (Panel A) or close times (Panel B) across the entire state.]}
\end{center}
\end{figure}

\begin{figure}[H]
\begin{center}
\caption{Comparison with CCES Data}
\label{fig:cces}
\vspace{-5pt}
\includegraphics[width=1\linewidth]{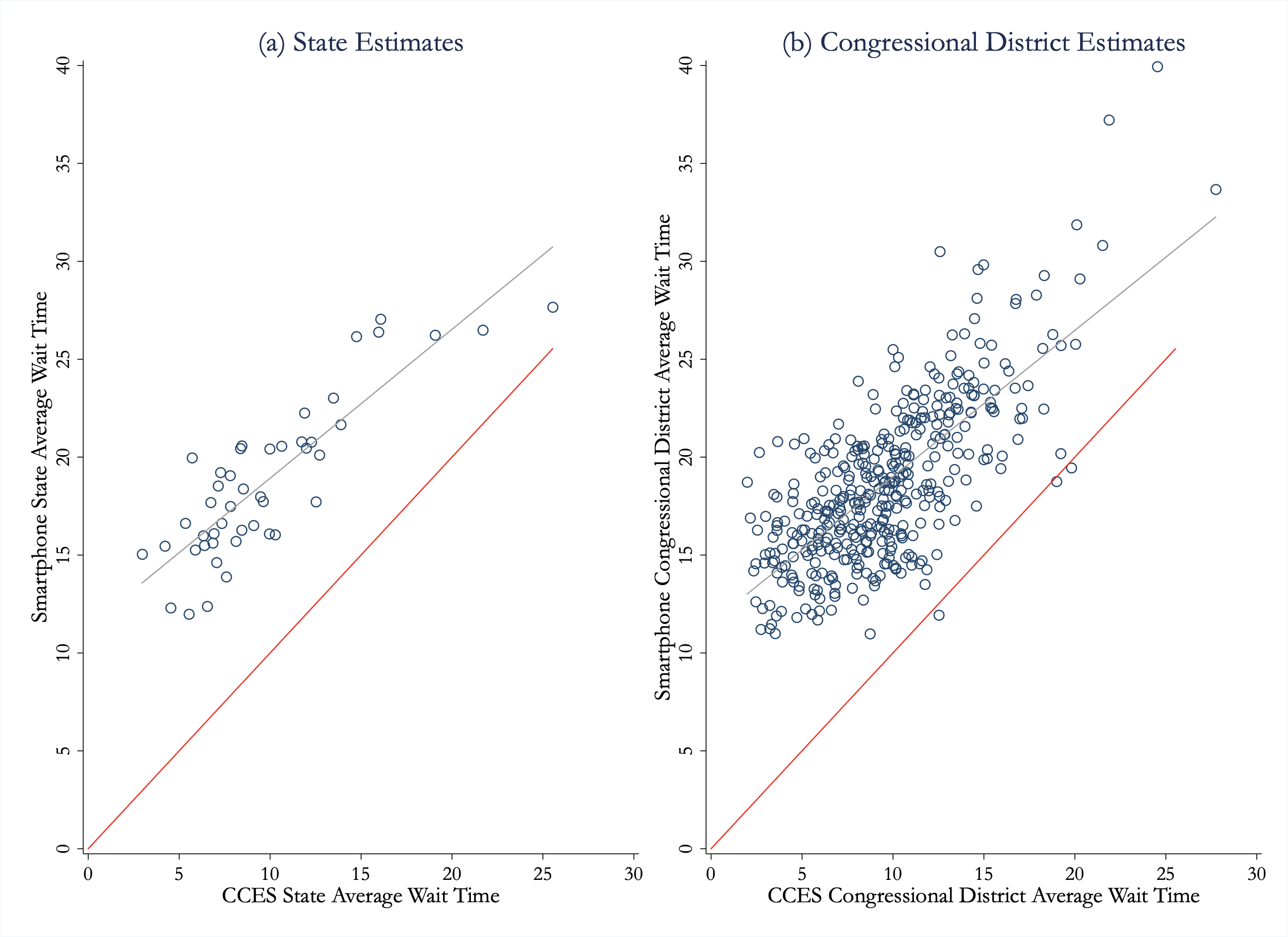}
\vspace{-10pt}
\caption*{\scriptsize \textbf{Notes}: The red line corresponds to the 45 degree line (lining up would indicate equality between the two measures). The gray line is produced with \texttt{lfit} in Stata, giving the prediction of the Smartphone measure given the CCES measure. Both measures are first independently empirical-Bayes-adjusted to account for measurement error.}
\end{center}
\end{figure}

\begin{figure}[H]
\begin{center}
\caption{Wait Time Disparities by Racial Categories}
\label{fig:app_otherraces}
\vspace{-5pt}
\includegraphics[width=1\linewidth]{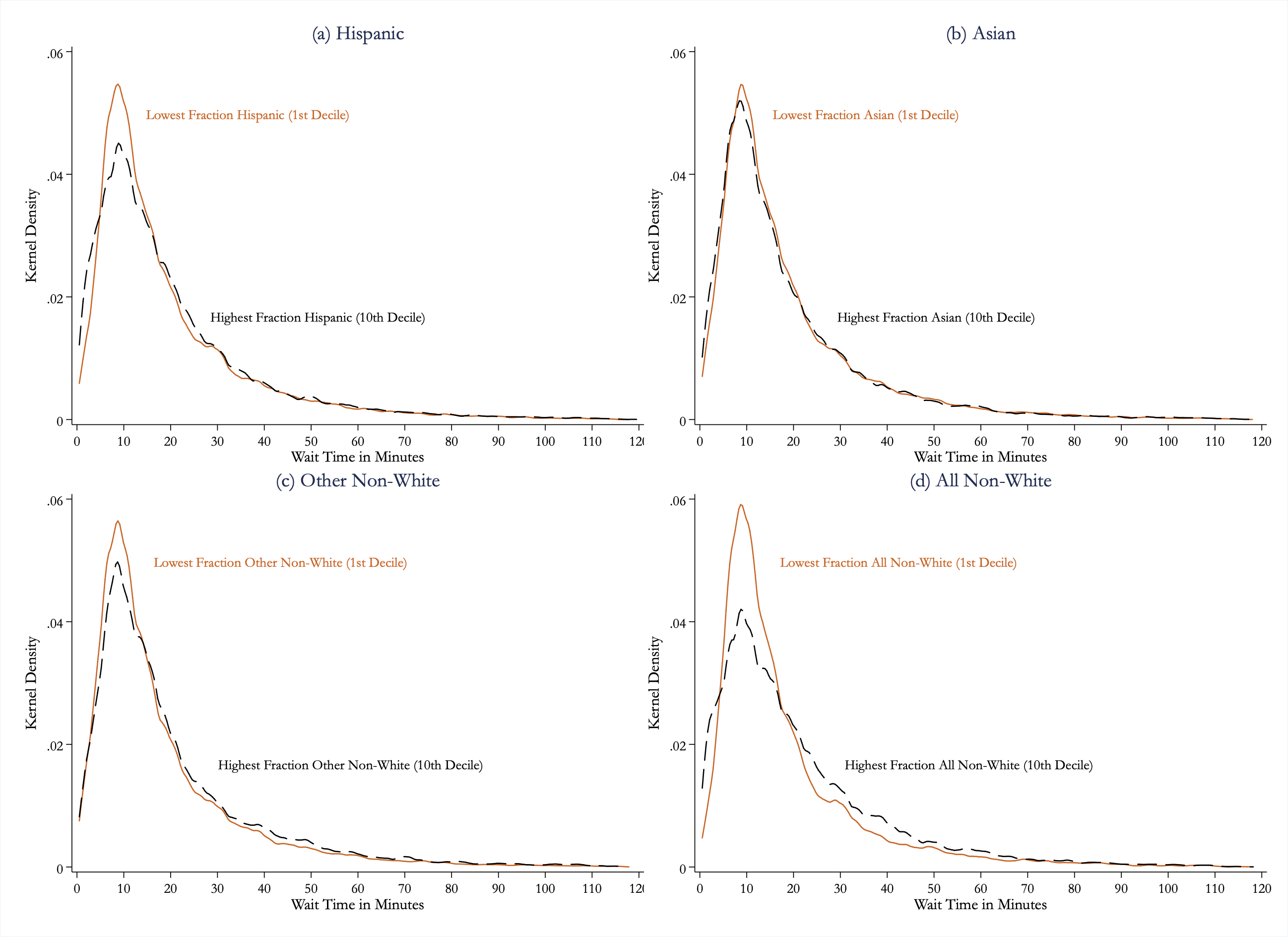}
\caption*{\scriptsize \textbf{Notes}: This figure repeats Figure \ref{fig:disparity} across other racial categories. We show the decile splits by Hispanic (Panel A), Asian (Panel B), and ``Other Non-White'' (Panel C), and then group these categories together with Black in Panel D. Note that ``Asian'' includes ``Pacific Islander.'' ``Other Non-White'' includes the ``Other,'' ``Native American,'' and ``Multiracial'' Census race categories. ``All Non-White'' includes Black, Hispanic, Asian, and Other Non-White.}
\end{center}
\end{figure}

\begin{figure}[H]
\begin{center}
\caption{Wait Time Disparities by Fraction Below Poverty Line}
\label{fig:app_poverty}
\vspace{-5pt}
\includegraphics[width=1\linewidth]{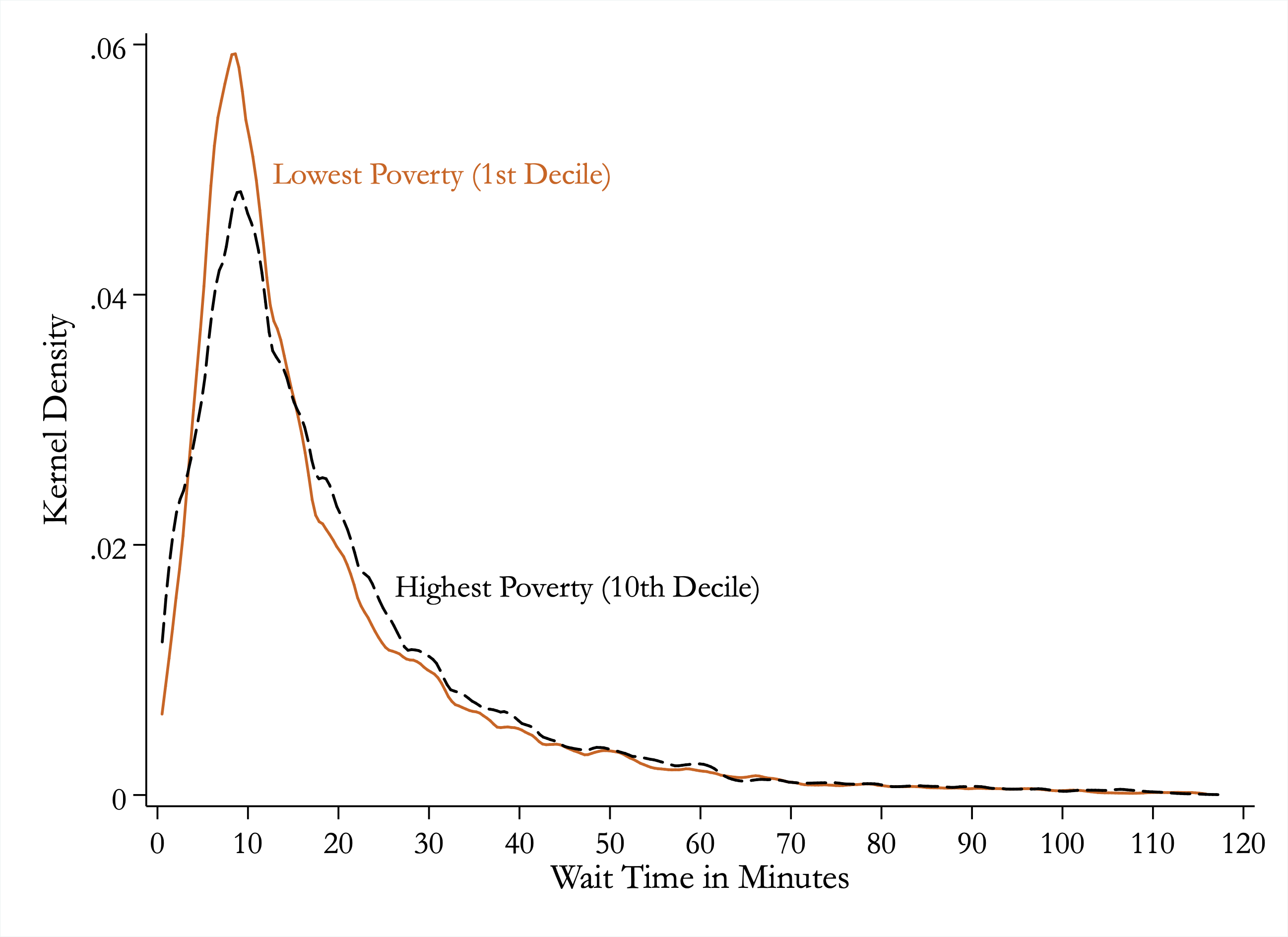}
\caption*{\scriptsize \textbf{Notes}: This figure repeats Figure \ref{fig:disparity} across the ``Fraction Below Poverty Line'' measure (top and bottom deciles).}
\end{center}
\end{figure}

\begin{figure}[H]
\begin{center}
\caption{Wait Time Disparities: Stricter Likely Voter Filter}
\label{fig:app_strictfigure}
\vspace{-5pt}
\includegraphics[width=1\linewidth]{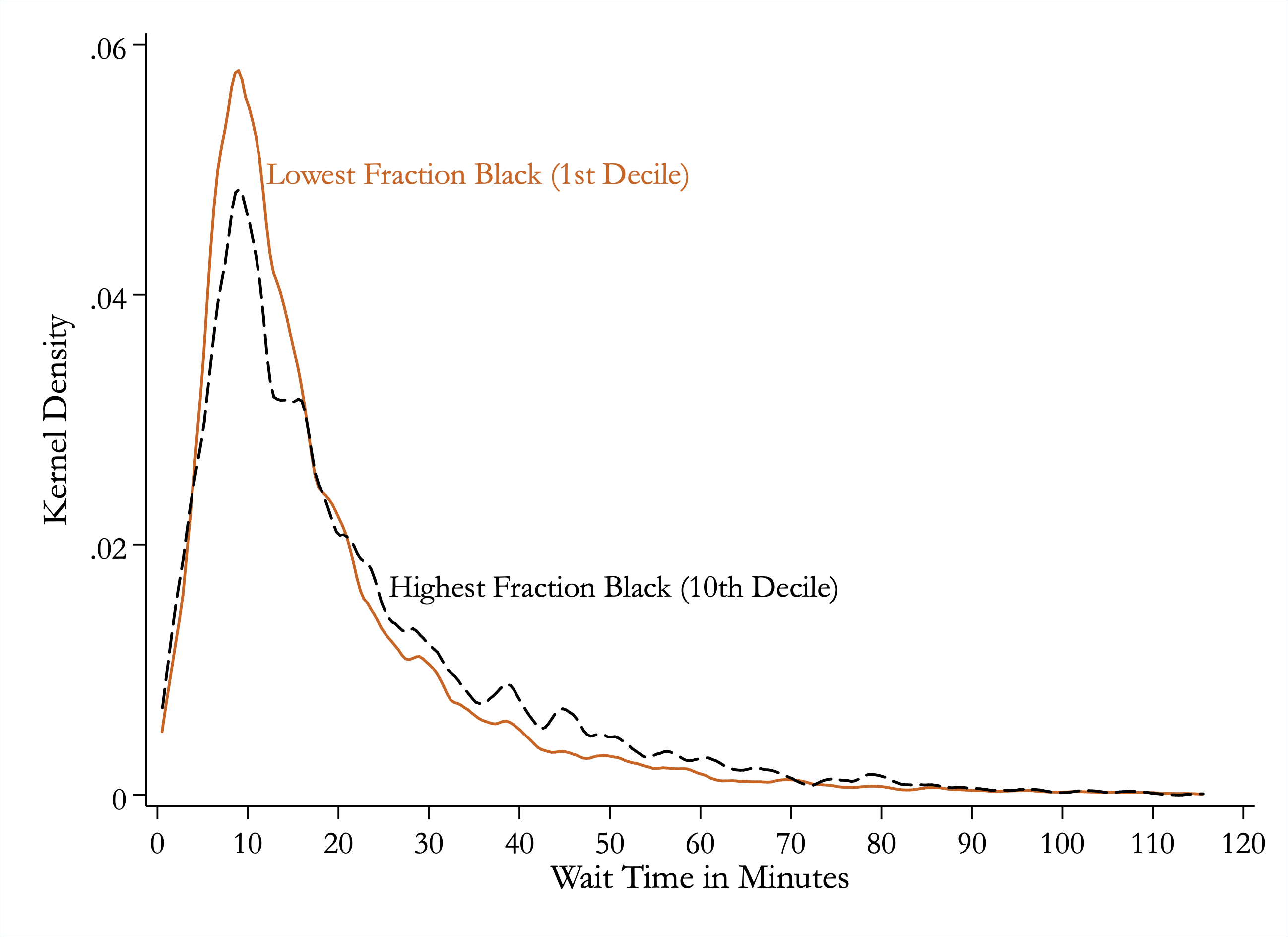}
\vspace{-10pt}
\caption*{\scriptsize \textbf{Notes}: In this figure, we repeat Figure \ref{fig:disparity} with a sub-sample of voters. Specifically, we use a more conservative first filter for identifying ``likely voters.'' Our primary analysis limited the sample to individuals who (a) spent at least one minute at a polling place, (b) did so at only one polling place on Election Day, and (c) did not spend more than one minute at that polling location in the week before or the week after Election Day. Here we make (c) stricter by dropping anyone who visited \textit{any other polling place} on any day in the week before or after Election Day, e.g. we would thus exclude a person who only visited a school polling place on Election Day, but who visited a church (that later serves a polling place) on the prior Sunday. This drops our primary analysis sample from 154,489 voters down to 68,812 voters. Kernel densities are estimated using 1 minute half-widths. The 1st decile corresponds to the 15,402 voters across 6,576 polling places with the lowest percent of black residents (mean = 0\%). The 10th decile corresponds to the 6,881 voters across the 3,229 polling places with the highest percent of black residents (mean = 54\%).}
\end{center}
\end{figure}

\begin{figure}[H]
\begin{center}
\caption{Main Specification Run on Placebo Days}
\label{fig:app_placebocoef}
\vspace{-5pt}
\includegraphics[width=1\linewidth]{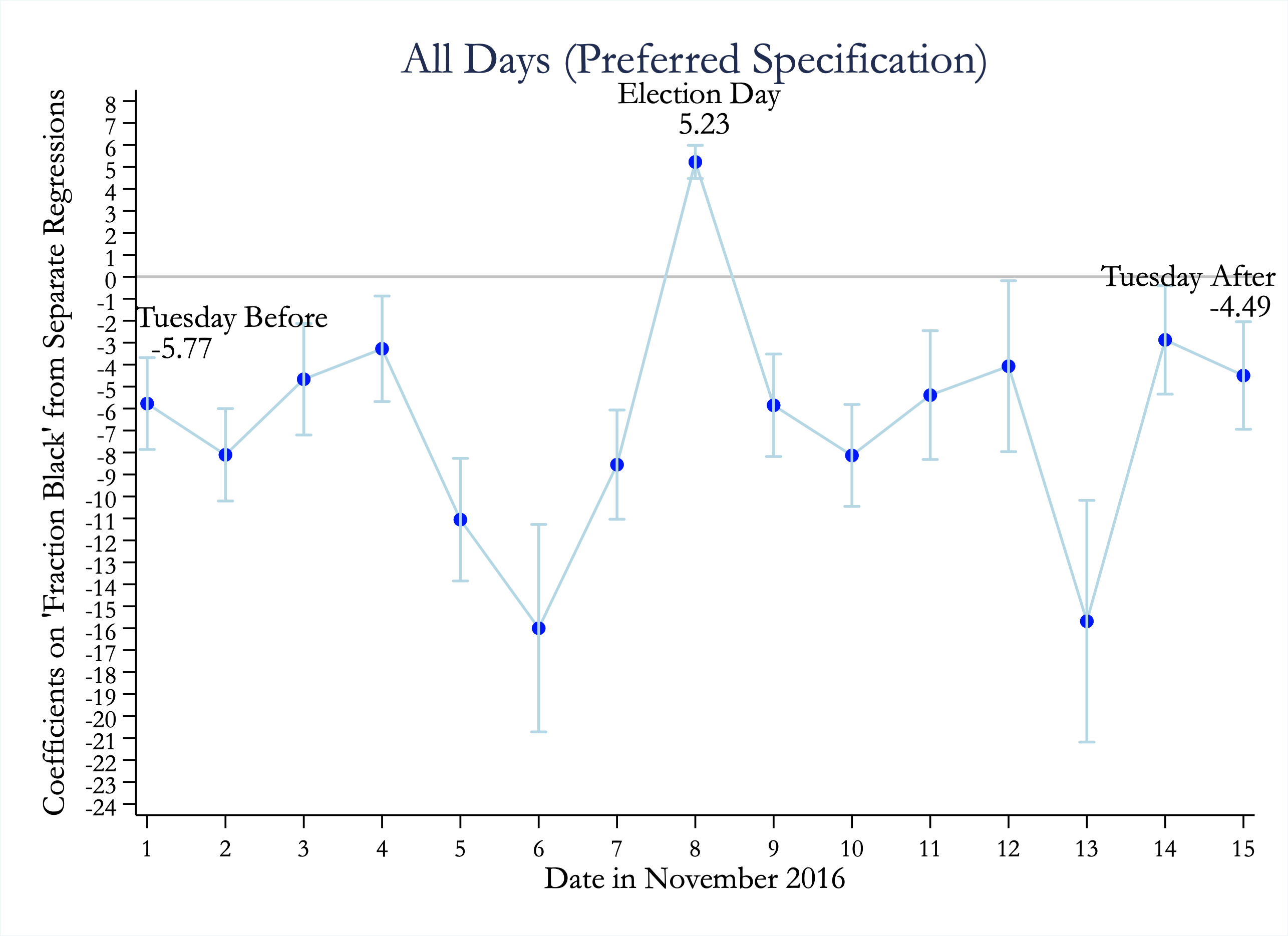}
\vspace{-10pt}
\caption*{\scriptsize \textbf{Notes}: In this figure, we replicate our sample construction for the 14 placebo days around Election Day, similar to \ref{fig:app_placebodist}. We then repeat the regression used in Table \ref{table:main}, Panel A, Column 1 for each of these days. We find that none of these alternative dates produces a positive coefficient, suggesting that our approach likely identifies a lower bound on the racial gap in wait times. \textit{Additional Notes:} Points correspond to coefficients on ``Fraction Black'' (+/- 1.96 standard errors) from separate regressions.}
\end{center}
\end{figure}

\begin{table}[H]
\begin{center}
\caption{Fraction Black and Voter Wait Time: OLS}
\label{table:app_ols}
\vspace{-10pt}
\scalebox{0.95}{{
\def\sym#1{\ifmmode^{#1}\else\(^{#1}\)\fi}
\begin{tabular}{l*{6}{c}}
\toprule
&\multicolumn{1}{c}{(1)}&\multicolumn{1}{c}{(2)}   &\multicolumn{1}{c}{(3)}&\multicolumn{1}{c}{(4)}&\multicolumn{1}{c}{(5)}   &\multicolumn{1}{c}{(6)}\\
\midrule
Fraction Black                &        5.23\sym{***}&        5.22\sym{***}&        4.96\sym{***}&        4.84\sym{***}&        3.27\sym{***}&        3.10\sym{***}\\
                              &      (0.39)         &      (0.39)         &      (0.42)         &      (0.42)         &      (0.45)         &      (0.44)         \\
\addlinespace
Fraction Asian                &                     &       -0.79         &       -2.48\sym{***}&        1.30\sym{*}  &       -1.10         &       -0.66         \\
                              &                     &      (0.72)         &      (0.74)         &      (0.76)         &      (0.81)         &      (0.81)         \\
\addlinespace
Fraction Hispanic             &                     &        1.15\sym{***}&        0.43         &        3.90\sym{***}&        1.50\sym{***}&        1.72\sym{***}\\
                              &                     &      (0.37)         &      (0.40)         &      (0.46)         &      (0.50)         &      (0.50)         \\
\addlinespace
Fraction Other Non-White      &                     &       12.01\sym{***}&       11.76\sym{***}&        1.66         &        2.04         &        1.75         \\
                              &                     &      (1.94)         &      (1.95)         &      (1.89)         &      (1.93)         &      (1.93)         \\
\addlinespace
Fraction Below Poverty Line   &                     &                     &        0.06         &       -2.03\sym{***}&        0.28         &        1.10         \\
                              &                     &                     &      (0.74)         &      (0.71)         &      (0.67)         &      (0.67)         \\
\addlinespace
Population (1000s)            &                     &                     &        0.43\sym{***}&        0.32\sym{***}&        0.28\sym{***}&        0.27\sym{***}\\
                              &                     &                     &      (0.06)         &      (0.05)         &      (0.05)         &      (0.05)         \\
\addlinespace
Population Per Sq Mile (1000s)&                     &                     &        0.04\sym{***}&        0.07\sym{***}&        0.06\sym{***}&        0.06\sym{***}\\
                              &                     &                     &      (0.01)         &      (0.01)         &      (0.01)         &      (0.01)         \\
\addlinespace
Android (0 = iPhone)          &                     &                     &                     &                     &                     &        0.38\sym{***}\\
                              &                     &                     &                     &                     &                     &      (0.10)         \\
\midrule
N                             &     154,411         &     154,411         &     154,260         &     154,260         &     154,260         &     154,260         \\
$R^2$                         &        0.00         &        0.00         &        0.01         &        0.06         &        0.13         &        0.17         \\
DepVarMean                    &       19.13         &       19.13         &       19.12         &       19.12         &       19.12         &       19.12         \\
Polling Area Controls? &No&No&Yes&Yes&Yes&Yes \\
State FE?                       &No&No&No&Yes&Yes&Yes \\
County FE?                &No&No&No&No&Yes&Yes \\
Hour of Day FE?                 &No&No&No&No&No&Yes \\
\hline\hline
\bottomrule
\multicolumn{7}{l}{\footnotesize \sym{*} \(p<0.10\),   \sym{**} \(p<0.05\), \sym{***} \(p<0.01\)} \\
\end{tabular}
}
}
\end{center}
\vspace{-20pt}
\caption*{\scriptsize \textit{Notes}: In this figure we repeat Table \ref{table:main}, Panel A, but display the coefficients on control variables (Fraction Below Poverty Line, Population, Population Per Sq Mile). We additionally add column 6 which adds two additional sets of control variables: fixed effects for each hour of the day (hour of arrival for a wait time) and whether the cellphone is Android (vs. iPhone). \textit{Additional Notes:} Robust standard errors, clustered at the polling place level, are in parentheses. Unit of  observation is a cellphone identifier on Election Day. \textit{DepVarMean} is the mean of the dependent variable. \textit{Polling Area Controls} includes the population, population per square mile, and fraction below poverty line for the block group of the polling station. ``Asian'' includes ``Pacific Islander.'' ``Other Non-White'' includes the ``Other,'' ``Native American,'' and ``Multiracial'' Census race categories. Column 6 adds an additional specification beyond Table 1; there we include fixed effects for the hour of arrival (i.e. the first ping of  a waiting spell within the 60 meters of the polling place centroid) and a dummy variable for whether the observation corresponds to an Android phone.}
\end{table}

\begin{table}[H]
\begin{center}
\caption{Fraction Black and Voter Wait Time: LPM}
\label{table:app_lpm}
\vspace{-10pt}
\scalebox{0.95}{{
\def\sym#1{\ifmmode^{#1}\else\(^{#1}\)\fi}
\begin{tabular}{l*{6}{c}}
\toprule
&\multicolumn{1}{c}{(1)}&\multicolumn{1}{c}{(2)}   &\multicolumn{1}{c}{(3)}&\multicolumn{1}{c}{(4)}&\multicolumn{1}{c}{(5)}   &\multicolumn{1}{c}{(6)}\\
\midrule
Fraction Black                &        0.12\sym{***}&        0.12\sym{***}&        0.11\sym{***}&        0.10\sym{***}&        0.07\sym{***}&        0.06\sym{***}\\
                              &      (0.01)         &      (0.01)         &      (0.01)         &      (0.01)         &      (0.01)         &      (0.01)         \\
\addlinespace
Fraction Asian                &                     &       -0.00         &       -0.04\sym{**} &        0.04\sym{**} &       -0.02         &       -0.01         \\
                              &                     &      (0.02)         &      (0.02)         &      (0.02)         &      (0.02)         &      (0.02)         \\
\addlinespace
Fraction Hispanic             &                     &        0.03\sym{***}&        0.01         &        0.08\sym{***}&        0.03\sym{***}&        0.04\sym{***}\\
                              &                     &      (0.01)         &      (0.01)         &      (0.01)         &      (0.01)         &      (0.01)         \\
\addlinespace
Fraction Other Non-White      &                     &        0.21\sym{***}&        0.21\sym{***}&        0.03         &        0.05         &        0.04         \\
                              &                     &      (0.04)         &      (0.04)         &      (0.04)         &      (0.04)         &      (0.04)         \\
\addlinespace
Fraction Below Poverty Line   &                     &                     &       -0.02         &       -0.05\sym{***}&        0.01         &        0.03\sym{*}  \\
                              &                     &                     &      (0.02)         &      (0.02)         &      (0.01)         &      (0.01)         \\
\addlinespace
Population (1000s)            &                     &                     &        0.01\sym{***}&        0.01\sym{***}&        0.01\sym{***}&        0.01\sym{***}\\
                              &                     &                     &      (0.00)         &      (0.00)         &      (0.00)         &      (0.00)         \\
\addlinespace
Population Per Sq Mile (1000s)&                     &                     &        0.00\sym{***}&        0.00\sym{***}&        0.00\sym{***}&        0.00\sym{***}\\
                              &                     &                     &      (0.00)         &      (0.00)         &      (0.00)         &      (0.00)         \\
\addlinespace
Android (0 = iPhone)          &                     &                     &                     &                     &                     &        0.01\sym{***}\\
                              &                     &                     &                     &                     &                     &      (0.00)         \\
\midrule
N                             &     154,411         &     154,411         &     154,260         &     154,260         &     154,260         &     154,260         \\
$R^2$                         &        0.00         &        0.00         &        0.01         &        0.04         &        0.10         &        0.14         \\
DepVarMean                    &        0.18         &        0.18         &        0.18         &        0.18         &        0.18         &        0.18         \\
Polling Area Controls? &No&No&Yes&Yes&Yes&Yes \\
State FE?                       &No&No&No&Yes&Yes&Yes \\
County FE?                &No&No&No&No&Yes&Yes \\
Hour of Day FE?                 &No&No&No&No&No&Yes \\
\hline\hline
\bottomrule
\multicolumn{7}{l}{\footnotesize \sym{*} \(p<0.10\),   \sym{**} \(p<0.05\), \sym{***} \(p<0.01\)} \\
\end{tabular}
}
}
\end{center}
\vspace{-20pt}
\caption*{\scriptsize \textit{Notes}: In this figure we repeat Table \ref{table:main}, Panel B, but display the coefficients on control variables (Fraction Below Poverty Line, Population, Population Per Sq Mile). We additionally add column 6 which adds two additional sets of control variables: fixed effects for each hour of the day (hour of arrival for a wait time) and whether the cellphone is Android (vs. iPhone). \textit{Additional Notes:}  Robust standard errors, clustered at the polling place level, are in parentheses. Unit of  observation is a cellphone identifier on Election Day. \textit{DepVarMean} is the mean of the dependent variable. The dependent variable is a binary variable equal to 1 if the wait time is greater than 30 minutes. \textit{Polling Area Controls} includes the population, population per square mile, and fraction below poverty line for the block group of the polling station. ``Asian'' includes ``Pacific Islander.'' ``Other Non-White'' includes the ``Other,'' ``Native American,'' and ``Multiracial'' Census race categories. Column 6 adds an additional specification beyond Table 1; there we include fixed effects for the hour of arrival (i.e. the first ping of  a waiting spell within the 60 meters of the polling place centroid) and a dummy variable for whether the observation corresponds to an Android phone.}
\end{table}

\begin{table}[H]
\begin{center}
\caption{Robustness: Regressions for Figure \ref{fig:robust}}
\label{table:app_robust}
\vspace{-10pt}
\scalebox{0.7}{{
\def\sym#1{\ifmmode^{#1}\else\(^{#1}\)\fi}
\begin{tabular}{l*{11}{c}}
\toprule
&\multicolumn{1}{c}{(1)}&\multicolumn{1}{c}{(2)}   &\multicolumn{1}{c}{(3)}&\multicolumn{1}{c}{(4)}&\multicolumn{1}{c}{(5)}   &\multicolumn{1}{c}{(6)}&\multicolumn{1}{c}{(7)}&\multicolumn{1}{c}{(8)}   &\multicolumn{1}{c}{(9)}&\multicolumn{1}{c}{(10)}&\multicolumn{1}{c}{(11)}\\
\multicolumn{11}{l}{\textbf{Panel A: Lower to Upper Bound Split (10\% increments)}} \\
\hline
                              &\multicolumn{1}{c}{Lower}&\multicolumn{1}{c}{S1}&\multicolumn{1}{c}{S2}&\multicolumn{1}{c}{S3}&\multicolumn{1}{c}{S4}&\multicolumn{1}{c}{Midpoint}&\multicolumn{1}{c}{S6}&\multicolumn{1}{c}{S7}&\multicolumn{1}{c}{S8}&\multicolumn{1}{c}{S9}&\multicolumn{1}{c}{Upper}\\
\midrule
Fraction Black                &        4.71\sym{***}&        4.82\sym{***}&        4.92\sym{***}&        5.02\sym{***}&        5.13\sym{***}&        5.23\sym{***}&        5.33\sym{***}&        5.44\sym{***}&        5.54\sym{***}&        5.65\sym{***}&        5.75\sym{***}\\
                              &      (0.35)         &      (0.36)         &      (0.36)         &      (0.37)         &      (0.38)         &      (0.39)         &      (0.40)         &      (0.41)         &      (0.42)         &      (0.43)         &      (0.45)         \\
\midrule
N                             &     154,411         &     154,411         &     154,411         &     154,411         &     154,411         &     154,411         &     154,411         &     154,411         &     154,411         &     154,411         &     154,411         \\
$R^2$                         &        0.00         &        0.00         &        0.00         &        0.00         &        0.00         &        0.00         &        0.00         &        0.00         &        0.00         &        0.00         &        0.00         \\
DepVarMean                    &       11.26         &       12.83         &       14.40         &       15.98         &       17.55         &       19.13         &       20.70         &       22.28         &       23.85         &       25.42         &       27.00         \\
\hline
\multicolumn{11}{l}{\textbf{Panel B: Reasonable Values (See Notes)}} \\
\hline
                              &\multicolumn{1}{c}{RV1}&\multicolumn{1}{c}{RV2}&\multicolumn{1}{c}{RV3}&\multicolumn{1}{c}{RV4}&\multicolumn{1}{c}{RV5}&\multicolumn{1}{c}{RV6}&\multicolumn{1}{c}{RV7}&\multicolumn{1}{c}{RV8}&\multicolumn{1}{c}{RV9}&\multicolumn{1}{c}{RV10}\\
\midrule
Fraction Black                &        5.78\sym{***}&        5.33\sym{***}&        5.23\sym{***}&        5.23\sym{***}&        5.28\sym{***}&        5.37\sym{***}&        3.26\sym{***}&        3.32\sym{***}&        3.39\sym{***}&        3.56\sym{***}\\
                              &      (0.54)         &      (0.49)         &      (0.45)         &      (0.39)         &      (0.39)         &      (0.39)         &      (0.23)         &      (0.23)         &      (0.23)         &      (0.23)         \\
\midrule
N                             &     159,046         &     158,167         &     156,937         &     154,411         &     154,014         &     153,433         &     141,170         &     140,470         &     139,788         &     138,452         \\
$R^2$                         &        0.00         &        0.00         &        0.00         &        0.00         &        0.00         &        0.00         &        0.00         &        0.00         &        0.00         &        0.00         \\
DepVarMean                    &       22.92         &       21.79         &       20.63         &       19.13         &       19.17         &       19.24         &       15.64         &       15.71         &       15.78         &       15.91         \\
\hline
\multicolumn{11}{l}{\textbf{Panel C: Radius Around Building (10 to 100 meters)}} \\
\hline
                              &\multicolumn{1}{c}{Rad10}&\multicolumn{1}{c}{Rad20}&\multicolumn{1}{c}{Rad30}&\multicolumn{1}{c}{Rad40}&\multicolumn{1}{c}{Rad50}&\multicolumn{1}{c}{Rad60}&\multicolumn{1}{c}{Rad70}&\multicolumn{1}{c}{Rad80}&\multicolumn{1}{c}{Rad90}&\multicolumn{1}{c}{Rad100}\\
\midrule
Fraction Black                &        1.43\sym{***}&        1.95\sym{***}&        2.86\sym{***}&        3.98\sym{***}&        4.53\sym{***}&        5.23\sym{***}&        5.68\sym{***}&        6.22\sym{***}&        6.72\sym{***}&        6.99\sym{***}\\
                              &      (0.39)         &      (0.32)         &      (0.33)         &      (0.35)         &      (0.37)         &      (0.39)         &      (0.41)         &      (0.43)         &      (0.46)         &      (0.48)         \\
\midrule
N                             &      60,822         &     120,921         &     150,994         &     161,728         &     161,140         &     154,411         &     144,880         &     134,133         &     123,417         &     113,797         \\
$R^2$                         &        0.00         &        0.00         &        0.00         &        0.00         &        0.00         &        0.00         &        0.00         &        0.00         &        0.00         &        0.00         \\
DepVarMean                    &       12.09         &       14.00         &       15.63         &       17.00         &       18.16         &       19.13         &       20.00         &       20.71         &       21.32         &       21.81         \\
\hline\hline
\bottomrule
\multicolumn{12}{l}{\footnotesize \sym{*} \(p<0.10\),    \sym{**} \(p<0.05\), \sym{***} \(p<0.01\)} \\
\end{tabular}
}
}
\end{center}
\vspace{-25pt}
\caption*{\scriptsize \textit{Notes}: Robust standard errors, clustered at the polling place level, are in parentheses. Unit of  observation is a cellphone identifier on Election Day. \textit{DepVarMean} is the mean of the dependent variable. All specifications are of the form used in Column 1 of Panel A, Table 1. See further notes on Figure \ref{fig:robust}.}
\end{table}

\begin{table}[H]
\begin{center}
\caption{Stricter Likely Voter Filter: Fraction Black and Voter Wait Time}
\label{table:app_stricttable}
\vspace{-10pt}
\scalebox{0.85}{{
\def\sym#1{\ifmmode^{#1}\else\(^{#1}\)\fi}
\begin{tabular}{l*{5}{c}}
\toprule
&\multicolumn{1}{c}{(1)}&\multicolumn{1}{c}{(2)}   &\multicolumn{1}{c}{(3)}&\multicolumn{1}{c}{(4)}&\multicolumn{1}{c}{(5)}\\
\multicolumn{5}{l}{\textbf{Panel A: Ordinary Least Squares (Y = Wait Time)}} \\
\hline
\midrule
Fraction Black                &        4.97\sym{***}&        4.93\sym{***}&        4.38\sym{***}&        4.31\sym{***}&        2.70\sym{***}\\
                              &      (0.53)         &      (0.53)         &      (0.56)         &      (0.57)         &      (0.63)         \\
\addlinespace
Fraction Asian                &                     &       -1.98\sym{*}  &       -3.80\sym{***}&        0.78         &       -2.21\sym{*}  \\
                              &                     &      (1.05)         &      (1.11)         &      (1.10)         &      (1.18)         \\
\addlinespace
Fraction Hispanic             &                     &        1.21\sym{**} &        0.23         &        4.27\sym{***}&        2.10\sym{***}\\
                              &                     &      (0.52)         &      (0.56)         &      (0.67)         &      (0.74)         \\
\addlinespace
Fraction Other Non-White      &                     &       12.54\sym{***}&       11.86\sym{***}&        0.85         &        2.05         \\
                              &                     &      (2.26)         &      (2.27)         &      (2.22)         &      (2.46)         \\
\midrule
N                             &      68,811         &      68,811         &      68,724         &      68,724         &      68,724         \\
$R^2$                         &        0.00         &        0.00         &        0.01         &        0.06         &        0.14         \\
DepVarMean                    &       19.38         &       19.38         &       19.36         &       19.36         &       19.36         \\
Polling Area Controls? &No&No&Yes&Yes&Yes \\
State FE?                       &No&No&No&Yes&Yes \\
County FE?                &No&No&No&No&Yes \\
\hline
\multicolumn{5}{l}{\textbf{Panel B: Linear Probability Model (Y = Wait Time $>$ 30min)}} \\
\hline
\midrule
Fraction Black                &        0.11\sym{***}&        0.11\sym{***}&        0.11\sym{***}&        0.09\sym{***}&        0.05\sym{***}\\
                              &      (0.01)         &      (0.01)         &      (0.01)         &      (0.01)         &      (0.01)         \\
\addlinespace
Fraction Asian                &                     &       -0.00         &       -0.04\sym{*}  &        0.05\sym{*}  &       -0.03         \\
                              &                     &      (0.02)         &      (0.02)         &      (0.02)         &      (0.03)         \\
\addlinespace
Fraction Hispanic             &                     &        0.03\sym{**} &        0.01         &        0.09\sym{***}&        0.04\sym{**} \\
                              &                     &      (0.01)         &      (0.01)         &      (0.02)         &      (0.02)         \\
\addlinespace
Fraction Other Non-White      &                     &        0.22\sym{***}&        0.21\sym{***}&        0.02         &        0.05         \\
                              &                     &      (0.05)         &      (0.05)         &      (0.05)         &      (0.06)         \\
\midrule
N                             &      68,811         &      68,811         &      68,724         &      68,724         &      68,724         \\
$R^2$                         &        0.00         &        0.00         &        0.01         &        0.05         &        0.12         \\
DepVarMean                    &        0.18         &        0.18         &        0.18         &        0.18         &        0.18         \\
Polling Area Controls? &No&No&Yes&Yes&Yes \\
State FE?                       &No&No&No&Yes&Yes \\
County FE?                &No&No&No&No&Yes \\
\hline\hline
\bottomrule
\multicolumn{6}{l}{\footnotesize \sym{*} \(p<0.10\),    \sym{**} \(p<0.05\), \sym{***} \(p<0.01\)} \\
\end{tabular}
}
}
\end{center}
\vspace{-20pt}
\caption*{\scriptsize \textit{Notes}: Robust standard errors, clustered at the polling place level, are in parentheses. Unit of  observation is a cellphone identifier on Election Day. \textit{DepVarMean} is the mean of the dependent variable. The dependent variable in Panel B is a binary variable equal to 1 if the wait time is greater than 30 minutes. \textit{Polling Area Controls} includes the population, population per square mile, and fraction below poverty line for the block group of the polling station. ``Asian'' includes ``Pacific Islander.'' ``Other Non-White'' includes the ``Other,'' ``Native American,'' and ``Multiracial'' Census race categories. See further notes on Figure \ref{fig:app_strictfigure}.}
\end{table}

\section*{Appendix B: Mechanisms}
\appendix
\renewcommand\thefigure{B.\arabic{figure}}    
\setcounter{figure}{0} 
\renewcommand\thetable{B.\arabic{table}}    
\setcounter{table}{0} 
\renewcommand\thesubsection{B.\arabic{subsection}}    
\setcounter{subsection}{0} 

In Section 4 of the paper, we documented large and persistent differences in wait times for areas with a larger fraction of black residents relative to white residents. In this section, we explore potential explanations for these differences. This descriptive exercise is important as different mechanisms may imply different corrective policies. For example, if wait time disparities are driven by differential job flexibility (and thus bunching in busy arrival hours), the best policy response might be to create Federal holidays for elections (e.g. as proposed in ``Democracy Day'' legislation). By contrast, if the disparity is driven by inequalities in provided resources, the optimal policy response might be to set up systems to monitor and ensure equal resources per voter across the nation. 

The nature of our data does not lend itself to a deep exploration of mechanism. A complete understanding of mechanism would likely need to include a large amount of investigative work including data for the quantity and quality of resources at the level of a polling place. There are also two measurement and identification issues to keep in mind. First, as noted in Section 3, our wait time measure may include voters who abandon the line after discovering a long wait time. Second, our estimates are conditional on a voter turning out. Each of the mechanisms below could affect either of these intermediate outcomes. For example, a Strict ID law could increase the amount of time it takes to process a single voter. However, it may also discourage potential voters from turning out to vote (decreasing the \textit{actual queue length} for the marginal voter) and it could increase the likelihood that a voter who does turnout would leave the line early (decreasing the \textit{average measured} time from our method). These two issues thus further caution against using this analysis in isolation to identify the causal effect of addressing these mechanisms. However, in our analysis below, we are able to cast doubt on a few potential mechanisms and draw some tentative conclusions that at the very least may help guide further work that attempts to pinpoint causal determinants of wait times.

\subsection{Inflexible Arrival Times}

One potential mechanism for the differences in wait times that we find is that areas differ in the intensity of voting that occurs at different times of day. For example, it is possible that polling stations in black and white areas are equally resourced and prepared to handle voters, but that voters in black areas are more likely to show up all at once. This could occur, for example, if black voters have less flexible jobs than white voters and therefore can only vote in the early morning or evening. This mechanism for differences in wait times is a bit more indirect than other potential mechanisms in that it is not driven by less attention or resources being devoted to black areas, but rather is a result of congestion caused by more general features of the economy (e.g. job flexibility).

\begin{figure}[H]
\begin{center}
\caption{Density of Voters by Hour of Day by Fraction Black}
\label{fig:bunching}
\vspace{-5pt}
\includegraphics[width=.85\linewidth]{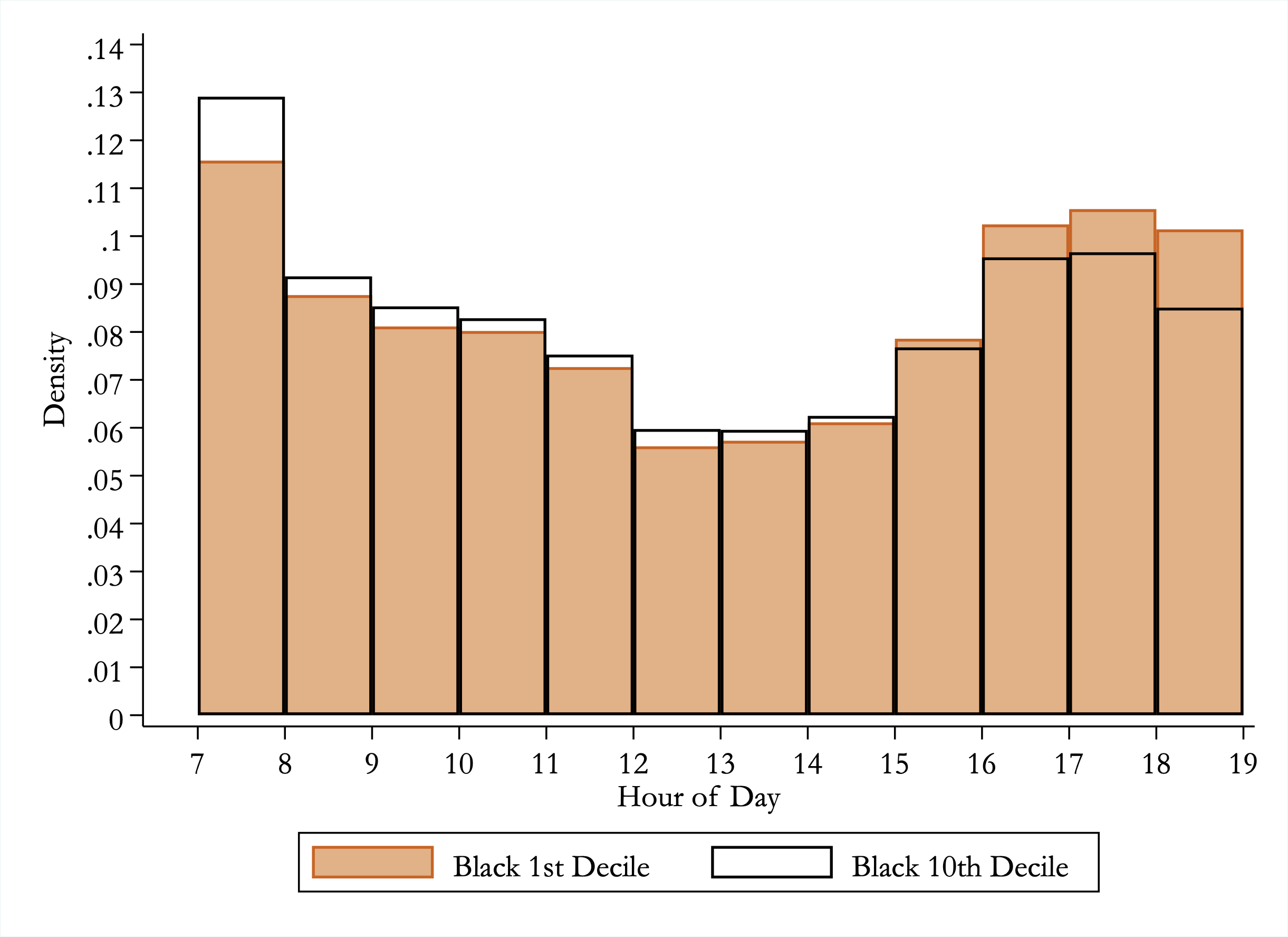}
\vspace{-10pt}
\caption*{\scriptsize \textbf{Notes}: Sample restricted to the 32 states that open no later than 7am and close no earlier than 7pm across all counties.}
\end{center}
\end{figure}

To test for evidence of this mechanism, Figure \ref{fig:bunching} plots the density of arrival time for voters from the most black areas (highest decile) and from the the least black areas (lowest decile).\footnote{We restrict the sample to the 32 states that opened no later than 7am and closed no earlier than 7pm, and restrict the range to be from 7am to 7pm in order to avoid having attrition in the graph due to the opening and closing times of different states. We thus exclude the following states from this figure: Arkansas, Georgia, Idaho, Kansas, Kentucky, Maine, Massachusetts, Minnesota, Nebraska, New Hampshire, North Dakota, Tennessee, Vermont. Despite this sample restriction, we find a similar disparity estimate in this restricted sample (coefficient = 5.43; t = 13; N = 124,950) as in the full sample (coefficient = 5.23; t = 14; N = 154,411).} A visual inspection of Figure \ref{fig:bunching} shows quite minor differences in bunching. Voters in black areas are slightly more likely to show up in the very early morning hours whereas voters in white areas are slightly more likely to show up in the evening. 

Figure \ref{fig:bunching} does not appear to make a particularly strong case for bunching in arrival times. However, as we showed in Panel B of Figure \ref{fig:openclose}, wait times are longer in the morning (when black voters are slightly more likely to show up). A simple test to see if these differences are large enough to explain the racial disparities we find is to include hour-of-the-day fixed effects in our main regression specification. These fixed effects account for any differences in wait times that are due to one group (e.g. voters from black areas) showing up disproportionately during hours that have longer wait times. We include hour-of-the-day fixed effects in Column 6 of Appendix Table \ref{table:app_ols}. The coefficient on fraction black drops from a disparity of 3.27 minutes to a disparity of 3.10 minutes, suggesting that hour-of-the-day differences are not a primary factor that contributes to the wait-time gap that we find.

A different way to show that bunching in arrival times is not sufficient to explain our results is to restrict the sample to hours that don't include the early morning. In Appendix Table \ref{table:app_tablebunching}, we replicate our main specification (Column 4 in Table 2), but only use data after 8am, 9am, and 10am. We continue to find strong evidence of a racial disparity in wait times despite the fact that this regression is including hours of the day (evening hours) when white areas may be more congested due to bunching. This table also provides estimates that exclude both morning and evening hours when there are differences in bunching by black and white areas and also restricts to just evening hours where white areas have higher relative volume in arrivals. Once again, we find strong black-white differences in voter wait times during these hours. 

We conclude that the evidence does not support congestion at the polls due to bunching of arrival times as a primary mechanism explaining the racial disparity in wait times that we document.

\subsection{Partisan Bias}

Another explanation for why voters in black areas may face longer wait times than voters in white areas is that election officials may provide fewer or lower quality resources to black areas. Using carefully-collected data by polling place across three states in the 2012 election (from \citealt{Famighetti2014}), \citet{Pettigrew2017} finds evidence of exactly this -- black areas were provided with fewer poll workers and machines than white areas. Thus, it seems likely that differential resources contribute to the effects that we find. An even deeper mechanism question though is why black areas might receive a lower quality or quantity of election resources. In this section, we explore whether partisanship is correlated with wait times.   

At the state level, the individual charged with being the chief elections officer is the secretary of state (although in some states it is the lieutenant governor or secretary of the commonwealth). The secretary of state often oversees the distribution of resources to individual polling places, although the process can vary substantially from state to state and much of the responsibility is at times passed down to thousands of more local officials (\citealt{Spencer2010}).\footnote{\citet{Spencer2010} provide a useful summary of the problem of identifying precisely who is responsible for election administration in each of the 116,990 polling places spread over the country:
\begin{quote}
    One major reason why polling place inefficiency has yet to be adequately studied is that the administration of elections in the United States is extremely complicated. Each state creates its own rules, budgets its own money, and constructs its own election processes. In some states, such as Wisconsin and Michigan, local jurisdictions have primary autonomy over election administration. In others, such as Oklahoma and Delaware, all election officials are state employees. Still others share administrative duties between state and local election officials. For example, in California, counties have significant authority, yet they operate within a broad framework established by the Secretary of State. On the federal level, the United States Constitution preserves the right of Congress to supersede state laws regulating congressional elections. The result is a complex web of overlapping jurisdictions and 10,071 government units that administer elections. To complicate matters further, authority in all jurisdictions is ceded to two million poll workers who control the success or failure of each election.
\end{quote}}

It could be that state and county officials uniformly have a bias against allocating resources to black areas and this creates racial disparities in wait times across the U.S. as a whole. Alternatively, some election officials may be especially unequal in the resources they provide. An observable factor that could proxy for how unfair an election official may be in allocating resource is party affiliation. In 2016, black voters were far more likely to vote for the Democratic candidate than the Republican candidate.\footnote{Exit polls suggested that 89\% of black voters cast their ballot for the Democratic candidate in 2016 whereas only 8\% voted for the Republican candidate (source: \url{https://www.cnn.com/election/2016/results/exit-polls}).} Given this large difference in vote share, it is possible that Republican party control or overall Republican party membership of an area predicts a motivation (either strategic or based in prejudice) for limiting resources to polling places in black areas.

\begin{figure}[H]
\begin{center}
\caption{Republican Vote Share and Racial Gaps}
\label{fig:partisan}
\vspace{-5pt}
\includegraphics[width=.85\linewidth]{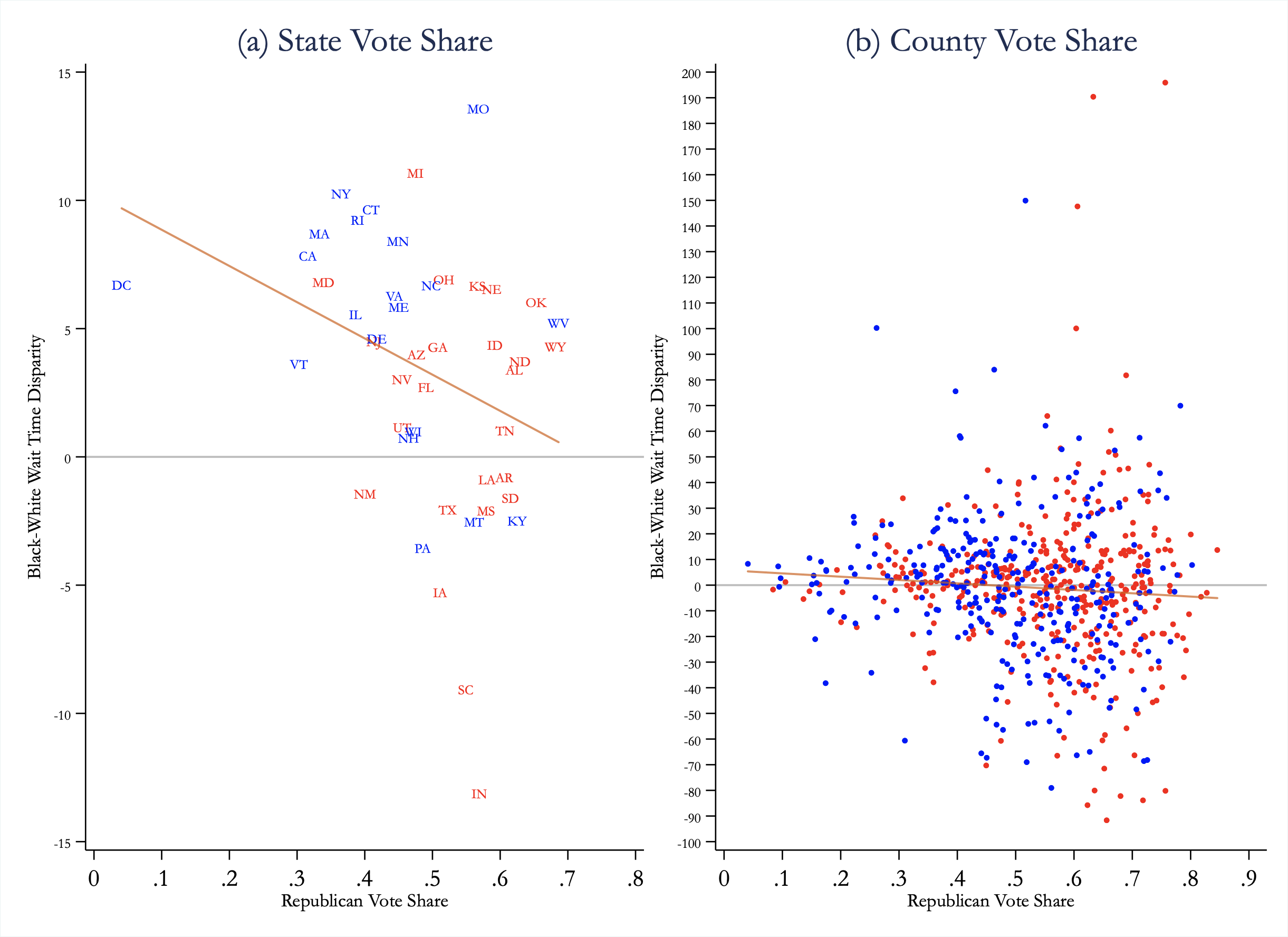}
\vspace{-10pt}
\caption*{\scriptsize \textbf{Notes}: Panel A shows a scatter plot of empirical-Bayes-adjusted state-level wait time disparities (i.e. the adjusted coefficient from a regression of wait time on ``Fraction Black'', with standard errors clustered at the pollling place level) against the 2016 Republican vote share for that state. Panel B shows the same relationship for county-level measures. Points are colored by the partisan affiliation of the chief elections officer in that State (Red = Republican). The fit lines are produced using lfit in \textit{Stata}.}
\end{center}
\end{figure}

To test for evidence of a partisan bias, we plot empirical-Bayes-adjusted state-level racial disparities in wait times against the 2016 Republican vote share at both the state (panel A of Figure \ref{fig:partisan}) and county level (panel B of Figure \ref{fig:partisan}).\footnote{The sample sizes for some counties are very small. Thus, we restrict the analysis to the 718 counties with at least 30 likely voters (and for which the disparity can be estimated) in order to avoid small-sample inference issues.} Panel A also color codes each state marker by the party affiliation of the chief elections officer in the state.\footnote{State and county Republican vote shares are taken from the MIT Election Data and Science Lab's County Presidential Election Returns 2000-2016 (\url{https://dataverse.harvard.edu/file.xhtml?persistentId=doi:10.7910/DVN/VOQCHQ/FQ9NBF&version=5.0}). We compute the Republican vote share as the number of votes cast at the County (or State) level divided by the total number of votes cast in that election, and thus states with a Republican vote share under 50\% may still have more votes for Trump over Clinton (e.g. Utah). The partisan affiliation of the chief elections officer in the state is taken from: \url{https://en.wikipedia.org/w/index.php?title=Secretary_of_state_(U.S._state_government)&oldid=746677873}} The fitted lines in both panels do not show evidence of positive correlation between Republican vote share and racial disparities in voter wait times. If anything we find larger disparities in areas that have a lower Republican vote share.

While this analysis is correlational in nature, it suggests that racial disparities in wait times are not primarily driven by how Republican the state/county is. Rather, both red and blue states and counties are susceptible to generating conditions that lead to black voters spending more time at the polls than their white counterparts.

\subsection{County-Level Correlates}

We do not find evidence of a correlation between party affiliation at the county level and racial disparities in wait times. However, there may be other characteristics of counties that correlate with our measure of racial disparities. In Figure \ref{fig:predictors}, we show estimates of a regression of our measure of racial disparities at the county-level (empirical-Bayes adjusted and limited to those counties with more than 30 observations) against a Social Capital Index, Top 1\% Income Share, Gini Coefficient, Theil Index of Racial Segregation, and two measures of social mobility from \citet{Chetty2018}. Each of these variables is taken from Figure 5 of \citet{Chetty2018}, corresponds to the 2000 Census, and has been standardized.\footnote{We source these variables from: \url{https://opportunityinsights.org/wp-content/uploads/2018/04/online_table4-2.dta} and merge on the Census County FIPS (taken from the 2000 Census in the \citet{Chetty2018} data and from the 2017 ACS in our data.} We find little evidence that voter wait time disparities are correlated with these additional measures. Overall, we argue that a clear pattern does not emerge where counties of a particular type are experiencing the largest disparities in voter wait time.

\subsection{State Voting Laws}

A large recent discussion has emerged regarding the impact of Strict ID laws (\citealt{Cantoni2019,Grimmer2019}) and unequal access to early voting (\citealt{Kaplan2019,Herron2014}) on the voting process. Both of these types of laws have the potential to produce racial inequalities in wait times. For example, Strict ID laws may disproportionately cause delays at polling places in minority areas. The effect of early voting laws is less clear. It is possible that early voting allows voters who would have otherwise faced long lines to take advantage of the early voting process and therefore release some of the pressure at the polling places with the longest waits. However, it is also possible that white voters are more likely to learn about and take advantage of early voting (or that early voting is more likely to be available in white areas within a State that has early voting) which could lead to even longer disparities in wait times if election officials don't adjust polling place resources to accommodate the new equilibrium. 

\begin{figure}[H]
\begin{center}
\caption{County Characteristics, State Laws, and Racial Disparities }
\label{fig:predictors}
\vspace{-5pt}
\includegraphics[width=.85\linewidth]{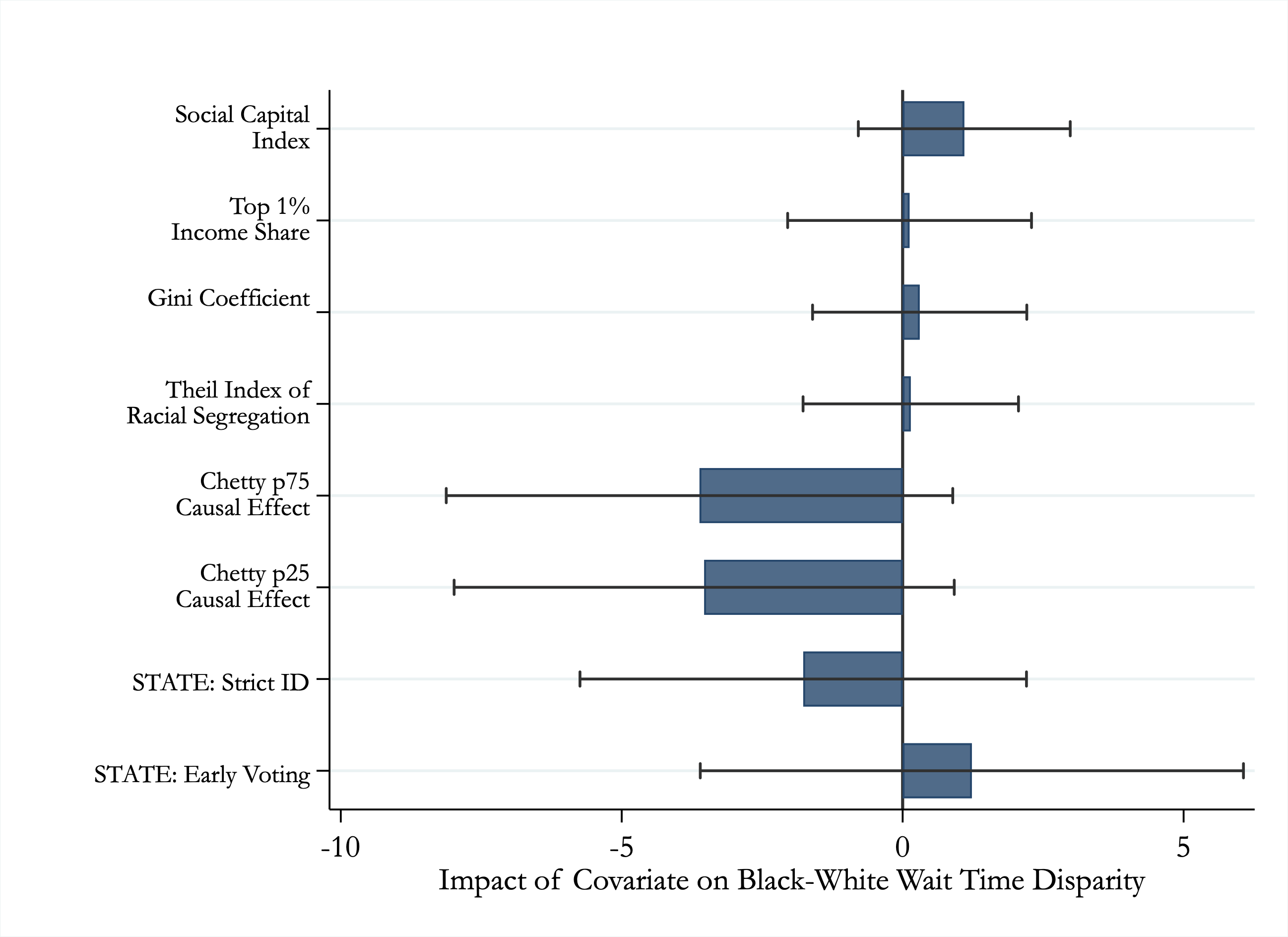}
\vspace{-10pt}
\caption*{\scriptsize \textbf{Notes}: Each row reports the coefficient from a bivariate regression of a county-level (empirical-Bayes-adjusted) wait time average on a county-level measure (rows 1-8) or of a state-level (empirical-Bayes-adjusted) wait time average on a state-level measure. See footnote 9 for further details on the county-level measures taken from \citet{Chetty2018}. States identified as having strict voter ID laws in 2016 are: Arizona, Georgia, Indiana, Kansas, Mississippi, North Dakota, Ohio, Tennessee, Virginia, and Wisconsin. States identified as \textit{not} having any type of early voting in 2016 are: Alabama, Delaware, Indiana, Kentucky, Michigan, Mississippi, Missouri, New York, Pennsylvania, Rhode Island, South Carolina, Virginia.}
\end{center}
\end{figure}

The final two bars in Figure \ref{fig:predictors} show how our measure of racial disparity at the state level interacts with states with early voting laws (N = 34) and states with Strict ID laws (N = 10).\footnote{Following \citet{Cantoni2019}, we source both of these measures from the National Conference of State Legislatures. We use Internet Archive snapshots from just before the 2016 Election to obtain measures relevant for that time period (e.g. for Strict ID laws we use the following link: \url{https://web.archive.org/web/20161113113845/http://www.ncsl.org/research/elections-and-campaigns/voter-id.aspx}). For the early-voting measure we define it as any state that has same-day voter registration, automatic voter registration, no-excuse absentee voting, or early voting (\citet{Cantoni2019} study multiple elections, and thus define this measure as the share of elections over which one of these was offered). States identified as having strict voter ID laws in 2016 are: Arizona, Georgia, Indiana, Kansas, Mississippi, North Dakota, Ohio, Tennessee, Virginia, and Wisconsin. States identified as \textit{not} having any type of early voting in 2016 are: Alabama, Delaware, Indiana, Kentucky, Michigan, Mississippi, Missouri, New York, Pennsylvania, Rhode Island, South Carolina, Virginia.} As can be seen in the figure, we do not find evidence that the variation in wait time disparities is being explained in a substantial way by these laws. 

\subsection{Congestion}

A final mechanism that we explore is congestion due to fewer or lower quality resources per voter at a polling place. Congestion may cause longer wait times and be more likely to be a factor at polling places with more black voters. We do not have a direct measure of resources or overall congestion at the polling place level, but a potential proxy for congestion is the number of registered voters who are assigned to each polling place. We use data from L2's 2016 General Election national voter file. These data allow us to determine the total number of registered voters who are assigned to vote at each polling place and also the number of actual votes cast. For most voters, their polling place was determined by the name of their assigned precinct; precincts were assigned to one or more polling places by their local election authority. In the rare case where voters were allowed their choice from multiple polling places, the polling place closest to their home address was used. Registered voters and votes cast by polling place are highly correlated (correlation = 0.96) and the analysis below is unchanged independent of what measure we use. We will therefore focus on the number of registered voters for each polling place. 

It is not obvious that polling places with more voters should have longer overall wait times. In a carefully-resourced system in equilibrium, high-volume polling places should have more machines and polling workers and therefore be set up to handle the higher number of voters. However, it is possible that the quality and quantity of polling resources is out of equilibrium and does not compensate for the higher volume. For example, polling-place closures or residential construction may increase the number of registered voters assigned to a given polling place and polling resources may not adjust fast enough to catch up to the changing volume. Alternatively, even if variable resource are in equilibrium, there may be fixed differences that lead to longer wait times in high volume areas (e.g. constrained building sizes leading to slower throughput, or a higher risk of technical issues).\footnote{In Appendix Table \ref{table:buildingtype}, we investigate these potential fixed building type differences directly by matching polling place buildings to information on size and types from Microsoft OpenStreetMap. We group building types into 6 categories (Commercial, Medical, Private, Public, Religious, School) and 76 sub-categories (e.g. Commercial is divided into Gym, Hotel, Shopping Center, and 7 other sub-categories). We show in Panel C that building categories and building size are only weakly predictive of fraction black. Panels A and B in turn show that controlling for a second-order polynomial in building size (Column 2), category fixed effects (Column 3), and sub-category fixed effects (Column 4) has little effect on estimates of the racial disparities. This analysis suggests that at least these coarse building characteristics, on their own, do not seem to mediate the relationship. Moreover, this analysis provides some reassurance that the rules for cleaning data---which may differentially affect different building types---do not skew our estimates of the racial disparities.} 

Following our baseline specifications, we regress voting wait time for each individual in our sample on the number of registered voters assigned to the polling place where they voted. These results can be found in Appendix Table \ref{table:congestion_overall}. We do indeed find a positive relationship across specifications with varied fixed effects suggesting that congestion may be an issue in high-volume polling locations. 

Given the above association, if polling places with a large fraction of black voters are also more likely to be high volume, this could help explain the black-white disparity in wait times that we have documented. The data, however, do not bear this out. There is not a strong correlation between volume and the fraction of black residents at a polling place (correlation = .03). One way to see this is we run our baseline regressions, but include the number of registered voters in each polling place as a control. The table indicates that this new control does not significantly diminish the racial disparity in wait times and if anything may cause the disparity to become a bit larger in some specifications.

\begin{figure}[H]
\begin{center}
\caption{Congestion and Wait Times by Fraction Black}
\label{fig:congestion}
\vspace{-5pt}
\includegraphics[width=.85\linewidth]{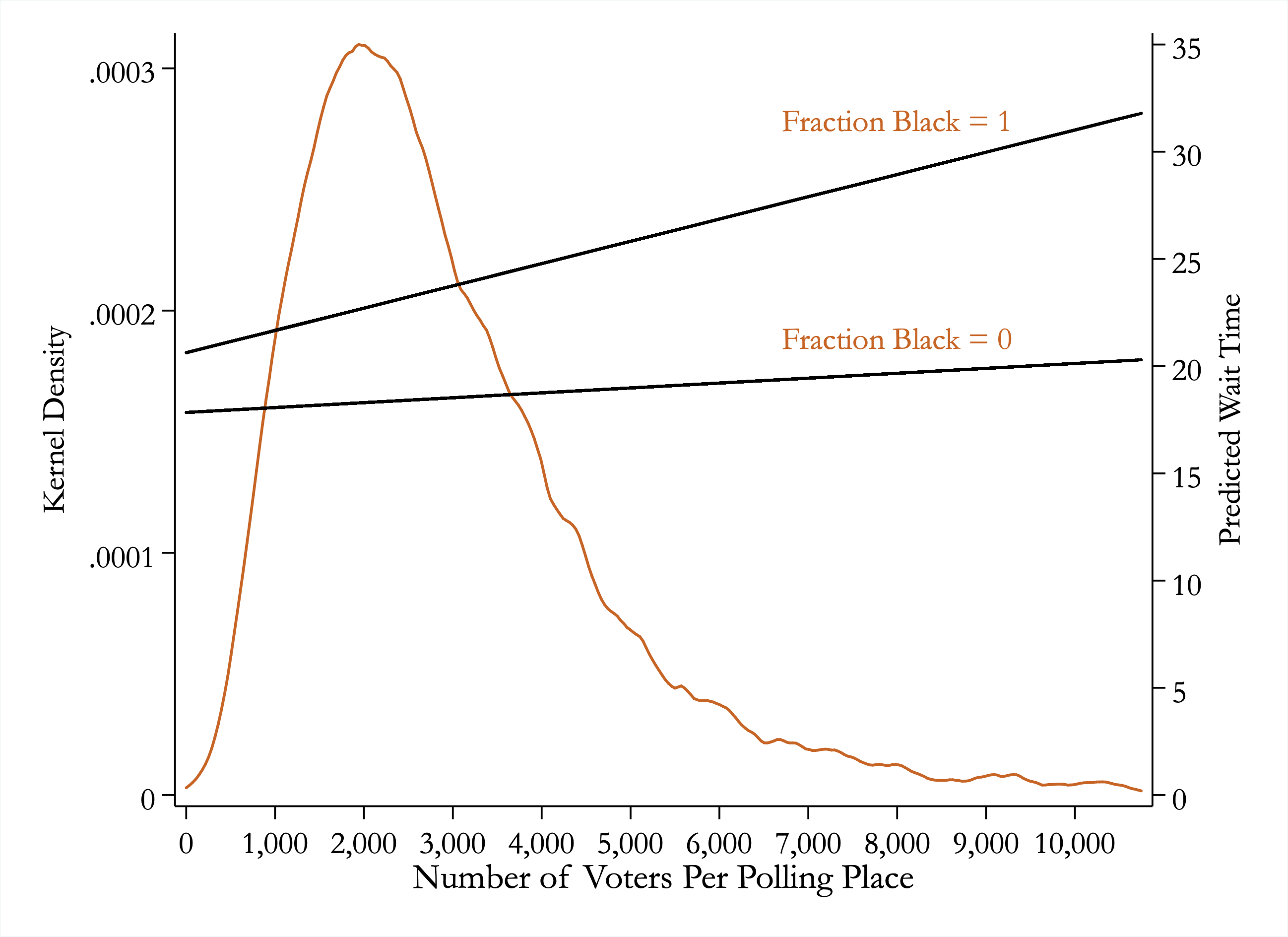}
\vspace{-10pt}
\caption*{\scriptsize \textbf{Notes}: The left y-axis corresponds to the kernel density (estimated using 100 person half-widths) of the Number of Registered Voters per Polling Place (after first dropping the top 1\% of observations, i.e. voters in polling places with more than 10,746 registered individuals). The right y-axis corresponds to the two regression lines (estimated on the full sample) -- both lines correspond to a voter (i.e. cellphone identifier)-level regression of wait time on ``Fraction Black'', the ``Number of Registered Voters Per Polling Place'', and the interaction. The top line reports the predicted regression line for ``Fraction Black'' = 1, while the bottom line reports this for ``Fraction Black'' = 0.}
\end{center}
\end{figure}

Lastly, we explore whether or not the racial disparity in voter wait times that we document interacts with our proxy for congestion. Is the racial gap in wait times larger or smaller in high-volume polling places? In Appendix Table \ref{table:congestion_interact} we run our baseline regressions and include the number of registered voters in each polling place and also an interaction between registered voters and the fraction of black residents. Across all specifications, we find a significant and robust interaction effect indicating larger racial disparities at higher volume polling places. Figure \ref{fig:congestion} helps put this interaction effect in perspective. In this figure, we plot the density function for the number of voters registered in each individual's polling place in our data (labeled on the left y-axis). We also plot the predicted wait time for an area composed entirely of black residents (fraction black = 1) as well as an area with no black residents (fraction black = 0) by the number of registered voters at the polling place (labeled on the right y-axis). The predicted lines indicate that the black-white disparity in wait times for individuals who vote at a low-volume polling location (10th percentile = 1,150 registered voters) is 3.7 minutes whereas the disparity in high-volume polling locations (90th percentile = 5,242 registered voters) is almost twice as large at 7 minutes.

Thus, we find that the largest racial disparities in voter wait times are in the highest volume polling places. This finding is consistent with several possible stories. For example, this pattern may reflect another dimension of the aforementioned inequality in polling machines, workers, and other support. Black areas may face persistent under-resourcing and these resourcing constraints may be especially harmful at higher volumes of voters. Relatedly, election officials may respond less quickly to adjustments in volume (e.g.  caused by polling closures or changes in voter-age population) in areas with higher concentrations of black residents. This off-equilibrium response may lead to the differential gradient we find in volume between black and white areas. Our analysis is correlational and thus does not allow us to make conclusive statements about the exact underlying mechanism. On the other hand, this descriptive exercise can provide guidance on potential sources for the disparity that are worthy of further exploration.

\begin{table}[H]
\begin{center}
\caption{Fraction Black and Voter Wait Time - Restricting Hour of Arrival Windows}
\label{table:app_tablebunching}
\vspace{-10pt}
\scalebox{0.9}{{
\def\sym#1{\ifmmode^{#1}\else\(^{#1}\)\fi}
\begin{tabular}{l*{6}{c}}
\toprule
&\multicolumn{1}{c}{(1)}&\multicolumn{1}{c}{(2)}   &\multicolumn{1}{c}{(3)}&\multicolumn{1}{c}{(4)}&\multicolumn{1}{c}{(5)}&\multicolumn{1}{c}{(6)} \\
\multicolumn{7}{l}{\textbf{Panel A: Ordinary Least Squares (Y = Wait Time)}} \\
\hline
\midrule
Fraction Black                &        4.84\sym{***}&        2.90\sym{***}&        2.21\sym{***}&        1.85\sym{***}&        1.71\sym{***}&        1.92\sym{***}\\
                              &      (0.42)         &      (0.43)         &      (0.43)         &      (0.44)         &      (0.54)         &      (0.51)         \\
\addlinespace
Fraction Asian                &        1.30\sym{*}  &        0.20         &       -0.14         &       -0.41         &        0.57         &       -1.08         \\
                              &      (0.76)         &      (0.77)         &      (0.79)         &      (0.80)         &      (1.01)         &      (0.98)         \\
\addlinespace
Fraction Hispanic             &        3.90\sym{***}&        3.23\sym{***}&        3.22\sym{***}&        3.26\sym{***}&        0.84         &        5.25\sym{***}\\
                              &      (0.46)         &      (0.48)         &      (0.50)         &      (0.51)         &      (0.62)         &      (0.63)         \\
\addlinespace
Fraction Other Non-White      &        1.66         &        1.14         &        1.40         &        2.20         &        0.53         &        3.02         \\
                              &      (1.89)         &      (1.92)         &      (2.00)         &      (2.06)         &      (2.44)         &      (2.55)         \\
\midrule
N                             &     154,260         &     124,367         &     111,480         &      99,858         &      57,863         &      52,995         \\
$R^2$                         &        0.06         &        0.04         &        0.04         &        0.04         &        0.04         &        0.04         \\
DepVarMean                    &       19.12         &       17.67         &       17.50         &       17.34         &       17.47         &       16.88         \\
Sample? &Full& $\geq8am$ & $\geq9am$ & $\geq10am$ & 10am-3pm & $\geq3pm$ \\
\hline
\multicolumn{7}{l}{\textbf{Panel B: LPM (Y = Wait Time $>$ 30min)}} \\
\hline
\midrule
Fraction Black                &        0.10\sym{***}&        0.06\sym{***}&        0.04\sym{***}&        0.03\sym{***}&        0.03\sym{**} &        0.03\sym{***}\\
                              &      (0.01)         &      (0.01)         &      (0.01)         &      (0.01)         &      (0.01)         &      (0.01)         \\
\addlinespace
Fraction Asian                &        0.04\sym{**} &        0.02         &        0.01         &        0.01         &        0.02         &        0.01         \\
                              &      (0.02)         &      (0.02)         &      (0.02)         &      (0.02)         &      (0.02)         &      (0.02)         \\
\addlinespace
Fraction Hispanic             &        0.08\sym{***}&        0.06\sym{***}&        0.06\sym{***}&        0.07\sym{***}&        0.01         &        0.11\sym{***}\\
                              &      (0.01)         &      (0.01)         &      (0.01)         &      (0.01)         &      (0.01)         &      (0.01)         \\
\addlinespace
Fraction Other Non-White      &        0.03         &        0.01         &        0.03         &        0.05         &        0.02         &        0.04         \\
                              &      (0.04)         &      (0.04)         &      (0.04)         &      (0.05)         &      (0.05)         &      (0.06)         \\
\midrule
N                             &     154,260         &     124,367         &     111,480         &      99,858         &      57,863         &      52,995         \\
$R^2$                         &        0.04         &        0.03         &        0.03         &        0.03         &        0.03         &        0.03         \\
DepVarMean                    &        0.18         &        0.14         &        0.14         &        0.14         &        0.14         &        0.13         \\
Sample? &Full& $\geq8am$ & $\geq9am$ & $\geq10am$ & 10am-3pm & $\geq3pm$ \\
\hline\hline
\bottomrule
\multicolumn{7}{l}{\footnotesize \sym{*} \(p<0.10\),    \sym{**} \(p<0.05\), \sym{***} \(p<0.01\)} \\
\end{tabular}
}
}
\end{center}
\vspace{-20pt}
\caption*{\scriptsize \textit{Notes}: Robust standard errors, clustered at the polling place level, are in parentheses. Unit of  observation is a cellphone identifier on Election Day. \textit{DepVarMean} is the mean of the dependent variable. Specifications match those of Table \ref{table:main}, Column 4. The dependent variable in Panel B is a binary variable equal to 1 if the wait time is greater than 30 minutes. All columns include state fixed effects and \textit{polling area controls} (includes the population, population per square mile, and fraction below poverty line for the block group of the polling station. ``Asian'' includes ``Pacific Islander.'' ``Other Non-White'' includes the ``Other,'' ``Native American,'' and ``Multiracial'' Census race categories).}
\end{table}

\begin{table}[H]
\begin{center}
\caption{Controlling for Building Type and Size}
\label{table:buildingtype}
\vspace{-10pt}
\scalebox{0.65}{{
\def\sym#1{\ifmmode^{#1}\else\(^{#1}\)\fi}
\begin{tabular}{l*{10}{c}}
\toprule
&\multicolumn{1}{c}{(1)}&\multicolumn{1}{c}{(2)}   &\multicolumn{1}{c}{(3)}&\multicolumn{1}{c}{(4)}&\multicolumn{1}{c}{(5)}   &\multicolumn{1}{c}{(6)}&\multicolumn{1}{c}{(7)}&\multicolumn{1}{c}{(8)}   &\multicolumn{1}{c}{(9)}&\multicolumn{1}{c}{(10)}\\
\multicolumn{10}{l}{\textbf{Panel A: Ordinary Least Squares (Y = Wait Time)}} \\
\hline
\midrule
Fraction Black                &        5.23\sym{***}&        5.41\sym{***}&        5.69\sym{***}&        5.55\sym{***}&        7.57         &       10.60         &        4.10\sym{***}&        4.54\sym{***}&        5.62\sym{***}&        6.36\sym{***}\\
                              &      (0.39)         &      (0.39)         &      (0.38)         &      (0.38)         &      (6.11)         &      (6.44)         &      (1.57)         &      (0.83)         &      (0.87)         &      (0.52)         \\
\midrule
N                             &     154,411         &     154,411         &     154,411         &     153,937         &       2,259         &         474         &      10,514         &      37,243         &      44,823         &      59,098         \\
$R^2$                         &        0.00         &        0.01         &        0.01         &        0.01         &        0.00         &        0.01         &        0.00         &        0.00         &        0.00         &        0.01         \\
DepVarMean                    &       19.13         &       19.13         &       19.13         &       19.12         &       19.60         &       20.18         &       19.35         &       19.42         &       20.33         &       17.96         \\
PollingPlaces                 &      43,385         &      43,385         &      43,385         &      43,220         &         628         &         165         &       3,962         &      12,630         &      12,173         &      13,827         \\
Category FE? &No&No&Yes&No&No&No&No&No&No&No \\
Subcategory FE?       &No&No&No&Yes&No&No&No&No&No&No \\
Subsample?            &All&All&All&All&Com&Med&Pri&Pub&Rel&Sch\\
\hline
\multicolumn{10}{l}{\textbf{Panel B: Linear Probability Model (Y = Wait Time $>$ 30min)}} \\
\hline
\midrule
Fraction Black                &        0.12\sym{***}&        0.12\sym{***}&        0.12\sym{***}&        0.12\sym{***}&        0.15         &        0.17         &        0.09\sym{***}&        0.10\sym{***}&        0.12\sym{***}&        0.14\sym{***}\\
                              &      (0.01)         &      (0.01)         &      (0.01)         &      (0.01)         &      (0.12)         &      (0.14)         &      (0.03)         &      (0.02)         &      (0.02)         &      (0.01)         \\
\midrule
N                             &     154,411         &     154,411         &     154,411         &     153,937         &       2,259         &         474         &      10,514         &      37,243         &      44,823         &      59,098         \\
$R^2$                         &        0.00         &        0.00         &        0.01         &        0.01         &        0.00         &        0.01         &        0.00         &        0.00         &        0.00         &        0.01         \\
DepVarMean                    &        0.18         &        0.18         &        0.18         &        0.18         &        0.20         &        0.19         &        0.18         &        0.18         &        0.20         &        0.16         \\
PollingPlaces                 &      43,385         &      43,385         &      43,385         &      43,220         &         628         &         165         &       3,962         &      12,630         &      12,173         &      13,827         \\
Category FE? &No&No&Yes&No&No&No&No&No&No&No \\
Subcategory FE?       &No&No&No&Yes&No&No&No&No&No&No \\
Subsample?            &All&All&All&All&Com&Med&Pri&Pub&Rel&Sch\\
\hline
\multicolumn{10}{l}{\textbf{Panel C: Do Building Characteristics Predict Race? (Y = Fraction Black)}} \\
\hline
\midrule
Poll: Medical                 &        0.03         &                     \\
                              &      (0.02)         &                     \\
\addlinespace
Poll: Private                 &        0.00         &                     \\
                              &      (0.01)         &                     \\
\addlinespace
Poll: Public                  &       -0.00         &                     \\
                              &      (0.01)         &                     \\
\addlinespace
Poll: Religious               &        0.00         &                     \\
                              &      (0.01)         &                     \\
\addlinespace
Poll: School                  &        0.03\sym{***}&                     \\
                              &      (0.01)         &                     \\
\addlinespace
Poll: Building Area           &                     &        0.01\sym{***}\\
                              &                     &      (0.00)         \\
\midrule
N                             &     154,411         &     154,411         \\
$R^2$                         &        0.01         &        0.00         \\
DepVarMean                    &       19.13         &       19.13         \\
PollingPlaces                 &      43,385         &      43,385         \\
\hline\hline
\bottomrule
\multicolumn{11}{l}{\footnotesize \sym{*} \(p<0.10\),    \sym{**} \(p<0.05\), \sym{***} \(p<0.01\)} \\
\end{tabular}
}
}
\end{center}
\vspace{-20pt}
\caption*{\scriptsize \textit{Notes}: Robust standard errors, clustered at the polling place level, are in parentheses. Unit of  observation is a cellphone identifier on Election Day. \textit{DepVarMean} is the mean of the dependent variable. The dependent variable in Panel B is a binary variable equal to 1 if the wait time is greater than 30 minutes. Column 1 of Panels A and B present the baseline specification shown in Column 1, Panel A, Table 1 in the text. Column 2 includes a second-order polynomial in building area, where building area is in 5,000 square meters units (close to the standard deviation of building area in our sample), Columns 3 includes building category (Commercial, Medical, Private, Public, Religious, School) fixed effects, and Column 4 includes building sub-category (76) fixed effects. Columns 5-10 show sub-sample estimates across the 6 building categories.The dependent variable in Panel C is ``Fraction Black,'' and the omitted category is ``Poll: Commercial''.}
\end{table}

\begin{table}[H]
\begin{center}
\caption{Congestion (Table 2 with added Volume Controls)}
\label{table:congestion_overall}
\vspace{-10pt}
\scalebox{0.65}{{
\def\sym#1{\ifmmode^{#1}\else\(^{#1}\)\fi}
\begin{tabular}{l*{10}{c}}
\toprule
&\multicolumn{1}{c}{(1)}&\multicolumn{1}{c}{(2)}   &\multicolumn{1}{c}{(3)}&\multicolumn{1}{c}{(4)}&\multicolumn{1}{c}{(5)}&\multicolumn{1}{c}{(6)}   &\multicolumn{1}{c}{(7)}&\multicolumn{1}{c}{(8)}&\multicolumn{1}{c}{(9)}&\multicolumn{1}{c}{(10)}\\
\multicolumn{10}{l}{\textbf{Panel A: Ordinary Least Squares (Y = Wait Time)}} \\
\hline
\midrule
Fraction Black                &        5.20\sym{***}&        5.21\sym{***}&        5.18\sym{***}&        5.18\sym{***}&        4.97\sym{***}&        4.96\sym{***}&        4.80\sym{***}&        4.82\sym{***}&        3.32\sym{***}&        3.36\sym{***}\\
                              &      (0.38)         &      (0.38)         &      (0.39)         &      (0.38)         &      (0.42)         &      (0.41)         &      (0.42)         &      (0.41)         &      (0.45)         &      (0.44)         \\
\addlinespace
Voters Per Polling Place      &                     &        0.29\sym{***}&                     &        0.30\sym{***}&                     &        0.25\sym{***}&                     &        0.51\sym{***}&                     &        0.61\sym{***}\\
                              &                     &      (0.07)         &                     &      (0.07)         &                     &      (0.07)         &                     &      (0.06)         &                     &      (0.05)         \\
\addlinespace
Fraction Asian                &                     &                     &       -0.80         &       -0.81         &       -2.51\sym{***}&       -2.34\sym{***}&        1.25\sym{*}  &        1.04         &       -1.13         &       -1.18         \\
                              &                     &                     &      (0.72)         &      (0.71)         &      (0.75)         &      (0.74)         &      (0.76)         &      (0.75)         &      (0.81)         &      (0.80)         \\
\addlinespace
Fraction Hispanic             &                     &                     &        1.01\sym{***}&        0.95\sym{**} &        0.31         &        0.30         &        3.81\sym{***}&        3.85\sym{***}&        1.53\sym{***}&        1.67\sym{***}\\
                              &                     &                     &      (0.37)         &      (0.37)         &      (0.40)         &      (0.40)         &      (0.46)         &      (0.47)         &      (0.50)         &      (0.51)         \\
\addlinespace
Fraction Other Non-White      &                     &                     &       12.49\sym{***}&       12.97\sym{***}&       12.32\sym{***}&       12.67\sym{***}&        1.96         &        2.26         &        1.95         &        1.90         \\
                              &                     &                     &      (1.96)         &      (1.97)         &      (1.97)         &      (1.98)         &      (1.90)         &      (1.89)         &      (1.95)         &      (1.95)         \\
\midrule
N                             &     152,317         &     152,317         &     152,317         &     152,317         &     152,167         &     152,167         &     152,167         &     152,167         &     152,167         &     152,167         \\
$R^2$                         &        0.00         &        0.00         &        0.00         &        0.01         &        0.01         &        0.01         &        0.06         &        0.06         &        0.13         &        0.13         \\
DepVarMean                    &       19.10         &       19.10         &       19.10         &       19.10         &       19.09         &       19.09         &       19.09         &       19.09         &       19.09         &       19.09         \\
Polling Area Controls?  &No&No&No&No&Yes&Yes&Yes&Yes&Yes&Yes \\
State FE?                        &No&No&No&No&No&No&Yes&Yes&Yes&Yes \\
County FE?                       &No&No&No&No&No&No&No&No&Yes&Yes \\
\hline
\multicolumn{10}{l}{\textbf{Panel B: Linear Probability Model (Y = Wait Time $>$ 30min)}} \\
\hline
\midrule
Fraction Black                &        0.12\sym{***}&        0.12\sym{***}&        0.12\sym{***}&        0.12\sym{***}&        0.11\sym{***}&        0.11\sym{***}&        0.10\sym{***}&        0.10\sym{***}&        0.07\sym{***}&        0.07\sym{***}\\
                              &      (0.01)         &      (0.01)         &      (0.01)         &      (0.01)         &      (0.01)         &      (0.01)         &      (0.01)         &      (0.01)         &      (0.01)         &      (0.01)         \\
\addlinespace
Voters Per Polling Place      &                     &        0.01\sym{***}&                     &        0.01\sym{***}&                     &        0.01\sym{***}&                     &        0.01\sym{***}&                     &        0.01\sym{***}\\
                              &                     &      (0.00)         &                     &      (0.00)         &                     &      (0.00)         &                     &      (0.00)         &                     &      (0.00)         \\
\addlinespace
Fraction Asian                &                     &                     &       -0.00         &       -0.00         &       -0.04\sym{**} &       -0.04\sym{**} &        0.04\sym{**} &        0.03\sym{**} &       -0.02         &       -0.02         \\
                              &                     &                     &      (0.02)         &      (0.02)         &      (0.02)         &      (0.02)         &      (0.02)         &      (0.02)         &      (0.02)         &      (0.02)         \\
\addlinespace
Fraction Hispanic             &                     &                     &        0.02\sym{***}&        0.02\sym{**} &        0.01         &        0.01         &        0.08\sym{***}&        0.08\sym{***}&        0.03\sym{***}&        0.04\sym{***}\\
                              &                     &                     &      (0.01)         &      (0.01)         &      (0.01)         &      (0.01)         &      (0.01)         &      (0.01)         &      (0.01)         &      (0.01)         \\
\addlinespace
Fraction Other Non-White      &                     &                     &        0.22\sym{***}&        0.23\sym{***}&        0.22\sym{***}&        0.23\sym{***}&        0.03         &        0.04         &        0.04         &        0.04         \\
                              &                     &                     &      (0.04)         &      (0.04)         &      (0.04)         &      (0.04)         &      (0.04)         &      (0.04)         &      (0.04)         &      (0.04)         \\
\midrule
N                             &     152,317         &     152,317         &     152,317         &     152,317         &     152,167         &     152,167         &     152,167         &     152,167         &     152,167         &     152,167         \\
$R^2$                         &        0.00         &        0.00         &        0.00         &        0.00         &        0.01         &        0.01         &        0.05         &        0.05         &        0.10         &        0.10         \\
DepVarMean                    &        0.18         &        0.18         &        0.18         &        0.18         &        0.18         &        0.18         &        0.18         &        0.18         &        0.18         &        0.18         \\
Polling Area Controls?  &No&No&No&No&Yes&Yes&Yes&Yes&Yes&Yes \\
State FE?                        &No&No&No&No&No&No&Yes&Yes&Yes&Yes \\
County FE?                       &No&No&No&No&No&No&No&No&Yes&Yes \\
\hline\hline
\bottomrule
\multicolumn{6}{l}{\footnotesize \sym{*} \(p<0.10\),    \sym{**} \(p<0.05\), \sym{***} \(p<0.01\)} \\
\end{tabular}
}
}
\end{center}
\vspace{-20pt}
\caption*{\scriptsize \textit{Notes}: Robust standard errors, clustered at the polling place level, are in parentheses. Unit of  observation is a cellphone identifier on Election Day. \textit{DepVarMean} is the mean of the dependent variable. The dependent variable in Panel B is a binary variable equal to 1 if the wait time is greater than 30 minutes. \textit{Polling Area Controls} includes the population, population per square mile, and fraction below poverty line for the block group of the polling station. ``Asian'' includes ``Pacific Islander.'' ``Other Non-White'' includes the ``Other,'' ``Native American,'' and ``Multiracial'' Census race categories. ``Voters per Polling Place'' is the number of registered individuals for that polling place in the National voterfile.}
\end{table}

\begin{table}[H]
\begin{center}
\caption{Congestion Heterogeneity (Table 2 with added Volume Interactions)}
\label{table:congestion_interact}
\vspace{-10pt}
\scalebox{0.65}{{
\def\sym#1{\ifmmode^{#1}\else\(^{#1}\)\fi}
\begin{tabular}{l*{5}{c}}
\toprule
&\multicolumn{1}{c}{(1)}&\multicolumn{1}{c}{(2)}   &\multicolumn{1}{c}{(3)}&\multicolumn{1}{c}{(4)}&\multicolumn{1}{c}{(5)}\\
\multicolumn{5}{l}{\textbf{Panel A: Ordinary Least Squares (Y = Wait Time)}} \\
\hline
\midrule
Fraction Black                &        2.79\sym{***}&        2.79\sym{***}&        3.01\sym{***}&        2.45\sym{***}&        1.08         \\
                              &      (0.79)         &      (0.78)         &      (0.82)         &      (0.75)         &      (0.74)         \\
\addlinespace
Voters Per Polling Place      &        0.23\sym{***}&        0.23\sym{***}&        0.20\sym{**} &        0.45\sym{***}&        0.54\sym{***}\\
                              &      (0.08)         &      (0.08)         &      (0.08)         &      (0.06)         &      (0.05)         \\
\addlinespace
Interaction: Black X VotersPerPoll&        0.81\sym{***}&        0.80\sym{***}&        0.65\sym{**} &        0.80\sym{***}&        0.76\sym{***}\\
                              &      (0.27)         &      (0.27)         &      (0.27)         &      (0.23)         &      (0.22)         \\
\addlinespace
Fraction Asian                &                     &       -0.88         &       -2.32\sym{***}&        1.10         &       -1.04         \\
                              &                     &      (0.71)         &      (0.74)         &      (0.75)         &      (0.80)         \\
\addlinespace
Fraction Hispanic             &                     &        0.93\sym{**} &        0.29         &        3.86\sym{***}&        1.68\sym{***}\\
                              &                     &      (0.37)         &      (0.40)         &      (0.46)         &      (0.51)         \\
\addlinespace
Fraction Other Non-White      &                     &       12.94\sym{***}&       12.62\sym{***}&        2.17         &        1.88         \\
                              &                     &      (1.97)         &      (1.98)         &      (1.89)         &      (1.95)         \\
\midrule
N                             &     152,317         &     152,317         &     152,167         &     152,167         &     152,167         \\
$R^2$                         &        0.00         &        0.01         &        0.01         &        0.06         &        0.13         \\
DepVarMean                    &       19.10         &       19.10         &       19.09         &       19.09         &       19.09         \\
Polling Area Controls? &No&No&Yes&Yes&Yes \\
State FE?                       &No&No&No&Yes&Yes \\
County FE?                       &No&No&No&No&Yes \\
\hline
\multicolumn{5}{l}{\textbf{Panel B: Linear Probability Model (Y = Wait Time $>$ 30min)}} \\
\hline
\midrule
Fraction Black                &        0.07\sym{***}&        0.07\sym{***}&        0.08\sym{***}&        0.06\sym{***}&        0.03         \\
                              &      (0.02)         &      (0.02)         &      (0.02)         &      (0.02)         &      (0.02)         \\
\addlinespace
Voters Per Polling Place      &        0.01\sym{***}&        0.01\sym{***}&        0.00\sym{***}&        0.01\sym{***}&        0.01\sym{***}\\
                              &      (0.00)         &      (0.00)         &      (0.00)         &      (0.00)         &      (0.00)         \\
\addlinespace
Interaction: Black X VotersPerPoll&        0.01\sym{**} &        0.01\sym{**} &        0.01\sym{*}  &        0.01\sym{***}&        0.01\sym{***}\\
                              &      (0.01)         &      (0.01)         &      (0.01)         &      (0.01)         &      (0.01)         \\
\addlinespace
Fraction Asian                &                     &       -0.00         &       -0.04\sym{**} &        0.03\sym{**} &       -0.02         \\
                              &                     &      (0.02)         &      (0.02)         &      (0.02)         &      (0.02)         \\
\addlinespace
Fraction Hispanic             &                     &        0.02\sym{**} &        0.01         &        0.08\sym{***}&        0.04\sym{***}\\
                              &                     &      (0.01)         &      (0.01)         &      (0.01)         &      (0.01)         \\
\addlinespace
Fraction Other Non-White      &                     &        0.23\sym{***}&        0.23\sym{***}&        0.04         &        0.04         \\
                              &                     &      (0.04)         &      (0.04)         &      (0.04)         &      (0.04)         \\
\midrule
N                             &     152,317         &     152,317         &     152,167         &     152,167         &     152,167         \\
$R^2$                         &        0.00         &        0.01         &        0.01         &        0.05         &        0.10         \\
DepVarMean                    &        0.18         &        0.18         &        0.18         &        0.18         &        0.18         \\
Polling Area Controls? &No&No&Yes&Yes&Yes \\
State FE?                       &No&No&No&Yes&Yes \\
County FE?                       &No&No&No&No&Yes \\
\hline\hline
\bottomrule
\multicolumn{6}{l}{\footnotesize \sym{*} \(p<0.10\),    \sym{**} \(p<0.05\), \sym{***} \(p<0.01\)} \\
\end{tabular}
}
}
\end{center}
\vspace{-20pt}
\caption*{\scriptsize \textit{Notes}: Robust standard errors, clustered at the polling place level, are in parentheses. Unit of  observation is a cellphone identifier on Election Day. \textit{DepVarMean} is the mean of the dependent variable. The dependent variable in Panel B is a binary variable equal to 1 if the wait time is greater than 30 minutes. \textit{Polling Area Controls} includes the population, population per square mile, and fraction below poverty line for the block group of the polling station. ``Asian'' includes ``Pacific Islander.'' ``Other Non-White'' includes the ``Other,'' ``Native American,'' and ``Multiracial'' Census race categories. ``Voters per Polling Place'' is the number of registered individuals for that polling place in the National voterfile.}
\end{table}

\appendix
\section*{Appendix C: Geographic Variation}
\appendix
\renewcommand\thefigure{C.\arabic{figure}}    
\setcounter{figure}{0} 
\renewcommand\thetable{C.\arabic{table}}    
\setcounter{table}{0} 

\begin{table}[H]
\begin{center}
\caption{State-Level Measures of Wait Time and Disparities}
\label{table:app_state_descriptives}
\vspace{-10pt}
\scalebox{.59}{{
\def\sym#1{\ifmmode^{#1}\else\(^{#1}\)\fi}
\begin{tabular}{l*{8}{c}}
\hline\hline
&\multicolumn{1}{c}{(1)}&\multicolumn{1}{c}{(2)}&\multicolumn{1}{c}{(3)}&\multicolumn{1}{c}{(4)}&\multicolumn{1}{c}{(5)}&\multicolumn{1}{c}{(6)}&\multicolumn{1}{c}{(7)}\\
              &                      &\multicolumn{2}{c}{\textbf{Unadjusted}}       & \textbf{Bayesian} &\multicolumn{2}{c}{\textbf{Unadjusted}}&\textbf{Bayesian}\\\cmidrule(lr){3-4}\cmidrule(lr){6-7}
\textbf{State}&            \textbf{N}&        \textbf{Mean}&        \textbf{Std Dev}&   \textbf{Adjusted Mean}&      \textbf{Disparity}  & \textbf{Std Error}  & \textbf{Adjusted Disparity} \\
\hline
Alabama &     4,410 &     23.04 &     17.25 &     23.03 &      3.46 &      1.86 &      3.46 \\ \hline 
Arizona &     2,069 &     20.80 &     18.75 &     20.78 &      4.67 &      6.56 &      4.05 \\ \hline 
Arkansas &       907 &     21.73 &     18.27 &     21.67 &     -1.84 &      3.36 &     -0.74 \\ \hline 
California &    11,743 &     20.43 &     17.37 &     20.43 &      8.32 &      2.04 &      7.89 \\ \hline 
Connecticut &     2,722 &     12.37 &     12.51 &     12.39 &     11.70 &      3.69 &      9.71 \\ \hline 
Delaware &       688 &     11.91 &     11.68 &     11.99 &      4.85 &      2.50 &      4.67 \\ \hline 
DistrictofColumbia &       179 &     27.49 &     22.94 &     26.24 &      8.38 &      4.56 &      6.77 \\ \hline 
Florida &     7,172 &     17.99 &     15.52 &     17.99 &      2.74 &      1.42 &      2.77 \\ \hline 
Georgia &     5,058 &     20.12 &     18.14 &     20.12 &      4.38 &      1.54 &      4.33 \\ \hline 
Idaho &     1,274 &     19.23 &     14.99 &     19.22 &     13.12 &     25.23 &      4.41 \\ \hline 
Illinois &     6,213 &     15.99 &     13.69 &     15.99 &      5.68 &      1.19 &      5.61 \\ \hline 
Indiana &     4,286 &     27.11 &     23.09 &     27.06 &    -15.93 &      2.73 &    -13.07 \\ \hline 
Iowa &     1,667 &     15.44 &     13.17 &     15.46 &     -9.66 &      4.72 &     -5.23 \\ \hline 
Kansas &     1,488 &     16.08 &     13.78 &     16.10 &      8.20 &      4.43 &      6.71 \\ \hline 
Kentucky &     3,166 &     14.61 &     12.64 &     14.63 &     -2.99 &      2.00 &     -2.44 \\ \hline 
Louisiana &     2,403 &     16.08 &     14.30 &     16.09 &     -0.96 &      1.13 &     -0.83 \\ \hline 
Maine &       463 &     17.66 &     15.13 &     17.69 &     27.35 &     24.83 &      5.90 \\ \hline 
Maryland &     4,949 &     20.48 &     16.97 &     20.47 &      7.03 &      1.41 &      6.87 \\ \hline 
Massachusetts &     2,655 &     12.29 &     10.94 &     12.31 &      9.75 &      2.82 &      8.76 \\ \hline 
Michigan &     9,776 &     22.27 &     16.44 &     22.26 &     11.48 &      1.42 &     11.12 \\ \hline 
Minnesota &     4,526 &     15.26 &     12.52 &     15.27 &     10.11 &      3.75 &      8.46 \\ \hline 
Mississippi &       999 &     17.73 &     15.87 &     17.74 &     -3.26 &      3.08 &     -2.05 \\ \hline 
Missouri &     6,231 &     26.20 &     20.70 &     26.17 &     15.00 &      2.40 &     13.63 \\ \hline 
Montana &       307 &     20.53 &     16.56 &     20.45 &   -117.11 &     92.15 &     -2.48 \\ \hline 
Nebraska &     1,355 &     16.60 &     16.02 &     16.63 &     13.22 &      9.91 &      6.60 \\ \hline 
Nevada &       976 &     15.67 &     14.15 &     15.71 &      2.57 &      8.31 &      3.08 \\ \hline 
NewHampshire &     1,325 &     15.48 &     12.10 &     15.50 &     -4.98 &     10.24 &      0.81 \\ \hline 
NewJersey &     4,446 &     13.89 &     13.24 &     13.90 &      4.64 &      1.58 &      4.57 \\ \hline 
NewMexico &       484 &     18.53 &     14.48 &     18.54 &    -35.21 &     21.06 &     -1.39 \\ \hline 
NewYork &     7,892 &     16.51 &     14.66 &     16.52 &     10.50 &      1.08 &     10.31 \\ \hline 
NorthCarolina &     4,061 &     20.58 &     16.81 &     20.57 &      6.99 &      1.78 &      6.74 \\ \hline 
NorthDakota &       424 &     20.03 &     17.76 &     19.97 &      8.97 &     42.73 &      3.79 \\ \hline 
Ohio &     8,343 &     17.49 &     14.27 &     17.49 &      7.10 &      1.22 &      6.98 \\ \hline 
Oklahoma &     3,445 &     26.45 &     20.96 &     26.39 &      7.29 &      4.41 &      6.09 \\ \hline 
Pennsylvania &     6,227 &     20.80 &     18.50 &     20.79 &     -4.34 &      2.29 &     -3.49 \\ \hline 
RhodeIsland &       785 &     19.07 &     15.77 &     19.07 &     35.78 &     15.56 &      9.30 \\ \hline 
SouthCarolina &     4,141 &     26.55 &     22.12 &     26.49 &    -11.68 &      3.03 &     -9.01 \\ \hline 
SouthDakota &       429 &     15.55 &     12.81 &     15.62 &    -12.56 &     10.26 &     -1.54 \\ \hline 
Tennessee &     2,418 &     16.27 &     15.40 &     16.28 &      0.94 &      1.60 &      1.09 \\ \hline 
Texas &     7,377 &     16.04 &     16.51 &     16.05 &     -2.25 &      1.38 &     -2.01 \\ \hline 
Utah &     1,201 &     27.89 &     22.96 &     27.67 &    -34.32 &     51.98 &      1.21 \\ \hline 
Vermont &       165 &     14.83 &     13.09 &     15.05 &      6.78 &     33.00 &      3.69 \\ \hline 
Virginia &     9,030 &     17.73 &     15.09 &     17.73 &      6.53 &      1.71 &      6.34 \\ \hline 
WestVirginia &       600 &     18.38 &     13.69 &     18.39 &      7.29 &      6.96 &      5.28 \\ \hline 
Wisconsin &     3,728 &     16.62 &     13.80 &     16.63 &      0.83 &      1.99 &      1.05 \\ \hline 
Wyoming &       286 &     20.65 &     13.15 &     20.59 &     17.90 &     44.06 &      4.35 \\ \hline 
\end{tabular}
}
}
\end{center}
\vspace{-20pt}
\caption*{\scriptsize \textit{Notes}: Columns 5-7 (\textit{Disparity}) correspond to the coefficients on the interaction between a state fixed effect and the ``Fraction Black'' variable from the voter-level regression of wait time on the full set of state fixed effects and the interaction of those fixed effects with ``Fraction Black'', omitting the constant and  clustering standard errors at the polling place level. Column 7 provides  empirical-Bayes-adjusted estimates of these state-level disparities to account for measurement error. Similarly, Column 4 provides empirical-Bayes-adjusted estimates of the unadjusted state-level means shown in Column 2.}
\end{table}

\begin{table}[H]
\begin{center}
\caption{Congressional District-Level Measures of Wait Time and Disparities (1)}
\label{table:app_cd_descriptives}
\vspace{-10pt}
\scalebox{.7}{{
\def\sym#1{\ifmmode^{#1}\else\(^{#1}\)\fi}
\begin{tabular}{l*{8}{c}}
\hline\hline
&\multicolumn{1}{c}{(1)}&\multicolumn{1}{c}{(2)}&\multicolumn{1}{c}{(3)}&\multicolumn{1}{c}{(4)}&\multicolumn{1}{c}{(5)}&\multicolumn{1}{c}{(6)}&\multicolumn{1}{c}{(7)}\\
              &                      &\multicolumn{2}{c}{\textbf{Unadjusted}}       & \textbf{Bayesian} &\multicolumn{2}{c}{\textbf{Unadjusted}}&\textbf{Bayesian}\\\cmidrule(lr){3-4}\cmidrule(lr){6-7}
\textbf{State \& District}&            \textbf{N}&        \textbf{Mean}&        \textbf{Std Dev}&   \textbf{Adjusted Mean}&      \textbf{Disparity}  & \textbf{Std Error}  & \textbf{Adjusted Disparity} \\
\hline
Alabama 01 &       518 &     22.38 &     16.63 &     22.28 &     -4.95 &      5.06 &     -3.40 \\ \hline 
Alabama 02 &       689 &     21.76 &     15.48 &     21.70 &     -4.26 &      6.77 &     -2.14 \\ \hline 
Alabama 03 &       468 &     22.66 &     17.40 &     22.53 &     10.10 &      8.05 &      6.66 \\ \hline 
Alabama 04 &       272 &     21.06 &     15.10 &     20.95 &     -1.92 &      8.59 &     -0.16 \\ \hline 
Alabama 05 &       956 &     22.05 &     17.49 &     21.99 &     13.94 &      5.80 &     10.64 \\ \hline 
Alabama 06 &     1,061 &     23.91 &     17.42 &     23.83 &      4.86 &      5.08 &      4.20 \\ \hline 
Alabama 07 &       446 &     27.49 &     19.78 &     27.09 &     -0.86 &      5.06 &     -0.23 \\ \hline 
Arizona 01 &       192 &     16.16 &     14.06 &     16.30 &    -10.97 &     13.22 &     -2.41 \\ \hline 
Arizona 02 &       193 &     20.02 &     20.16 &     19.88 &     54.76 &     59.49 &      3.49 \\ \hline 
Arizona 03 &       150 &     23.88 &     19.66 &     23.24 &    -75.87 &     39.55 &     -2.54 \\ \hline 
Arizona 04 &       226 &     18.28 &     15.76 &     18.30 &    -28.90 &     28.87 &     -1.12 \\ \hline 
Arizona 05 &       375 &     21.59 &     19.48 &     21.44 &    -14.55 &     68.23 &      1.57 \\ \hline 
Arizona 06 &       252 &     21.37 &     19.95 &     21.16 &     76.05 &     39.51 &      6.25 \\ \hline 
Arizona 07 &       133 &     23.03 &     18.27 &     22.50 &      9.03 &      5.27 &      7.34 \\ \hline 
Arizona 08 &       334 &     19.98 &     18.04 &     19.92 &     -6.32 &     16.88 &     -0.03 \\ \hline 
Arizona 09 &       214 &     23.97 &     20.95 &     23.44 &    -36.35 &     24.47 &     -3.11 \\ \hline 
Arkansas 01 &       127 &     19.85 &     16.68 &     19.72 &     -2.12 &      8.10 &     -0.39 \\ \hline 
Arkansas 02 &       415 &     20.80 &     17.22 &     20.72 &      1.72 &      3.43 &      1.75 \\ \hline 
Arkansas 03 &       234 &     23.63 &     20.17 &     23.20 &     14.72 &     47.25 &      2.50 \\ \hline 
Arkansas 04 &       131 &     23.13 &     19.22 &     22.54 &     -5.72 &     11.96 &     -0.98 \\ \hline 
California 01 &       220 &     16.17 &     13.87 &     16.28 &      5.56 &     83.94 &      2.02 \\ \hline 
California 02 &       125 &     16.96 &     15.52 &     17.12 &    -53.21 &     35.46 &     -1.90 \\ \hline 
California 03 &       264 &     19.31 &     14.91 &     19.28 &      0.84 &      8.83 &      1.36 \\ \hline 
California 04 &       290 &     18.73 &     18.47 &     18.73 &    -31.90 &     69.72 &      1.19 \\ \hline 
California 05 &       184 &     18.76 &     16.51 &     18.75 &     16.49 &      4.45 &     13.83 \\ \hline 
California 06 &       205 &     18.03 &     15.69 &     18.07 &      2.43 &     10.04 &      2.18 \\ \hline 
California 07 &       287 &     17.66 &     15.64 &     17.70 &     -2.79 &     12.09 &      0.16 \\ \hline 
California 08 &       164 &     23.89 &     21.22 &     23.21 &     51.09 &     34.65 &      5.53 \\ \hline 
California 09 &       257 &     16.83 &     14.00 &     16.90 &     18.14 &     11.60 &      8.40 \\ \hline 
California 10 &       247 &     16.91 &     14.61 &     16.99 &     27.82 &     28.69 &      4.57 \\ \hline 
California 11 &       274 &     18.64 &     15.72 &     18.64 &      6.48 &      7.40 &      4.75 \\ \hline 
California 12 &       145 &     17.46 &     20.08 &     17.62 &     17.74 &     29.13 &      3.51 \\ \hline 
California 13 &       133 &     21.35 &     20.38 &     20.96 &      7.39 &      8.95 &      4.81 \\ \hline 
California 14 &       174 &     21.43 &     18.91 &     21.15 &    -32.29 &     39.81 &      0.00 \\ \hline 
California 15 &       253 &     18.08 &     15.41 &     18.11 &      2.37 &     11.02 &      2.13 \\ \hline 
California 16 &       175 &     20.32 &     18.58 &     20.16 &     36.56 &     16.42 &     10.59 \\ \hline 
California 17 &       219 &     17.76 &     16.05 &     17.81 &      7.62 &     35.34 &      2.36 \\ \hline 
California 18 &       220 &     19.29 &     16.45 &     19.25 &    -35.51 &     33.45 &     -0.94 \\ \hline 
California 19 &       205 &     17.95 &     16.88 &     18.00 &    -17.27 &     36.32 &      0.67 \\ \hline 
California 20 &       112 &     19.68 &     18.48 &     19.54 &     89.54 &     19.14 &     19.24 \\ \hline 
California 21 &        74 &     17.97 &     14.37 &     18.06 &     -6.19 &     16.92 &      0.01 \\ \hline 
\hline
\end{tabular}
}
}
\end{center}
\vspace{-20pt}
\caption*{\scriptsize \textit{Notes}: Columns 5-7 (\textit{Disparity}) correspond to the coefficients on the interaction between a congressional district fixed effect and the ``Fraction Black'' variable from the voter-level regression of wait time on the full set of congressional district fixed effects and the interaction of those fixed effects with ``Fraction Black'', omitting the constant and  clustering standard errors at the polling place level. Column 7 provides  empirical-Bayes-adjusted estimates of these congressional-district-level disparities to account for measurement error. Similarly, Column 4 provides empirical-Bayes-adjusted estimates of the unadjusted congressional-district-level means shown in Column 2.}
\end{table}

\setcounter{table}{1} 
\begin{table}[H]
\begin{center}
\caption{Congressional District-Level Measures of Wait Time and Disparities (2)}
\vspace{-10pt}
\scalebox{.7}{{
\def\sym#1{\ifmmode^{#1}\else\(^{#1}\)\fi}
\begin{tabular}{l*{8}{c}}
\hline\hline
&\multicolumn{1}{c}{(1)}&\multicolumn{1}{c}{(2)}&\multicolumn{1}{c}{(3)}&\multicolumn{1}{c}{(4)}&\multicolumn{1}{c}{(5)}&\multicolumn{1}{c}{(6)}&\multicolumn{1}{c}{(7)}\\
              &                      &\multicolumn{2}{c}{\textbf{Unadjusted}}       & \textbf{Bayesian} &\multicolumn{2}{c}{\textbf{Unadjusted}}&\textbf{Bayesian}\\\cmidrule(lr){3-4}\cmidrule(lr){6-7}
\textbf{State \& District}&            \textbf{N}&        \textbf{Mean}&        \textbf{Std Dev}&   \textbf{Adjusted Mean}&      \textbf{Disparity}  & \textbf{Std Error}  & \textbf{Adjusted Disparity} \\
\hline
California 22 &       285 &     21.97 &     18.49 &     21.77 &    -63.14 &     24.93 &     -6.38 \\ \hline 
California 23 &       268 &     18.79 &     13.90 &     18.79 &      7.52 &     14.14 &      3.67 \\ \hline 
California 24 &       171 &     20.40 &     16.93 &     20.25 &    122.44 &     73.82 &      4.45 \\ \hline 
California 25 &       348 &     22.55 &     17.37 &     22.37 &     31.20 &     12.57 &     12.49 \\ \hline 
California 26 &       275 &     20.32 &     18.65 &     20.21 &     -7.38 &     22.44 &      0.53 \\ \hline 
California 27 &       214 &     19.71 &     15.37 &     19.65 &     17.63 &     11.57 &      8.21 \\ \hline 
California 28 &       189 &     22.92 &     19.69 &     22.49 &     -3.66 &     52.23 &      1.76 \\ \hline 
California 29 &       161 &     27.40 &     21.16 &     26.25 &      4.99 &     47.28 &      2.09 \\ \hline 
California 30 &       271 &     22.53 &     17.21 &     22.31 &     22.10 &     27.85 &      4.09 \\ \hline 
California 31 &        78 &     26.65 &     22.08 &     24.63 &     43.66 &     47.82 &      3.70 \\ \hline 
California 32 &       196 &     21.07 &     16.94 &     20.89 &      7.75 &     30.66 &      2.48 \\ \hline 
California 33 &       234 &     24.59 &     20.65 &     24.06 &    -39.32 &     28.35 &     -2.29 \\ \hline 
California 34 &       121 &     23.55 &     19.86 &     22.81 &      9.87 &     38.17 &      2.44 \\ \hline 
California 35 &       259 &     22.50 &     17.19 &     22.28 &      1.54 &     40.33 &      1.93 \\ \hline 
California 36 &       250 &     23.53 &     18.64 &     23.19 &     27.06 &     16.95 &      7.93 \\ \hline 
California 37 &       162 &     24.20 &     20.23 &     23.54 &      7.83 &      6.36 &      5.99 \\ \hline 
California 38 &       188 &     19.75 &     16.63 &     19.67 &     23.97 &     46.88 &      2.91 \\ \hline 
California 39 &       286 &     20.45 &     16.40 &     20.36 &    -61.49 &     31.34 &     -3.53 \\ \hline 
California 40 &       129 &     21.28 &     16.00 &     21.02 &    -42.47 &     23.12 &     -4.50 \\ \hline 
California 41 &       308 &     20.42 &     15.64 &     20.35 &     24.92 &     14.17 &      9.02 \\ \hline 
California 42 &       496 &     21.04 &     17.86 &     20.96 &     27.76 &     27.67 &      4.73 \\ \hline 
California 43 &       177 &     23.39 &     18.27 &     22.95 &      0.99 &      5.21 &      1.22 \\ \hline 
California 44 &       119 &     24.61 &     19.30 &     23.75 &    -20.00 &      6.73 &    -12.57 \\ \hline 
California 45 &       378 &     20.62 &     15.45 &     20.55 &    -26.77 &     22.53 &     -2.40 \\ \hline 
California 46 &       154 &     26.46 &     24.46 &     25.10 &     51.88 &    125.13 &      2.48 \\ \hline 
California 47 &       208 &     18.78 &     14.73 &     18.78 &      1.43 &     10.22 &      1.72 \\ \hline 
California 48 &       277 &     21.19 &     15.79 &     21.07 &    -47.61 &     40.00 &     -0.85 \\ \hline 
California 49 &       291 &     21.07 &     17.12 &     20.94 &    -12.97 &     67.91 &      1.61 \\ \hline 
California 50 &       357 &     18.31 &     15.20 &     18.32 &     51.96 &     34.81 &      5.56 \\ \hline 
California 51 &       141 &     22.20 &     19.01 &     21.77 &      0.06 &      9.88 &      1.06 \\ \hline 
California 52 &       286 &     20.97 &     19.78 &     20.81 &    103.50 &     55.78 &      5.23 \\ \hline 
California 53 &       239 &     17.21 &     14.28 &     17.28 &     20.12 &     19.98 &      5.31 \\ \hline 
Connecticut 01 &       590 &     10.91 &     10.57 &     10.99 &      4.77 &      2.59 &      4.57 \\ \hline 
Connecticut 02 &       529 &     11.38 &     10.97 &     11.47 &     -4.51 &      5.84 &     -2.71 \\ \hline 
Connecticut 03 &       508 &     12.60 &     13.40 &     12.71 &     18.98 &      7.79 &     12.06 \\ \hline 
Connecticut 04 &       545 &     13.67 &     12.75 &     13.75 &     16.69 &      6.49 &     11.94 \\ \hline 
Connecticut 05 &       550 &     13.37 &     14.39 &     13.48 &     20.93 &      9.29 &     11.57 \\ \hline 
Delaware 01 &       688 &     11.91 &     11.68 &     11.98 &      4.85 &      2.51 &      4.66 \\ \hline 
DistrictofColumbia 01 &       179 &     27.49 &     22.94 &     26.27 &      8.38 &      4.57 &      7.15 \\ \hline 
Florida 01 &       321 &     16.25 &     13.98 &     16.33 &      8.88 &      4.62 &      7.53 \\ \hline 
Florida 02 &       173 &     14.52 &     12.70 &     14.72 &      3.44 &      5.89 &      3.02 \\ \hline  
\hline
\end{tabular}
}
}
\end{center}
\vspace{-20pt}
\caption*{\scriptsize \textit{Notes}: Columns 5-7 (\textit{Disparity}) correspond to the coefficients on the interaction between a congressional district fixed effect and the ``Fraction Black'' variable from the voter-level regression of wait time on the full set of congressional district fixed effects and the interaction of those fixed effects with ``Fraction Black'', omitting the constant and  clustering standard errors at the polling place level. Column 7 provides  empirical-Bayes-adjusted estimates of these congressional-district-level disparities to account for measurement error. Similarly, Column 4 provides empirical-Bayes-adjusted estimates of the unadjusted congressional-district-level means shown in Column 2}
\end{table}

\setcounter{table}{1} 
\begin{table}[H]
\begin{center}
\caption{Congressional District-Level Measures of Wait Time and Disparities (3)}
\vspace{-10pt}
\scalebox{.7}{{
\def\sym#1{\ifmmode^{#1}\else\(^{#1}\)\fi}
\begin{tabular}{l*{8}{c}}
\hline\hline
&\multicolumn{1}{c}{(1)}&\multicolumn{1}{c}{(2)}&\multicolumn{1}{c}{(3)}&\multicolumn{1}{c}{(4)}&\multicolumn{1}{c}{(5)}&\multicolumn{1}{c}{(6)}&\multicolumn{1}{c}{(7)}\\
              &                      &\multicolumn{2}{c}{\textbf{Unadjusted}}       & \textbf{Bayesian} &\multicolumn{2}{c}{\textbf{Unadjusted}}&\textbf{Bayesian}\\\cmidrule(lr){3-4}\cmidrule(lr){6-7}
\textbf{State \& District}&            \textbf{N}&        \textbf{Mean}&        \textbf{Std Dev}&   \textbf{Adjusted Mean}&      \textbf{Disparity}  & \textbf{Std Error}  & \textbf{Adjusted Disparity} \\
\hline
Florida 03 &       288 &     17.20 &     15.61 &     17.27 &      0.10 &     11.53 &      1.21 \\ \hline 
Florida 04 &       285 &     13.07 &     10.86 &     13.20 &     -0.64 &      9.89 &      0.72 \\ \hline 
Florida 05 &       170 &     13.69 &     13.61 &     13.96 &      0.13 &      5.41 &      0.58 \\ \hline 
Florida 06 &       299 &     17.94 &     15.36 &     17.97 &      8.52 &      7.08 &      6.15 \\ \hline 
Florida 07 &       277 &     15.61 &     13.57 &     15.72 &     -3.66 &     12.73 &     -0.03 \\ \hline 
Florida 08 &       341 &     15.50 &     14.66 &     15.60 &      0.95 &      4.51 &      1.14 \\ \hline 
Florida 09 &       292 &     18.28 &     15.31 &     18.30 &    -18.70 &      7.96 &    -10.09 \\ \hline 
Florida 10 &       249 &     19.59 &     15.15 &     19.55 &     -5.24 &      4.70 &     -3.80 \\ \hline 
Florida 11 &       300 &     16.23 &     12.74 &     16.30 &     -8.30 &      8.83 &     -3.50 \\ \hline 
Florida 12 &       499 &     17.85 &     13.74 &     17.87 &     -1.81 &     16.83 &      1.05 \\ \hline 
Florida 13 &       261 &     18.75 &     15.52 &     18.75 &      5.58 &      5.05 &      4.76 \\ \hline 
Florida 14 &       215 &     17.35 &     13.36 &     17.40 &     -0.79 &      4.16 &     -0.34 \\ \hline 
Florida 15 &       397 &     17.41 &     13.34 &     17.44 &     -0.17 &      8.50 &      0.79 \\ \hline 
Florida 16 &       346 &     17.68 &     14.68 &     17.71 &     20.91 &     14.74 &      7.48 \\ \hline 
Florida 17 &       261 &     16.23 &     14.24 &     16.33 &     18.88 &     15.56 &      6.52 \\ \hline 
Florida 18 &       304 &     19.42 &     17.48 &     19.38 &     -3.28 &     11.49 &     -0.15 \\ \hline 
Florida 19 &       215 &     18.26 &     16.86 &     18.28 &     22.18 &     15.83 &      7.28 \\ \hline 
Florida 20 &       152 &     20.92 &     18.13 &     20.68 &     -6.21 &      6.94 &     -3.33 \\ \hline 
Florida 21 &       348 &     20.65 &     17.86 &     20.56 &     -5.90 &      5.20 &     -4.06 \\ \hline 
Florida 22 &       305 &     20.27 &     17.63 &     20.18 &      9.94 &      9.64 &      5.86 \\ \hline 
Florida 23 &       248 &     23.11 &     19.70 &     22.76 &      6.53 &     13.91 &      3.40 \\ \hline 
Florida 24 &       120 &     21.20 &     16.45 &     20.92 &      6.95 &      6.19 &      5.44 \\ \hline 
Florida 25 &       193 &     22.54 &     18.68 &     22.19 &    -30.83 &     36.47 &     -0.22 \\ \hline 
Florida 26 &       173 &     18.64 &     14.80 &     18.65 &     21.82 &      8.50 &     12.90 \\ \hline 
Florida 27 &       138 &     22.97 &     20.99 &     22.33 &     12.87 &     15.71 &      4.86 \\ \hline 
Georgia 01 &       291 &     25.32 &     20.86 &     24.82 &      8.96 &      6.85 &      6.53 \\ \hline 
Georgia 02 &       255 &     15.21 &     12.71 &     15.32 &      5.79 &      3.11 &      5.41 \\ \hline 
Georgia 03 &       385 &     16.06 &     14.22 &     16.13 &     -3.36 &      3.20 &     -2.81 \\ \hline 
Georgia 04 &       294 &     20.11 &     18.03 &     20.03 &     -0.85 &      3.78 &     -0.46 \\ \hline 
Georgia 05 &       273 &     23.84 &     19.13 &     23.49 &    -11.33 &      3.36 &     -9.82 \\ \hline 
Georgia 06 &       644 &     17.45 &     15.59 &     17.48 &      3.30 &      5.60 &      2.95 \\ \hline 
Georgia 07 &       676 &     28.59 &     24.64 &     28.12 &     31.24 &      9.64 &     16.26 \\ \hline 
Georgia 08 &       207 &     15.55 &     11.80 &     15.66 &     -9.62 &      5.87 &     -6.37 \\ \hline 
Georgia 09 &       324 &     16.29 &     12.35 &     16.35 &     -1.94 &     11.15 &      0.33 \\ \hline 
Georgia 10 &       316 &     21.63 &     20.05 &     21.44 &     10.10 &     10.72 &      5.51 \\ \hline 
Georgia 11 &       655 &     18.85 &     15.42 &     18.84 &     14.01 &      4.98 &     11.37 \\ \hline 
Georgia 12 &       199 &     14.38 &     13.10 &     14.57 &      2.04 &      3.00 &      2.03 \\ \hline 
Georgia 13 &       310 &     23.77 &     20.21 &     23.43 &      6.10 &      7.85 &      4.40 \\ \hline 
Georgia 14 &       229 &     15.10 &     12.00 &     15.22 &      2.32 &      7.38 &      2.18 \\ \hline 
Idaho 01 &       665 &     20.07 &     14.84 &     20.04 &    -15.27 &     23.49 &     -0.48 \\ \hline 
Idaho 02 &       609 &     18.31 &     15.11 &     18.32 &     51.18 &     57.40 &      3.47 \\ \hline 
\hline
\end{tabular}
}
}
\end{center}
\vspace{-20pt}
\caption*{\scriptsize \textit{Notes}: Columns 5-7 (\textit{Disparity}) correspond to the coefficients on the interaction between a congressional district fixed effect and the ``Fraction Black'' variable from the voter-level regression of wait time on the full set of congressional district fixed effects and the interaction of those fixed effects with ``Fraction Black'', omitting the constant and  clustering standard errors at the polling place level. Column 7 provides  empirical-Bayes-adjusted estimates of these congressional-district-level disparities to account for measurement error. Similarly, Column 4 provides empirical-Bayes-adjusted estimates of the unadjusted congressional-district-level means shown in Column 2}
\end{table}

\setcounter{table}{1} 
\begin{table}[H]
\begin{center}
\caption{Congressional District-Level Measures of Wait Time and Disparities (4)}
\vspace{-10pt}
\scalebox{.7}{{
\def\sym#1{\ifmmode^{#1}\else\(^{#1}\)\fi}
\begin{tabular}{l*{8}{c}}
\hline\hline
&\multicolumn{1}{c}{(1)}&\multicolumn{1}{c}{(2)}&\multicolumn{1}{c}{(3)}&\multicolumn{1}{c}{(4)}&\multicolumn{1}{c}{(5)}&\multicolumn{1}{c}{(6)}&\multicolumn{1}{c}{(7)}\\
              &                      &\multicolumn{2}{c}{\textbf{Unadjusted}}       & \textbf{Bayesian} &\multicolumn{2}{c}{\textbf{Unadjusted}}&\textbf{Bayesian}\\\cmidrule(lr){3-4}\cmidrule(lr){6-7}
\textbf{State \& District}&            \textbf{N}&        \textbf{Mean}&        \textbf{Std Dev}&   \textbf{Adjusted Mean}&      \textbf{Disparity}  & \textbf{Std Error}  & \textbf{Adjusted Disparity} \\
\hline
Illinois 01 &       295 &     17.07 &     14.99 &     17.14 &      6.56 &      2.83 &      6.17 \\ \hline 
Illinois 02 &       224 &     18.22 &     14.32 &     18.24 &      4.82 &      2.60 &      4.61 \\ \hline 
Illinois 03 &       272 &     18.10 &     13.38 &     18.12 &     11.08 &     11.55 &      5.61 \\ \hline 
Illinois 04 &       113 &     20.94 &     17.38 &     20.65 &    -48.99 &     23.93 &     -5.03 \\ \hline 
Illinois 05 &       183 &     22.94 &     20.49 &     22.47 &     32.84 &     69.70 &      2.65 \\ \hline 
Illinois 06 &       546 &     15.67 &     14.82 &     15.73 &     66.64 &     41.00 &      5.47 \\ \hline 
Illinois 07 &       174 &     22.46 &     20.80 &     22.01 &      1.61 &      5.42 &      1.69 \\ \hline 
Illinois 08 &       412 &     16.28 &     11.54 &     16.32 &    -17.80 &      8.66 &     -8.74 \\ \hline 
Illinois 09 &       270 &     17.33 &     13.99 &     17.38 &     -5.20 &     11.30 &     -0.98 \\ \hline 
Illinois 10 &       416 &     16.49 &     13.66 &     16.54 &     19.07 &     14.37 &      7.12 \\ \hline 
Illinois 11 &       588 &     14.74 &     11.43 &     14.79 &      7.47 &      7.64 &      5.28 \\ \hline 
Illinois 12 &       366 &     13.75 &     11.24 &     13.84 &      7.27 &      3.51 &      6.62 \\ \hline 
Illinois 13 &       403 &     15.10 &     13.22 &     15.18 &     -5.28 &      4.02 &     -4.16 \\ \hline 
Illinois 14 &       669 &     14.04 &     11.85 &     14.09 &     -0.57 &     12.51 &      1.04 \\ \hline 
Illinois 15 &       222 &     14.04 &     12.16 &     14.20 &      2.33 &      4.75 &      2.26 \\ \hline 
Illinois 16 &       361 &     14.90 &     13.19 &     14.99 &     19.88 &      7.84 &     12.54 \\ \hline 
Illinois 17 &       210 &     16.57 &     14.09 &     16.67 &     10.21 &      9.22 &      6.17 \\ \hline 
Illinois 18 &       488 &     14.03 &     11.03 &     14.10 &      5.18 &      9.26 &      3.59 \\ \hline 
Indiana 01 &       289 &     16.18 &     15.27 &     16.28 &      2.35 &      3.80 &      2.29 \\ \hline 
Indiana 02 &       484 &     26.12 &     19.25 &     25.82 &    -25.16 &      7.96 &    -13.85 \\ \hline 
Indiana 03 &       588 &     29.81 &     23.12 &     29.29 &    -24.60 &     11.62 &     -8.58 \\ \hline 
Indiana 04 &       412 &     30.73 &     24.71 &     29.83 &    -10.91 &     25.30 &      0.35 \\ \hline 
Indiana 05 &       823 &     38.27 &     29.21 &     37.23 &    -65.07 &     10.93 &    -26.63 \\ \hline 
Indiana 06 &       329 &     22.17 &     17.77 &     22.00 &      6.34 &     16.75 &      3.02 \\ \hline 
Indiana 07 &       532 &     23.76 &     21.34 &     23.53 &    -14.32 &      9.25 &     -6.33 \\ \hline 
Indiana 08 &       324 &     21.95 &     16.61 &     21.81 &      3.32 &      4.29 &      3.08 \\ \hline 
Indiana 09 &       505 &     20.12 &     16.81 &     20.08 &    -14.71 &     19.85 &     -1.15 \\ \hline 
Iowa 01 &       368 &     14.52 &     13.46 &     14.62 &    -18.53 &      6.90 &    -11.36 \\ \hline 
Iowa 02 &       374 &     15.60 &     13.09 &     15.67 &    -17.36 &      9.45 &     -7.67 \\ \hline 
Iowa 03 &       610 &     15.89 &     12.45 &     15.92 &     -6.02 &      7.23 &     -3.06 \\ \hline 
Iowa 04 &       315 &     15.49 &     14.22 &     15.60 &     34.82 &     28.39 &      5.33 \\ \hline 
Kansas 01 &       220 &     15.96 &     13.53 &     16.08 &    -34.21 &     26.88 &     -2.12 \\ \hline 
Kansas 02 &       305 &     16.43 &     14.54 &     16.51 &     -0.77 &      9.67 &      0.63 \\ \hline 
Kansas 03 &       582 &     14.85 &     14.88 &     14.92 &     11.41 &      8.84 &      6.98 \\ \hline 
Kansas 04 &       381 &     17.76 &     11.17 &     17.78 &      9.70 &      4.91 &      8.04 \\ \hline 
Kentucky 01 &       277 &     12.69 &      9.34 &     12.79 &      2.75 &      4.87 &      2.58 \\ \hline 
Kentucky 02 &       627 &     15.25 &     13.04 &     15.30 &     21.00 &     20.37 &      5.36 \\ \hline 
Kentucky 03 &       775 &     11.65 &      9.17 &     11.69 &      1.08 &      1.73 &      1.11 \\ \hline 
Kentucky 04 &       720 &     17.54 &     14.65 &     17.55 &      5.64 &     16.91 &      2.84 \\ \hline 
Kentucky 05 &       170 &     15.06 &     15.60 &     15.32 &     -6.24 &     30.51 &      1.21 \\ \hline 
Kentucky 06 &       597 &     15.04 &     12.94 &     15.09 &      1.07 &      6.59 &      1.36 \\ \hline 
\hline
\end{tabular}
}
}
\end{center}
\vspace{-20pt}
\caption*{\scriptsize \textit{Notes}: Columns 5-7 (\textit{Disparity}) correspond to the coefficients on the interaction between a congressional district fixed effect and the ``Fraction Black'' variable from the voter-level regression of wait time on the full set of congressional district fixed effects and the interaction of those fixed effects with ``Fraction Black'', omitting the constant and  clustering standard errors at the polling place level. Column 7 provides  empirical-Bayes-adjusted estimates of these congressional-district-level disparities to account for measurement error. Similarly, Column 4 provides empirical-Bayes-adjusted estimates of the unadjusted congressional-district-level means shown in Column 2}
\end{table}

\setcounter{table}{1} 
\begin{table}[H]
\begin{center}
\caption{Congressional District-Level Measures of Wait Time and Disparities (5)}
\vspace{-10pt}
\scalebox{.7}{{
\def\sym#1{\ifmmode^{#1}\else\(^{#1}\)\fi}
\begin{tabular}{l*{8}{c}}
\hline\hline
&\multicolumn{1}{c}{(1)}&\multicolumn{1}{c}{(2)}&\multicolumn{1}{c}{(3)}&\multicolumn{1}{c}{(4)}&\multicolumn{1}{c}{(5)}&\multicolumn{1}{c}{(6)}&\multicolumn{1}{c}{(7)}\\
              &                      &\multicolumn{2}{c}{\textbf{Unadjusted}}       & \textbf{Bayesian} &\multicolumn{2}{c}{\textbf{Unadjusted}}&\textbf{Bayesian}\\\cmidrule(lr){3-4}\cmidrule(lr){6-7}
\textbf{State \& District}&            \textbf{N}&        \textbf{Mean}&        \textbf{Std Dev}&   \textbf{Adjusted Mean}&      \textbf{Disparity}  & \textbf{Std Error}  & \textbf{Adjusted Disparity} \\
\hline
Louisiana 01 &       547 &     16.39 &     15.97 &     16.45 &      9.22 &      7.71 &      6.30 \\ \hline 
Louisiana 02 &       350 &     17.31 &     15.47 &     17.35 &      1.29 &      3.86 &      1.39 \\ \hline 
Louisiana 03 &       506 &     15.84 &     13.13 &     15.89 &     -5.65 &      2.82 &     -5.03 \\ \hline 
Louisiana 04 &       370 &     15.92 &     14.27 &     15.99 &     -4.23 &      1.97 &     -3.97 \\ \hline 
Louisiana 05 &       148 &     17.86 &     15.38 &     17.92 &     -5.16 &      3.66 &     -4.22 \\ \hline 
Louisiana 06 &       482 &     14.66 &     12.05 &     14.72 &     -1.84 &      3.65 &     -1.34 \\ \hline 
Maine 01 &       334 &     17.49 &     13.91 &     17.53 &     16.58 &     20.95 &      4.46 \\ \hline 
Maine 02 &       129 &     18.09 &     17.96 &     18.16 &    142.74 &    176.10 &      3.04 \\ \hline 
Maryland 01 &       705 &     15.88 &     13.13 &     15.91 &      4.40 &      5.27 &      3.82 \\ \hline 
Maryland 02 &       674 &     26.57 &     20.54 &     26.31 &      5.70 &      4.31 &      5.05 \\ \hline 
Maryland 03 &       672 &     24.44 &     19.76 &     24.26 &     -4.15 &      9.02 &     -1.23 \\ \hline 
Maryland 04 &       555 &     23.69 &     17.87 &     23.54 &     -1.60 &      3.78 &     -1.10 \\ \hline 
Maryland 05 &       583 &     18.00 &     14.15 &     18.01 &      0.32 &      2.70 &      0.44 \\ \hline 
Maryland 06 &       695 &     16.30 &     12.97 &     16.33 &      9.31 &      6.14 &      7.11 \\ \hline 
Maryland 07 &       445 &     22.49 &     18.25 &     22.34 &      9.23 &      3.78 &      8.22 \\ \hline 
Maryland 08 &       620 &     17.48 &     13.89 &     17.50 &     13.27 &      6.79 &      9.39 \\ \hline 
Massachusetts 01 &       270 &     12.42 &     13.01 &     12.62 &      9.62 &     13.56 &      4.46 \\ \hline 
Massachusetts 02 &       376 &     12.07 &      9.65 &     12.16 &      2.97 &      7.51 &      2.58 \\ \hline 
Massachusetts 03 &       355 &     11.78 &     10.83 &     11.90 &     13.01 &     14.38 &      5.29 \\ \hline 
Massachusetts 04 &       278 &     12.33 &      9.45 &     12.44 &      3.45 &      9.43 &      2.70 \\ \hline 
Massachusetts 05 &       241 &     11.70 &      8.88 &     11.82 &      6.74 &      4.49 &      5.85 \\ \hline 
Massachusetts 06 &       336 &     10.89 &      9.33 &     11.00 &     22.43 &     12.30 &      9.54 \\ \hline 
Massachusetts 07 &       179 &     18.02 &     15.84 &     18.07 &     -4.32 &      4.49 &     -3.15 \\ \hline 
Massachusetts 08 &       331 &     12.77 &     11.77 &     12.90 &     10.33 &      9.15 &      6.26 \\ \hline 
Massachusetts 09 &       289 &     11.09 &      9.32 &     11.21 &     49.75 &     26.89 &      7.34 \\ \hline 
Michigan 01 &       316 &     19.24 &     15.51 &     19.22 &    -18.95 &     25.02 &     -0.71 \\ \hline 
Michigan 02 &       777 &     19.66 &     13.62 &     19.65 &      2.35 &      8.66 &      2.17 \\ \hline 
Michigan 03 &       667 &     21.97 &     15.90 &     21.90 &      5.80 &      6.72 &      4.50 \\ \hline 
Michigan 04 &       450 &     20.28 &     15.21 &     20.23 &    -13.66 &      7.45 &     -7.64 \\ \hline 
Michigan 05 &       589 &     23.29 &     17.00 &     23.17 &      7.51 &      3.78 &      6.73 \\ \hline 
Michigan 06 &       559 &     24.80 &     17.53 &     24.62 &      3.54 &      6.79 &      3.00 \\ \hline 
Michigan 07 &       603 &     20.62 &     14.20 &     20.59 &     11.11 &      8.62 &      6.94 \\ \hline 
Michigan 08 &     1,022 &     21.62 &     16.50 &     21.58 &     -2.36 &      8.49 &     -0.42 \\ \hline 
Michigan 09 &       874 &     20.22 &     14.19 &     20.20 &      2.14 &      5.99 &      2.09 \\ \hline 
Michigan 10 &       854 &     18.98 &     13.54 &     18.98 &      8.32 &      7.08 &      6.02 \\ \hline 
Michigan 11 &     1,154 &     23.13 &     16.65 &     23.07 &     20.40 &     11.38 &      9.46 \\ \hline 
Michigan 12 &       722 &     24.32 &     17.89 &     24.19 &     18.07 &     11.31 &      8.56 \\ \hline 
Michigan 13 &       538 &     26.03 &     20.30 &     25.73 &     13.96 &      3.30 &     12.64 \\ \hline 
Michigan 14 &       651 &     28.15 &     19.34 &     27.86 &      6.67 &      3.30 &      6.16 \\ \hline 
Minnesota 01 &       347 &     14.31 &     10.08 &     14.38 &     -2.88 &      8.75 &     -0.64 \\ \hline 
Minnesota 02 &       903 &     13.60 &     10.42 &     13.63 &     14.82 &     10.52 &      7.68 \\ \hline 
\hline
\end{tabular}
}
}
\end{center}
\vspace{-20pt}
\caption*{\scriptsize \textit{Notes}: Columns 5-7 (\textit{Disparity}) correspond to the coefficients on the interaction between a congressional district fixed effect and the ``Fraction Black'' variable from the voter-level regression of wait time on the full set of congressional district fixed effects and the interaction of those fixed effects with ``Fraction Black'', omitting the constant and  clustering standard errors at the polling place level. Column 7 provides  empirical-Bayes-adjusted estimates of these congressional-district-level disparities to account for measurement error. Similarly, Column 4 provides empirical-Bayes-adjusted estimates of the unadjusted congressional-district-level means shown in Column 2}
\end{table}

\setcounter{table}{1} 
\begin{table}[H]
\begin{center}
\caption{Congressional District-Level Measures of Wait Time and Disparities (6)}
\vspace{-10pt}
\scalebox{.7}{{
\def\sym#1{\ifmmode^{#1}\else\(^{#1}\)\fi}
\begin{tabular}{l*{8}{c}}
\hline\hline
&\multicolumn{1}{c}{(1)}&\multicolumn{1}{c}{(2)}&\multicolumn{1}{c}{(3)}&\multicolumn{1}{c}{(4)}&\multicolumn{1}{c}{(5)}&\multicolumn{1}{c}{(6)}&\multicolumn{1}{c}{(7)}\\
              &                      &\multicolumn{2}{c}{\textbf{Unadjusted}}       & \textbf{Bayesian} &\multicolumn{2}{c}{\textbf{Unadjusted}}&\textbf{Bayesian}\\\cmidrule(lr){3-4}\cmidrule(lr){6-7}
\textbf{State \& District}&            \textbf{N}&        \textbf{Mean}&        \textbf{Std Dev}&   \textbf{Adjusted Mean}&      \textbf{Disparity}  & \textbf{Std Error}  & \textbf{Adjusted Disparity} \\
\hline
Minnesota 03 &       874 &     16.80 &     13.69 &     16.82 &     24.64 &     12.34 &     10.33 \\ \hline 
Minnesota 04 &       642 &     16.19 &     13.73 &     16.23 &     -0.06 &      4.62 &      0.33 \\ \hline 
Minnesota 05 &       384 &     17.50 &     16.45 &     17.54 &     -0.72 &      6.16 &      0.08 \\ \hline 
Minnesota 06 &       855 &     14.43 &     11.08 &     14.47 &     -1.96 &      8.82 &     -0.13 \\ \hline 
Minnesota 07 &       235 &     14.44 &     13.34 &     14.60 &     20.56 &     38.51 &      3.08 \\ \hline 
Minnesota 08 &       286 &     15.02 &     10.84 &     15.10 &     63.07 &     24.95 &      9.77 \\ \hline 
Mississippi 01 &       332 &     16.95 &     13.65 &     17.00 &     -8.22 &      6.37 &     -5.03 \\ \hline 
Mississippi 02 &       153 &     16.29 &     15.70 &     16.48 &      8.30 &      5.66 &      6.61 \\ \hline 
Mississippi 03 &       248 &     14.71 &     13.86 &     14.87 &     -6.50 &      3.49 &     -5.47 \\ \hline 
Mississippi 04 &       266 &     22.34 &     19.06 &     22.08 &     -4.54 &      5.30 &     -2.97 \\ \hline 
Missouri 01 &       634 &     29.49 &     20.63 &     29.11 &     10.67 &      3.11 &      9.81 \\ \hline 
Missouri 02 &     1,408 &     22.82 &     16.61 &     22.78 &      2.38 &     13.41 &      2.10 \\ \hline 
Missouri 03 &       814 &     20.43 &     16.33 &     20.40 &     18.00 &     14.16 &      6.90 \\ \hline 
Missouri 04 &       422 &     20.65 &     17.37 &     20.57 &     31.80 &     20.47 &      7.25 \\ \hline 
Missouri 05 &       830 &     40.97 &     27.05 &     39.95 &     -2.95 &      6.50 &     -1.36 \\ \hline 
Missouri 06 &       980 &     30.85 &     23.01 &     30.51 &     26.29 &     28.53 &      4.43 \\ \hline 
Missouri 07 &       906 &     20.72 &     14.54 &     20.70 &     75.62 &     32.40 &      7.98 \\ \hline 
Missouri 08 &       237 &     17.03 &     12.86 &     17.09 &    -13.13 &     10.37 &     -4.86 \\ \hline 
Montana 01 &       307 &     20.53 &     16.56 &     20.45 &   -117.11 &     92.37 &      0.17 \\ \hline 
Nebraska 01 &       485 &     17.52 &     17.83 &     17.56 &     78.02 &     34.86 &      7.43 \\ \hline 
Nebraska 02 &       615 &     16.06 &     15.14 &     16.11 &      8.69 &     10.25 &      5.04 \\ \hline 
Nebraska 03 &       255 &     16.17 &     14.35 &     16.27 &    -10.56 &     22.53 &      0.06 \\ \hline 
Nevada 01 &       163 &     15.01 &     12.14 &     15.18 &    -13.11 &     10.09 &     -5.06 \\ \hline 
Nevada 02 &       291 &     16.06 &     14.62 &     16.16 &     -2.79 &     19.71 &      1.06 \\ \hline 
Nevada 03 &       294 &     14.15 &     11.79 &     14.26 &     -5.37 &     14.39 &     -0.25 \\ \hline 
Nevada 04 &       228 &     17.62 &     17.18 &     17.69 &      3.20 &     12.74 &      2.40 \\ \hline 
NewHampshire 01 &       755 &     16.16 &     12.77 &     16.19 &     -3.35 &     11.18 &     -0.25 \\ \hline 
NewHampshire 02 &       570 &     14.58 &     11.09 &     14.62 &     -8.54 &     21.88 &      0.29 \\ \hline 
NewJersey 01 &       432 &     12.18 &     10.03 &     12.26 &      3.56 &      3.71 &      3.34 \\ \hline 
NewJersey 02 &       324 &     14.45 &     12.68 &     14.56 &     12.78 &      7.42 &      8.63 \\ \hline 
NewJersey 03 &       411 &     13.06 &     14.13 &     13.21 &     -3.95 &      2.49 &     -3.56 \\ \hline 
NewJersey 04 &       458 &     11.17 &      9.65 &     11.25 &     -3.30 &      5.38 &     -2.01 \\ \hline 
NewJersey 05 &       415 &     13.93 &     13.78 &     14.04 &     -7.09 &      9.67 &     -2.45 \\ \hline 
NewJersey 06 &       426 &     13.98 &     13.25 &     14.09 &     13.04 &      9.82 &      7.27 \\ \hline 
NewJersey 07 &       566 &     14.28 &     13.59 &     14.35 &     12.02 &     11.34 &      6.07 \\ \hline 
NewJersey 08 &        64 &     21.06 &     21.45 &     20.39 &      9.94 &     10.50 &      5.52 \\ \hline 
NewJersey 09 &       252 &     13.16 &     11.61 &     13.32 &     16.17 &      5.74 &     12.31 \\ \hline 
NewJersey 10 &       194 &     18.30 &     14.51 &     18.32 &     -5.83 &      4.06 &     -4.60 \\ \hline 
NewJersey 11 &       447 &     12.84 &     14.02 &     12.98 &     51.94 &     20.36 &     10.90 \\ \hline 
NewJersey 12 &       457 &     16.47 &     14.40 &     16.53 &     -9.00 &      3.25 &     -7.83 \\ \hline 
NewMexico 01 &       171 &     19.62 &     14.14 &     19.57 &    -43.72 &     29.29 &     -2.49 \\ \hline 
\hline
\end{tabular}
}
}
\end{center}
\vspace{-20pt}
\caption*{\scriptsize \textit{Notes}: Columns 5-7 (\textit{Disparity}) correspond to the coefficients on the interaction between a congressional district fixed effect and the ``Fraction Black'' variable from the voter-level regression of wait time on the full set of congressional district fixed effects and the interaction of those fixed effects with ``Fraction Black'', omitting the constant and  clustering standard errors at the polling place level. Column 7 provides  empirical-Bayes-adjusted estimates of these congressional-district-level disparities to account for measurement error. Similarly, Column 4 provides empirical-Bayes-adjusted estimates of the unadjusted congressional-district-level means shown in Column 2}
\end{table}

\setcounter{table}{1} 
\begin{table}[H]
\begin{center}
\caption{Congressional District-Level Measures of Wait Time and Disparities (7)}
\vspace{-10pt}
\scalebox{.7}{{
\def\sym#1{\ifmmode^{#1}\else\(^{#1}\)\fi}
\begin{tabular}{l*{8}{c}}
\hline\hline
&\multicolumn{1}{c}{(1)}&\multicolumn{1}{c}{(2)}&\multicolumn{1}{c}{(3)}&\multicolumn{1}{c}{(4)}&\multicolumn{1}{c}{(5)}&\multicolumn{1}{c}{(6)}&\multicolumn{1}{c}{(7)}\\
              &                      &\multicolumn{2}{c}{\textbf{Unadjusted}}       & \textbf{Bayesian} &\multicolumn{2}{c}{\textbf{Unadjusted}}&\textbf{Bayesian}\\\cmidrule(lr){3-4}\cmidrule(lr){6-7}
\textbf{State \& District}&            \textbf{N}&        \textbf{Mean}&        \textbf{Std Dev}&   \textbf{Adjusted Mean}&      \textbf{Disparity}  & \textbf{Std Error}  & \textbf{Adjusted Disparity} \\
\hline
NewMexico 02 &       160 &     17.94 &     14.56 &     17.99 &    -35.96 &     32.87 &     -1.06 \\ \hline 
NewMexico 03 &       153 &     17.93 &     14.79 &     17.98 &      4.21 &     69.60 &      2.01 \\ \hline 
NewYork 01 &       743 &     14.97 &     13.46 &     15.02 &     15.31 &     20.04 &      4.41 \\ \hline 
NewYork 02 &       615 &     13.90 &     11.93 &     13.95 &     17.19 &      4.66 &     14.19 \\ \hline 
NewYork 03 &       469 &     13.35 &     10.60 &     13.41 &     22.03 &      9.53 &     11.87 \\ \hline 
NewYork 05 &       379 &     25.07 &     18.19 &     24.78 &      2.50 &      3.10 &      2.45 \\ \hline 
NewYork 06 &       327 &     18.98 &     14.34 &     18.97 &     24.18 &     12.33 &     10.17 \\ \hline 
NewYork 07 &       147 &     19.50 &     16.49 &     19.43 &     17.89 &     20.99 &      4.67 \\ \hline 
NewYork 08 &       260 &     20.29 &     15.11 &     20.21 &      4.75 &      3.28 &      4.45 \\ \hline 
NewYork 09 &       218 &     22.81 &     17.72 &     22.50 &      4.34 &      3.09 &      4.11 \\ \hline 
NewYork 10 &       236 &     19.51 &     15.92 &     19.46 &    -24.72 &     22.98 &     -1.96 \\ \hline 
NewYork 11 &       413 &     14.70 &     11.88 &     14.77 &     -0.13 &      1.87 &     -0.05 \\ \hline 
NewYork 12 &       277 &     20.29 &     18.41 &     20.19 &    -19.39 &      7.52 &    -11.07 \\ \hline 
NewYork 13 &       145 &     22.41 &     18.98 &     21.97 &      6.72 &      6.31 &      5.24 \\ \hline 
NewYork 14 &       205 &     22.52 &     19.45 &     22.17 &     -3.87 &     16.19 &      0.47 \\ \hline 
NewYork 15 &       174 &     20.66 &     16.69 &     20.50 &     -0.55 &      7.67 &      0.45 \\ \hline 
NewYork 16 &        73 &     21.98 &     17.93 &     21.34 &     -9.23 &      8.91 &     -3.94 \\ \hline 
NewYork 17 &       159 &     14.83 &     13.00 &     15.04 &     -5.32 &     10.97 &     -1.13 \\ \hline 
NewYork 18 &       402 &     13.87 &     11.36 &     13.95 &      3.31 &      4.08 &      3.10 \\ \hline 
NewYork 19 &       216 &     14.32 &     11.77 &     14.46 &     10.70 &     12.73 &      5.06 \\ \hline 
NewYork 20 &       291 &     11.98 &     11.19 &     12.14 &      5.93 &     11.29 &      3.59 \\ \hline 
NewYork 21 &       141 &     14.79 &     15.69 &     15.12 &     -8.80 &     16.68 &     -0.66 \\ \hline 
NewYork 22 &       255 &     14.41 &     13.42 &     14.57 &     -8.62 &     16.68 &     -0.62 \\ \hline 
NewYork 23 &       135 &     12.04 &      9.72 &     12.28 &     38.25 &     22.84 &      7.34 \\ \hline 
NewYork 24 &       535 &     16.70 &     14.23 &     16.74 &     11.57 &      4.89 &      9.52 \\ \hline 
NewYork 25 &       545 &     15.25 &     15.10 &     15.32 &      0.59 &      6.17 &      1.00 \\ \hline 
NewYork 26 &       253 &     13.08 &     13.02 &     13.28 &      6.15 &      5.45 &      5.10 \\ \hline 
NewYork 27 &       279 &     13.50 &     11.90 &     13.63 &     44.88 &     36.55 &      4.80 \\ \hline 
NorthCarolina 01 &       178 &     19.49 &     15.17 &     19.44 &     -4.35 &      5.10 &     -2.91 \\ \hline 
NorthCarolina 02 &       558 &     24.58 &     19.29 &     24.37 &     -0.10 &      7.07 &      0.64 \\ \hline 
NorthCarolina 03 &       168 &     19.24 &     16.62 &     19.19 &     -8.91 &      7.76 &     -4.51 \\ \hline 
NorthCarolina 04 &       418 &     22.62 &     18.38 &     22.45 &     19.61 &      7.05 &     13.26 \\ \hline 
NorthCarolina 05 &       263 &     18.70 &     16.35 &     18.70 &     15.29 &      8.64 &      9.19 \\ \hline 
NorthCarolina 06 &       306 &     18.30 &     13.12 &     18.31 &      3.74 &      5.27 &      3.31 \\ \hline 
NorthCarolina 07 &       239 &     17.63 &     13.44 &     17.67 &      3.19 &      6.32 &      2.81 \\ \hline 
NorthCarolina 08 &       381 &     19.74 &     15.93 &     19.70 &      0.04 &      7.30 &      0.76 \\ \hline 
NorthCarolina 09 &       372 &     20.20 &     15.74 &     20.15 &      4.74 &      6.73 &      3.80 \\ \hline 
NorthCarolina 10 &       256 &     15.44 &     12.51 &     15.54 &     11.24 &     12.18 &      5.44 \\ \hline 
NorthCarolina 11 &       176 &     17.66 &     16.80 &     17.74 &    -31.10 &     19.22 &     -4.53 \\ \hline 
NorthCarolina 12 &       405 &     25.52 &     19.28 &     25.19 &      3.11 &      5.23 &      2.84 \\ \hline 
NorthCarolina 13 &       341 &     19.15 &     15.76 &     19.13 &     -3.33 &      3.56 &     -2.67 \\ \hline 
\hline
\end{tabular}
}
}
\end{center}
\vspace{-20pt}
\caption*{\scriptsize \textit{Notes}: Columns 5-7 (\textit{Disparity}) correspond to the coefficients on the interaction between a congressional district fixed effect and the ``Fraction Black'' variable from the voter-level regression of wait time on the full set of congressional district fixed effects and the interaction of those fixed effects with ``Fraction Black'', omitting the constant and  clustering standard errors at the polling place level. Column 7 provides  empirical-Bayes-adjusted estimates of these congressional-district-level disparities to account for measurement error. Similarly, Column 4 provides empirical-Bayes-adjusted estimates of the unadjusted congressional-district-level means shown in Column 2}
\end{table}

\setcounter{table}{1} 
\begin{table}[H]
\begin{center}
\caption{Congressional District-Level Measures of Wait Time and Disparities (8)}
\vspace{-10pt}
\scalebox{.7}{{
\def\sym#1{\ifmmode^{#1}\else\(^{#1}\)\fi}
\begin{tabular}{l*{8}{c}}
\hline\hline
&\multicolumn{1}{c}{(1)}&\multicolumn{1}{c}{(2)}&\multicolumn{1}{c}{(3)}&\multicolumn{1}{c}{(4)}&\multicolumn{1}{c}{(5)}&\multicolumn{1}{c}{(6)}&\multicolumn{1}{c}{(7)}\\
              &                      &\multicolumn{2}{c}{\textbf{Unadjusted}}       & \textbf{Bayesian} &\multicolumn{2}{c}{\textbf{Unadjusted}}&\textbf{Bayesian}\\\cmidrule(lr){3-4}\cmidrule(lr){6-7}
\textbf{State \& District}&            \textbf{N}&        \textbf{Mean}&        \textbf{Std Dev}&   \textbf{Adjusted Mean}&      \textbf{Disparity}  & \textbf{Std Error}  & \textbf{Adjusted Disparity} \\
\hline
NorthDakota 01 &       424 &     20.03 &     17.76 &     19.97 &      8.97 &     42.83 &      2.31 \\ \hline 
Ohio 01 &       672 &     20.96 &     14.85 &     20.92 &     11.17 &      3.15 &     10.23 \\ \hline 
Ohio 02 &       589 &     18.73 &     14.07 &     18.73 &      6.73 &      6.18 &      5.29 \\ \hline 
Ohio 03 &       542 &     22.11 &     15.98 &     22.02 &      1.51 &      3.70 &      1.57 \\ \hline 
Ohio 04 &       310 &     12.09 &      9.82 &     12.20 &     -8.88 &      3.50 &     -7.57 \\ \hline 
Ohio 05 &       575 &     15.42 &     13.03 &     15.47 &      7.90 &      9.38 &      4.94 \\ \hline 
Ohio 06 &       260 &     15.19 &     11.01 &     15.27 &     35.86 &     21.81 &      7.38 \\ \hline 
Ohio 07 &       368 &     15.30 &     13.41 &     15.39 &    -13.77 &      7.85 &     -7.32 \\ \hline 
Ohio 08 &       669 &     13.78 &     11.89 &     13.84 &     -8.52 &      3.29 &     -7.37 \\ \hline 
Ohio 09 &       383 &     16.21 &     10.98 &     16.26 &      8.64 &      2.74 &      8.12 \\ \hline 
Ohio 10 &       563 &     24.47 &     19.55 &     24.26 &    -13.94 &      3.89 &    -11.62 \\ \hline 
Ohio 11 &       331 &     19.02 &     14.19 &     19.01 &      4.47 &      2.56 &      4.30 \\ \hline 
Ohio 12 &       774 &     16.74 &     14.03 &     16.77 &     18.83 &     13.19 &      7.67 \\ \hline 
Ohio 13 &       456 &     16.07 &     12.94 &     16.12 &     17.79 &      6.04 &     13.17 \\ \hline 
Ohio 14 &       507 &     14.42 &     11.07 &     14.48 &    -11.11 &      6.59 &     -6.81 \\ \hline 
Ohio 15 &       701 &     19.05 &     15.20 &     19.04 &     19.26 &     15.02 &      6.87 \\ \hline 
Ohio 16 &       643 &     15.83 &     13.34 &     15.87 &     11.80 &     14.78 &      4.81 \\ \hline 
Oklahoma 01 &       968 &     24.53 &     19.67 &     24.41 &      1.42 &      8.22 &      1.65 \\ \hline 
Oklahoma 02 &       192 &     20.49 &     17.37 &     20.34 &     32.36 &     12.17 &     13.37 \\ \hline 
Oklahoma 03 &       591 &     25.77 &     20.51 &     25.51 &     11.18 &     29.58 &      2.84 \\ \hline 
Oklahoma 04 &       728 &     28.65 &     22.58 &     28.29 &     -0.68 &     14.28 &      1.16 \\ \hline 
Oklahoma 05 &       966 &     28.31 &     21.48 &     28.07 &      4.67 &      5.89 &      3.91 \\ \hline 
Pennsylvania 01 &       132 &     16.60 &     17.65 &     16.83 &     11.11 &      4.88 &      9.17 \\ \hline 
Pennsylvania 02 &       141 &     18.99 &     20.51 &     18.95 &     -0.32 &      4.89 &      0.16 \\ \hline 
Pennsylvania 03 &       292 &     19.04 &     17.57 &     19.02 &    -34.09 &     14.26 &     -9.04 \\ \hline 
Pennsylvania 04 &       479 &     26.11 &     22.47 &     25.71 &     -3.17 &     19.10 &      0.94 \\ \hline 
Pennsylvania 05 &       209 &     24.52 &     21.54 &     23.89 &    118.97 &     40.17 &      8.54 \\ \hline 
Pennsylvania 06 &       571 &     21.79 &     19.16 &     21.68 &      0.05 &     18.90 &      1.57 \\ \hline 
Pennsylvania 07 &       512 &     17.64 &     17.11 &     17.67 &    -14.90 &      7.70 &     -8.13 \\ \hline 
Pennsylvania 08 &       821 &     22.55 &     18.29 &     22.47 &    -27.55 &     11.57 &     -9.82 \\ \hline 
Pennsylvania 09 &       173 &     18.31 &     14.70 &     18.34 &    -18.23 &     16.57 &     -3.01 \\ \hline 
Pennsylvania 10 &       214 &     19.63 &     16.74 &     19.57 &      3.65 &     25.74 &      2.16 \\ \hline 
Pennsylvania 11 &       279 &     23.60 &     22.15 &     23.17 &    -15.56 &     18.16 &     -1.80 \\ \hline 
Pennsylvania 12 &       339 &     19.01 &     17.50 &     18.99 &      6.13 &     30.83 &      2.33 \\ \hline 
Pennsylvania 13 &       326 &     17.68 &     17.50 &     17.73 &      1.25 &      6.56 &      1.48 \\ \hline 
Pennsylvania 14 &       179 &     16.09 &     13.24 &     16.22 &      7.10 &      4.97 &      5.98 \\ \hline 
Pennsylvania 15 &       469 &     23.32 &     18.29 &     23.15 &    -44.29 &     19.63 &     -6.81 \\ \hline 
Pennsylvania 16 &       405 &     17.49 &     14.55 &     17.52 &      8.14 &     15.10 &      3.70 \\ \hline 
Pennsylvania 17 &       263 &     22.21 &     16.99 &     22.01 &    -10.02 &      8.06 &     -4.95 \\ \hline 
Pennsylvania 18 &       423 &     20.23 &     17.95 &     20.16 &    -20.47 &     24.44 &     -1.01 \\ \hline 
RhodeIsland 01 &       354 &     21.33 &     18.65 &     21.19 &     37.64 &     21.09 &      7.98 \\ \hline 
\hline
\end{tabular}
}
}
\end{center}
\vspace{-20pt}
\caption*{\scriptsize \textit{Notes}: Columns 5-7 (\textit{Disparity}) correspond to the coefficients on the interaction between a congressional district fixed effect and the ``Fraction Black'' variable from the voter-level regression of wait time on the full set of congressional district fixed effects and the interaction of those fixed effects with ``Fraction Black'', omitting the constant and  clustering standard errors at the polling place level. Column 7 provides  empirical-Bayes-adjusted estimates of these congressional-district-level disparities to account for measurement error. Similarly, Column 4 provides empirical-Bayes-adjusted estimates of the unadjusted congressional-district-level means shown in Column 2}
\end{table}

\setcounter{table}{1} 
\begin{table}[H]
\begin{center}
\caption{Congressional District-Level Measures of Wait Time and Disparities (9)}
\vspace{-10pt}
\scalebox{.7}{{
\def\sym#1{\ifmmode^{#1}\else\(^{#1}\)\fi}
\begin{tabular}{l*{8}{c}}
\hline\hline
&\multicolumn{1}{c}{(1)}&\multicolumn{1}{c}{(2)}&\multicolumn{1}{c}{(3)}&\multicolumn{1}{c}{(4)}&\multicolumn{1}{c}{(5)}&\multicolumn{1}{c}{(6)}&\multicolumn{1}{c}{(7)}\\
              &                      &\multicolumn{2}{c}{\textbf{Unadjusted}}       & \textbf{Bayesian} &\multicolumn{2}{c}{\textbf{Unadjusted}}&\textbf{Bayesian}\\\cmidrule(lr){3-4}\cmidrule(lr){6-7}
\textbf{State \& District}&            \textbf{N}&        \textbf{Mean}&        \textbf{Std Dev}&   \textbf{Adjusted Mean}&      \textbf{Disparity}  & \textbf{Std Error}  & \textbf{Adjusted Disparity} \\
\hline
RhodeIsland 02 &       431 &     17.22 &     12.66 &     17.25 &      9.56 &     14.58 &      4.21 \\ \hline 
SouthCarolina 01 &       715 &     34.44 &     25.83 &     33.68 &    -26.27 &     12.07 &     -8.75 \\ \hline 
SouthCarolina 02 &       737 &     23.82 &     20.64 &     23.66 &     -6.41 &      6.24 &     -3.86 \\ \hline 
SouthCarolina 03 &       449 &     20.12 &     17.60 &     20.06 &    -12.50 &      8.35 &     -6.14 \\ \hline 
SouthCarolina 04 &       749 &     25.99 &     20.53 &     25.77 &     -2.18 &      8.35 &     -0.36 \\ \hline 
SouthCarolina 05 &       588 &     22.77 &     19.20 &     22.63 &    -16.31 &      7.17 &     -9.60 \\ \hline 
SouthCarolina 06 &       315 &     23.37 &     21.02 &     23.04 &      3.53 &      8.24 &      2.85 \\ \hline 
SouthCarolina 07 &       588 &     31.46 &     23.78 &     30.82 &    -31.43 &      7.45 &    -18.56 \\ \hline 
SouthDakota 01 &       429 &     15.55 &     12.81 &     15.62 &    -12.56 &     10.28 &     -4.67 \\ \hline 
Tennessee 01 &       286 &     17.26 &     15.53 &     17.32 &    -18.24 &     23.23 &     -0.95 \\ \hline 
Tennessee 02 &       279 &     15.30 &     14.64 &     15.43 &      8.72 &      9.45 &      5.33 \\ \hline 
Tennessee 03 &       344 &     19.39 &     15.95 &     19.36 &      5.48 &      6.74 &      4.29 \\ \hline 
Tennessee 04 &       264 &     13.85 &     12.34 &     13.99 &     -3.60 &      9.93 &     -0.68 \\ \hline 
Tennessee 05 &       287 &     15.01 &     14.75 &     15.15 &      4.55 &      3.88 &      4.17 \\ \hline 
Tennessee 06 &       301 &     18.68 &     18.16 &     18.67 &     28.58 &      7.90 &     17.58 \\ \hline 
Tennessee 07 &       242 &     14.65 &     12.59 &     14.78 &      2.04 &      6.05 &      2.01 \\ \hline 
Tennessee 08 &       241 &     15.66 &     17.61 &     15.86 &      4.04 &      5.98 &      3.44 \\ \hline 
Tennessee 09 &       174 &     14.68 &     14.20 &     14.91 &      1.92 &      3.37 &      1.93 \\ \hline 
Texas 01 &       114 &     13.75 &     12.06 &     14.07 &      1.53 &     10.89 &      1.78 \\ \hline 
Texas 02 &       228 &     14.21 &     17.46 &     14.51 &     14.14 &     14.17 &      5.70 \\ \hline 
Texas 03 &       355 &     14.12 &     16.41 &     14.29 &      8.47 &     11.24 &      4.65 \\ \hline 
Texas 04 &       160 &     12.83 &     11.02 &     13.06 &     -8.24 &      6.64 &     -4.85 \\ \hline 
Texas 05 &       162 &     16.39 &     19.22 &     16.64 &     -3.67 &     10.71 &     -0.50 \\ \hline 
Texas 06 &       285 &     14.22 &     12.10 &     14.34 &      0.09 &      4.40 &      0.42 \\ \hline 
Texas 07 &       246 &     13.22 &     15.83 &     13.51 &    -14.29 &      8.87 &     -6.65 \\ \hline 
Texas 08 &       270 &     15.85 &     16.40 &     16.00 &    -11.01 &     16.76 &     -1.18 \\ \hline 
Texas 09 &       134 &     16.33 &     17.47 &     16.59 &     -1.14 &      7.46 &      0.06 \\ \hline 
Texas 10 &       203 &     16.50 &     18.06 &     16.67 &     -0.80 &      8.69 &      0.47 \\ \hline 
Texas 11 &       156 &     15.75 &     17.36 &     16.02 &     12.40 &     27.92 &      3.06 \\ \hline 
Texas 12 &       246 &     13.24 &     13.18 &     13.44 &      6.08 &     11.82 &      3.56 \\ \hline 
Texas 13 &       164 &     16.22 &     18.19 &     16.47 &     -5.93 &     29.13 &      1.18 \\ \hline 
Texas 14 &       181 &     16.33 &     17.49 &     16.52 &      7.57 &      7.36 &      5.44 \\ \hline 
Texas 15 &       135 &     19.37 &     19.61 &     19.27 &    -48.73 &     17.28 &     -9.75 \\ \hline 
Texas 16 &       176 &     15.01 &     14.51 &     15.23 &     46.67 &     46.53 &      3.91 \\ \hline 
Texas 17 &       261 &     20.43 &     17.08 &     20.33 &      3.11 &     13.38 &      2.34 \\ \hline 
Texas 18 &       184 &     14.18 &     16.06 &     14.50 &     -6.57 &      4.66 &     -4.89 \\ \hline 
Texas 19 &       175 &     13.45 &     12.70 &     13.70 &      7.45 &     10.34 &      4.45 \\ \hline 
Texas 20 &       215 &     16.67 &     15.38 &     16.78 &     34.88 &     20.25 &      7.90 \\ \hline 
Texas 21 &       242 &     18.86 &     20.60 &     18.85 &    108.93 &     52.16 &      5.81 \\ \hline 
Texas 22 &       264 &     16.89 &     16.15 &     16.98 &     18.52 &      9.50 &     10.16 \\ \hline 
Texas 23 &       133 &     22.56 &     22.39 &     21.90 &     64.90 &     56.35 &      3.95 \\ \hline 
\hline
\end{tabular}
}
}
\end{center}
\vspace{-20pt}
\caption*{\scriptsize \textit{Notes}: Columns 5-7 (\textit{Disparity}) correspond to the coefficients on the interaction between a congressional district fixed effect and the ``Fraction Black'' variable from the voter-level regression of wait time on the full set of congressional district fixed effects and the interaction of those fixed effects with ``Fraction Black'', omitting the constant and  clustering standard errors at the polling place level. Column 7 provides  empirical-Bayes-adjusted estimates of these congressional-district-level disparities to account for measurement error. Similarly, Column 4 provides empirical-Bayes-adjusted estimates of the unadjusted congressional-district-level means shown in Column 2}
\end{table}

\setcounter{table}{1} 
\begin{table}[H]
\begin{center}
\caption{Congressional District-Level Measures of Wait Time and Disparities (10)}
\vspace{-10pt}
\scalebox{.7}{{
\def\sym#1{\ifmmode^{#1}\else\(^{#1}\)\fi}
\begin{tabular}{l*{8}{c}}
\hline\hline
&\multicolumn{1}{c}{(1)}&\multicolumn{1}{c}{(2)}&\multicolumn{1}{c}{(3)}&\multicolumn{1}{c}{(4)}&\multicolumn{1}{c}{(5)}&\multicolumn{1}{c}{(6)}&\multicolumn{1}{c}{(7)}\\
              &                      &\multicolumn{2}{c}{\textbf{Unadjusted}}       & \textbf{Bayesian} &\multicolumn{2}{c}{\textbf{Unadjusted}}&\textbf{Bayesian}\\\cmidrule(lr){3-4}\cmidrule(lr){6-7}
\textbf{State \& District}&            \textbf{N}&        \textbf{Mean}&        \textbf{Std Dev}&   \textbf{Adjusted Mean}&      \textbf{Disparity}  & \textbf{Std Error}  & \textbf{Adjusted Disparity} \\
\hline
Texas 24 &       236 &     13.88 &     14.04 &     14.09 &     -2.55 &      7.31 &     -0.85 \\ \hline 
Texas 25 &       217 &     18.09 &     16.65 &     18.13 &      2.24 &     10.40 &      2.08 \\ \hline 
Texas 26 &       410 &     18.40 &     16.74 &     18.41 &     38.07 &     24.20 &      6.82 \\ \hline 
Texas 27 &       157 &     24.32 &     23.42 &     23.41 &    -75.44 &     24.77 &     -8.07 \\ \hline 
Texas 28 &       141 &     18.65 &     17.15 &     18.65 &     -7.88 &     16.05 &     -0.58 \\ \hline 
Texas 29 &       147 &     11.72 &      9.40 &     11.94 &      2.51 &      5.20 &      2.38 \\ \hline 
Texas 30 &       203 &     14.83 &     14.81 &     15.04 &     -6.35 &      4.16 &     -4.99 \\ \hline 
Texas 31 &       218 &     15.52 &     15.65 &     15.70 &     12.08 &     18.69 &      4.03 \\ \hline 
Texas 32 &       312 &     14.74 &     14.62 &     14.88 &      1.60 &      7.16 &      1.73 \\ \hline 
Texas 33 &       145 &     14.77 &     13.93 &     15.04 &      7.06 &     11.81 &      3.94 \\ \hline 
Texas 34 &       112 &     19.22 &     19.36 &     19.13 &     11.09 &     44.74 &      2.38 \\ \hline 
Texas 35 &       166 &     17.69 &     19.91 &     17.80 &    -13.70 &      8.79 &     -6.41 \\ \hline 
Texas 36 &       224 &     13.63 &     12.69 &     13.82 &     -9.44 &      7.10 &     -5.30 \\ \hline 
Utah 01 &       119 &     18.77 &     14.51 &     18.76 &    -57.57 &      9.26 &    -28.30 \\ \hline 
Utah 02 &       253 &     33.75 &     25.70 &     31.88 &     30.66 &     72.88 &      2.56 \\ \hline 
Utah 03 &       594 &     25.83 &     20.48 &     25.56 &   -158.01 &    192.90 &      0.80 \\ \hline 
Utah 04 &       235 &     31.41 &     26.84 &     29.59 &     63.87 &     87.85 &      2.95 \\ \hline 
Vermont 01 &       165 &     14.83 &     13.09 &     15.03 &      6.78 &     33.08 &      2.34 \\ \hline 
Virginia 01 &     1,053 &     16.08 &     13.81 &     16.10 &     -4.14 &      8.19 &     -1.52 \\ \hline 
Virginia 02 &     1,022 &     18.78 &     15.44 &     18.78 &      4.57 &      7.22 &      3.60 \\ \hline 
Virginia 03 &       674 &     21.25 &     19.32 &     21.17 &     -1.26 &      3.84 &     -0.80 \\ \hline 
Virginia 04 &       824 &     19.96 &     16.86 &     19.94 &     -0.56 &      4.08 &     -0.16 \\ \hline 
Virginia 05 &       535 &     17.89 &     13.99 &     17.90 &     15.11 &      7.00 &     10.42 \\ \hline 
Virginia 06 &       562 &     18.61 &     15.96 &     18.61 &     -7.10 &      8.88 &     -2.83 \\ \hline 
Virginia 07 &     1,049 &     20.08 &     15.48 &     20.07 &     -1.75 &      5.48 &     -0.81 \\ \hline 
Virginia 08 &       569 &     17.49 &     15.32 &     17.51 &     14.38 &      6.85 &     10.07 \\ \hline 
Virginia 09 &       444 &     16.52 &     15.36 &     16.58 &      2.08 &     12.75 &      2.00 \\ \hline 
Virginia 10 &     1,347 &     14.54 &     11.45 &     14.57 &     13.47 &     10.77 &      6.95 \\ \hline 
Virginia 11 &       951 &     15.98 &     13.55 &     16.00 &     14.18 &      5.49 &     11.07 \\ \hline 
WestVirginia 01 &       141 &     15.85 &     13.88 &     16.04 &      2.34 &     12.68 &      2.09 \\ \hline 
WestVirginia 02 &       333 &     19.50 &     12.00 &     19.48 &      7.08 &      7.20 &      5.19 \\ \hline 
WestVirginia 03 &       126 &     18.27 &     17.00 &     18.31 &     -7.86 &     27.65 &      0.90 \\ \hline 
Wisconsin 01 &       536 &     17.11 &     15.52 &     17.14 &      2.79 &      6.60 &      2.52 \\ \hline 
Wisconsin 02 &       525 &     16.82 &     14.17 &     16.86 &     14.09 &     15.11 &      5.37 \\ \hline 
Wisconsin 03 &       394 &     17.22 &     13.62 &     17.26 &     -0.27 &     23.26 &      1.64 \\ \hline 
Wisconsin 04 &       377 &     15.50 &     13.48 &     15.58 &      3.33 &      2.78 &      3.22 \\ \hline 
Wisconsin 05 &       662 &     15.67 &     13.22 &     15.71 &    -19.14 &     14.96 &     -4.06 \\ \hline 
Wisconsin 06 &       516 &     16.57 &     12.93 &     16.61 &    -32.77 &     18.73 &     -5.13 \\ \hline 
Wisconsin 07 &       261 &     17.58 &     13.78 &     17.62 &    -30.10 &     58.11 &      0.99 \\ \hline 
Wisconsin 08 &       455 &     17.15 &     13.44 &     17.18 &      7.23 &     17.75 &      3.13 \\ \hline 
Wyoming 01 &       286 &     20.65 &     13.15 &     20.59 &     17.90 &     44.17 &      2.72 \\ \hline 
\hline
\end{tabular}
}
}
\end{center}
\vspace{-20pt}
\caption*{\scriptsize \textit{Notes}: Columns 5-7 (\textit{Disparity}) correspond to the coefficients on the interaction between a congressional district fixed effect and the ``Fraction Black'' variable from the voter-level regression of wait time on the full set of congressional district fixed effects and the interaction of those fixed effects with ``Fraction Black'', omitting the constant and  clustering standard errors at the polling place level. Column 7 provides  empirical-Bayes-adjusted estimates of these congressional-district-level disparities to account for measurement error. Similarly, Column 4 provides empirical-Bayes-adjusted estimates of the unadjusted congressional-district-level means shown in Column 2}
\end{table}

\setcounter{table}{2} 
\begin{table}[H]
\begin{center}
\caption{(100 Most Populous) County-Level Measures of Wait Time and Disparities (1)}
\label{table:app_county_descriptives}
\vspace{-10pt}
\scalebox{.65}{{
\def\sym#1{\ifmmode^{#1}\else\(^{#1}\)\fi}
\begin{tabular}{l*{9}{c}}
\hline\hline
&\multicolumn{1}{c}{(1)}&\multicolumn{1}{c}{(2)}&\multicolumn{1}{c}{(3)}&\multicolumn{1}{c}{(4)}&\multicolumn{1}{c}{(5)}&\multicolumn{1}{c}{(6)}&\multicolumn{1}{c}{(7)}&\multicolumn{1}{c}{(8)}\\
              & &                      &\multicolumn{2}{c}{\textbf{Unadjusted}}       & \textbf{Bayesian} &\multicolumn{2}{c}{\textbf{Unadjusted}}&\textbf{Bayesian}\\\cmidrule(lr){4-5}\cmidrule(lr){7-8}
\textbf{County \& State}& \textbf{Population}&  \textbf{N}&        \textbf{Mean}&        \textbf{Std Dev}&   \textbf{Adjusted Mean}&      \textbf{Disparity}  & \textbf{Std Error}  & \textbf{Adjusted Disparity} \\
\hline
Alameda California & 1,629,615 &       430 &     19.31 &     16.95 &     19.29 &     10.76 &      6.89 &     10.54 \\ \hline 
Allegheny Pennsylvania & 1,229,605 &       572 &     19.02 &     16.73 &     19.01 &     -0.91 &      5.13 &     -0.91 \\ \hline 
BaltimoreCity Maryland &   619,796 &       220 &     24.41 &     17.88 &     24.08 &      1.23 &      4.29 &      1.21 \\ \hline 
Baltimore Maryland &   828,637 &       806 &     31.25 &     20.88 &     30.99 &      1.49 &      4.17 &      1.47 \\ \hline 
Bergen NewJersey &   937,920 &       433 &     11.45 &     11.12 &     11.52 &      0.87 &      5.80 &      0.85 \\ \hline 
Bernalillo NewMexico &   674,855 &       161 &     20.30 &     15.10 &     20.19 &    -42.40 &     28.36 &    -32.35 \\ \hline 
Bexar Texas & 1,892,004 &       530 &     18.26 &     18.60 &     18.26 &      5.37 &      8.72 &      5.18 \\ \hline 
Bronx NewYork & 1,455,846 &       355 &     20.59 &     16.59 &     20.52 &     -0.64 &      3.99 &     -0.65 \\ \hline 
Broward Florida & 1,890,416 &       560 &     21.55 &     17.64 &     21.48 &     -0.08 &      4.67 &     -0.08 \\ \hline 
Bucks Pennsylvania &   626,486 &       712 &     22.94 &     18.60 &     22.85 &    -31.53 &     13.08 &    -29.58 \\ \hline 
Clark Nevada & 2,112,436 &       670 &     15.33 &     13.57 &     15.36 &      6.17 &      9.10 &      5.94 \\ \hline 
Cobb Georgia &   739,072 &       759 &     20.29 &     17.14 &     20.26 &      5.06 &      5.35 &      4.99 \\ \hline 
Collin Texas &   914,075 &       388 &     14.36 &     16.28 &     14.45 &      7.22 &     10.49 &      6.88 \\ \hline 
ContraCosta California & 1,123,678 &       471 &     17.71 &     14.66 &     17.72 &      7.23 &      6.92 &      7.08 \\ \hline 
Cook Illinois & 5,238,541 &     1,603 &     20.10 &     16.25 &     20.09 &      1.26 &      1.52 &      1.26 \\ \hline 
Cuyahoga Ohio & 1,257,401 &       754 &     16.88 &     13.83 &     16.89 &      6.87 &      1.67 &      6.86 \\ \hline 
DC DistrictofColumbia &   672,391 &       179 &     27.49 &     22.94 &     26.56 &      8.38 &      4.62 &      8.30 \\ \hline 
Dallas Texas & 2,552,213 &       767 &     14.64 &     15.16 &     14.68 &     -2.28 &      2.11 &     -2.28 \\ \hline 
Davidson Tennessee &   678,322 &       255 &     14.82 &     13.85 &     14.91 &      5.20 &      3.86 &      5.16 \\ \hline 
Dekalb Georgia &   736,066 &       335 &     18.25 &     16.91 &     18.25 &      0.22 &      2.50 &      0.22 \\ \hline 
Denton Texas &   781,321 &       346 &     18.62 &     16.93 &     18.60 &     21.07 &     24.12 &     16.94 \\ \hline 
DuPage Illinois &   931,826 &       697 &     14.37 &     13.09 &     14.40 &     33.82 &     27.92 &     25.61 \\ \hline 
Duval Florida &   912,043 &       275 &     12.62 &     11.77 &     12.72 &      6.00 &      5.98 &      5.90 \\ \hline 
ElPaso Texas &   834,825 &       194 &     15.42 &     15.17 &     15.53 &     36.92 &     46.46 &     19.46 \\ \hline 
Erie NewYork &   923,995 &       407 &     12.58 &     12.49 &     12.66 &      6.57 &      5.33 &      6.49 \\ \hline 
Essex Massachusetts &   775,860 &       292 &     11.77 &     11.76 &     11.88 &     22.51 &     13.63 &     20.90 \\ \hline 
Essex NewJersey &   800,401 &       293 &     17.52 &     16.33 &     17.54 &     -2.76 &      3.25 &     -2.75 \\ \hline 
Fairfax Virginia & 1,142,004 &     1,262 &     14.75 &     12.45 &     14.77 &     24.18 &      6.77 &     23.73 \\ \hline 
Fairfield Connecticut &   947,328 &       708 &     12.81 &     11.92 &     12.85 &     13.23 &      5.94 &     13.03 \\ \hline 
FortBend Texas &   711,421 &       134 &     17.11 &     16.33 &     17.18 &      5.01 &      9.16 &      4.82 \\ \hline 
Franklin Ohio & 1,253,507 &     1,238 &     20.93 &     16.17 &     20.91 &      4.54 &      3.21 &      4.52 \\ \hline 
Fresno California &   971,616 &       291 &     19.79 &     16.96 &     19.73 &     14.02 &     14.65 &     12.85 \\ \hline 
Fulton Georgia & 1,010,420 &       483 &     20.90 &     17.40 &     20.84 &      4.72 &      3.01 &      4.70 \\ \hline 
Gwinnett Georgia &   889,954 &       779 &     30.16 &     24.64 &     29.82 &     17.67 &      7.87 &     17.22 \\ \hline 
Hamilton Ohio &   808,703 &       654 &     22.33 &     15.89 &     22.27 &      7.05 &      3.25 &      7.02 \\ \hline 
Harris Texas & 4,525,519 &     1,282 &     13.61 &     15.20 &     13.64 &      1.31 &      2.72 &      1.30 \\ \hline 
Hartford Connecticut &   897,417 &       768 &     11.40 &     12.04 &     11.45 &      6.18 &      2.68 &      6.16 \\ \hline 
Hennepin Minnesota & 1,224,763 &     1,063 &     17.54 &     14.92 &     17.55 &     10.20 &      6.47 &     10.02 \\ \hline 
Hidalgo Texas &   839,539 &       119 &     20.06 &     20.08 &     19.84 &  1,032.15 &    431.93 &     15.63 \\ \hline 
Hillsborough Florida & 1,351,087 &       468 &     18.05 &     14.63 &     18.05 &      0.87 &      5.04 &      0.85 \\ \hline 
\hline
\end{tabular}
}
}
\end{center}
\vspace{-20pt}
\caption*{\scriptsize \textit{Notes}: Columns 6-8 (\textit{Disparity}) correspond to the coefficients on the interaction between a county fixed effect and the ``Fraction Black'' variable from the voter-level regression of wait time on the full set of county fixed effects and the interaction of those fixed effects with ``Fraction Black'', omitting the constant and  clustering standard errors at the polling place level. Column 7 provides  empirical-Bayes-adjusted estimates of these congressional-district-level disparities to account for measurement error. Similarly, Column 5 provides empirical-Bayes-adjusted estimates of the unadjusted congressional-district-level means shown in Column 3. Column 2 displays the population of each listed county; we just show the 100 largest counties (by population in the 2017 American Community Survey's five-year estimates).}
\end{table}

\setcounter{table}{2} 
\begin{table}[H]
\begin{center}
\caption{(100 Most Populous) County-Level Measures of Wait Time and Disparities (2)}
\vspace{-10pt}
\scalebox{.65}{{
\def\sym#1{\ifmmode^{#1}\else\(^{#1}\)\fi}
\begin{tabular}{l*{9}{c}}
\hline\hline
&\multicolumn{1}{c}{(1)}&\multicolumn{1}{c}{(2)}&\multicolumn{1}{c}{(3)}&\multicolumn{1}{c}{(4)}&\multicolumn{1}{c}{(5)}&\multicolumn{1}{c}{(6)}&\multicolumn{1}{c}{(7)}&\multicolumn{1}{c}{(8)}\\
              & &                      &\multicolumn{2}{c}{\textbf{Unadjusted}}       & \textbf{Bayesian} &\multicolumn{2}{c}{\textbf{Unadjusted}}&\textbf{Bayesian}\\\cmidrule(lr){4-5}\cmidrule(lr){7-8}
\textbf{County \& State}& \textbf{Population}&  \textbf{N}&        \textbf{Mean}&        \textbf{Std Dev}&   \textbf{Adjusted Mean}&      \textbf{Disparity}  & \textbf{Std Error}  & \textbf{Adjusted Disparity} \\
\hline
Hudson NewJersey &   679,756 &        10 &     17.93 &     21.48 &  &      5.01 &    154.65 &  \\ \hline 
Jackson Missouri &   688,554 &       950 &     42.65 &     26.94 &     41.96 &     -7.08 &      6.63 &     -6.97 \\ \hline 
Jefferson Alabama &   659,460 &       854 &     26.41 &     18.03 &     26.29 &      1.94 &      2.58 &      1.94 \\ \hline 
Jefferson Kentucky &   764,378 &       833 &     12.25 &     10.49 &     12.28 &     -0.24 &      2.29 &     -0.24 \\ \hline 
Kent Michigan &   636,376 &       646 &     22.67 &     16.64 &     22.60 &      6.91 &      7.33 &      6.75 \\ \hline 
Kern California &   878,744 &       259 &     17.76 &     12.33 &     17.77 &    -23.78 &      9.84 &    -22.93 \\ \hline 
Kings NewYork & 2,635,121 &       693 &     20.52 &     15.79 &     20.49 &      5.91 &      1.79 &      5.90 \\ \hline 
Lake Illinois &   704,476 &       522 &     16.03 &     13.07 &     16.06 &     22.98 &     12.30 &     21.62 \\ \hline 
Lee Florida &   700,165 &       185 &     19.00 &     17.61 &     18.95 &     23.38 &     19.01 &     20.33 \\ \hline 
LosAngeles California &  10105722 &     2,719 &     22.62 &     18.41 &     22.60 &      4.23 &      2.87 &      4.21 \\ \hline 
Macomb Michigan &   864,019 &     1,248 &     19.38 &     13.64 &     19.37 &      5.34 &      5.18 &      5.28 \\ \hline 
Maricopa Arizona & 4,155,501 &     1,378 &     21.61 &     19.29 &     21.58 &      4.01 &      6.58 &      3.93 \\ \hline 
Marion Indiana &   939,964 &       726 &     23.54 &     20.81 &     23.42 &    -15.18 &      7.67 &    -14.85 \\ \hline 
Mecklenburg NorthCarolina & 1,034,290 &       574 &     25.05 &     18.44 &     24.90 &      5.25 &      4.28 &      5.20 \\ \hline 
Miami-Dade Florida & 2,702,602 &       537 &     21.02 &     17.99 &     20.96 &      4.71 &      3.95 &      4.68 \\ \hline 
Middlesex Massachusetts & 1,582,857 &       642 &     11.48 &      8.98 &     11.51 &      3.59 &      4.26 &      3.56 \\ \hline 
Middlesex NewJersey &   837,288 &       558 &     16.95 &     15.04 &     16.97 &      1.01 &     10.70 &      0.93 \\ \hline 
Milwaukee Wisconsin &   956,586 &       600 &     16.08 &     13.86 &     16.10 &      0.86 &      2.82 &      0.86 \\ \hline 
Monmouth NewJersey &   627,551 &       448 &     11.00 &      9.89 &     11.06 &     -4.22 &      4.49 &     -4.20 \\ \hline 
Monroe NewYork &   748,680 &       556 &     15.24 &     15.05 &     15.29 &      0.45 &      6.19 &      0.43 \\ \hline 
Montgomery Maryland & 1,039,198 &       829 &     19.90 &     14.41 &     19.89 &      5.99 &      4.98 &      5.92 \\ \hline 
Montgomery Pennsylvania &   818,677 &       714 &     20.81 &     18.54 &     20.76 &      7.95 &     10.09 &      7.61 \\ \hline 
NewHaven Connecticut &   862,127 &       536 &     13.46 &     14.59 &     13.53 &     19.11 &      7.62 &     18.66 \\ \hline 
NewYork NewYork & 1,653,877 &       524 &     20.49 &     18.31 &     20.43 &      2.73 &      5.26 &      2.69 \\ \hline 
Norfolk Massachusetts &   694,389 &       290 &     12.04 &      9.28 &     12.11 &      3.30 &      4.86 &      3.26 \\ \hline 
Oakland Michigan & 1,241,860 &     1,843 &     23.23 &     16.57 &     23.20 &      7.12 &      3.24 &      7.09 \\ \hline 
Ocean NewJersey &   589,699 &       254 &     13.18 &     14.36 &     13.33 &     17.05 &     31.32 &     12.01 \\ \hline 
Oklahoma Oklahoma &   774,203 &       975 &     28.88 &     21.65 &     28.69 &      5.21 &      5.89 &      5.12 \\ \hline 
Orange California & 3,155,816 &     1,202 &     21.92 &     17.47 &     21.88 &    -19.00 &     22.43 &    -15.99 \\ \hline 
Orange Florida & 1,290,216 &       443 &     19.13 &     14.99 &     19.11 &     -4.22 &      4.34 &     -4.20 \\ \hline 
PalmBeach Florida & 1,426,772 &       662 &     20.87 &     18.95 &     20.82 &     -2.41 &      4.65 &     -2.40 \\ \hline 
Philadelphia Pennsylvania & 1,569,657 &       286 &     16.08 &     18.40 &     16.16 &      3.75 &      3.58 &      3.72 \\ \hline 
Pima Arizona & 1,007,257 &       247 &     16.89 &     15.01 &     16.93 &      6.60 &     19.21 &      5.64 \\ \hline 
Pinellas Florida &   949,842 &       396 &     19.59 &     16.05 &     19.55 &      2.99 &      5.16 &      2.95 \\ \hline 
Polk Florida &   652,256 &       290 &     16.60 &     14.51 &     16.64 &     -5.28 &      6.03 &     -5.22 \\ \hline 
PrinceGeorge'S Maryland &   905,161 &       547 &     21.76 &     16.30 &     21.70 &     -1.74 &      3.35 &     -1.74 \\ \hline 
Providence RhodeIsland &   633,704 &       403 &     21.05 &     18.16 &     20.97 &     29.43 &     17.35 &     26.19 \\ \hline 
Queens NewYork & 2,339,280 &     1,056 &     21.59 &     17.05 &     21.55 &      6.83 &      2.18 &      6.81 \\ \hline 
Riverside California & 2,355,002 &     1,137 &     21.13 &     17.28 &     21.10 &     26.14 &     10.10 &     25.08 \\ \hline 
Sacramento California & 1,495,400 &       482 &     17.83 &     15.87 &     17.84 &      0.98 &      7.73 &      0.93 \\ \hline 
\hline
\end{tabular}
}
}
\end{center}
\vspace{-20pt}
\caption*{\scriptsize \textit{Notes}: Columns 6-8 (\textit{Disparity}) correspond to the coefficients on the interaction between a county fixed effect and the ``Fraction Black'' variable from the voter-level regression of wait time on the full set of county fixed effects and the interaction of those fixed effects with ``Fraction Black'', omitting the constant and  clustering standard errors at the polling place level. Column 7 provides  empirical-Bayes-adjusted estimates of these congressional-district-level disparities to account for measurement error. Similarly, Column 5 provides empirical-Bayes-adjusted estimates of the unadjusted congressional-district-level means shown in Column 3. Column 2 displays the population of each listed county; we just show the 100 largest counties (by population in the 2017 American Community Survey's five-year estimates).}
\end{table}

\setcounter{table}{2} 
\begin{table}[H]
\begin{center}
\caption{(100 Most Populous) County-Level Measures of Wait Time and Disparities (3)}
\vspace{-10pt}
\scalebox{.65}{{
\def\sym#1{\ifmmode^{#1}\else\(^{#1}\)\fi}
\begin{tabular}{l*{9}{c}}
\hline\hline
&\multicolumn{1}{c}{(1)}&\multicolumn{1}{c}{(2)}&\multicolumn{1}{c}{(3)}&\multicolumn{1}{c}{(4)}&\multicolumn{1}{c}{(5)}&\multicolumn{1}{c}{(6)}&\multicolumn{1}{c}{(7)}&\multicolumn{1}{c}{(8)}\\
              & &                      &\multicolumn{2}{c}{\textbf{Unadjusted}}       & \textbf{Bayesian} &\multicolumn{2}{c}{\textbf{Unadjusted}}&\textbf{Bayesian}\\\cmidrule(lr){4-5}\cmidrule(lr){7-8}
\textbf{County \& State}& \textbf{Population}&  \textbf{N}&        \textbf{Mean}&        \textbf{Std Dev}&   \textbf{Adjusted Mean}&      \textbf{Disparity}  & \textbf{Std Error}  & \textbf{Adjusted Disparity} \\
\hline
SaltLake Utah & 1,106,700 &       226 &     40.18 &     30.62 &     37.21 &    -15.85 &     85.33 &     -4.77 \\ \hline 
SanBernardino California & 2,121,220 &       472 &     23.39 &     19.17 &     23.25 &     32.00 &     25.55 &     25.22 \\ \hline 
SanDiego California & 3,283,665 &     1,085 &     19.22 &     16.80 &     19.21 &     23.30 &     10.77 &     22.24 \\ \hline 
SanFrancisco California &   864,263 &       169 &     17.67 &     20.16 &     17.71 &     10.04 &     28.83 &      7.32 \\ \hline 
SanJoaquin California &   724,153 &       172 &     16.75 &     15.40 &     16.81 &     27.33 &     15.04 &     25.00 \\ \hline 
SanMateo California &   763,450 &       186 &     22.46 &     18.54 &     22.18 &    -17.39 &     45.89 &     -9.89 \\ \hline 
SantaClara California & 1,911,226 &       534 &     17.89 &     16.58 &     17.89 &    -13.97 &     18.76 &    -12.37 \\ \hline 
Shelby Tennessee &   937,847 &       319 &     14.83 &     14.95 &     14.92 &      1.24 &      2.65 &      1.23 \\ \hline 
StLouis Missouri &   999,539 &     1,418 &     27.12 &     19.09 &     27.03 &     13.33 &      2.81 &     13.28 \\ \hline 
Suffolk Massachusetts &   780,685 &       182 &     19.56 &     17.70 &     19.47 &     -3.33 &      4.98 &     -3.30 \\ \hline 
Suffolk NewYork & 1,497,595 &     1,707 &     14.01 &     12.22 &     14.02 &     16.65 &      4.00 &     16.54 \\ \hline 
Tarrant Texas & 1,983,675 &       708 &     14.34 &     13.51 &     14.38 &      3.86 &      4.44 &      3.82 \\ \hline 
Travis Texas & 1,176,584 &       419 &     21.41 &     20.45 &     21.29 &     26.42 &     11.93 &     24.96 \\ \hline 
Tulsa Oklahoma &   637,123 &       811 &     23.98 &     19.47 &     23.88 &      1.48 &      8.53 &      1.41 \\ \hline 
Ventura California &   847,834 &       341 &     20.01 &     17.35 &     19.95 &      1.16 &     19.72 &      0.88 \\ \hline 
Wake NorthCarolina & 1,023,811 &       720 &     24.40 &     19.21 &     24.29 &     14.55 &      6.45 &     14.30 \\ \hline 
Wayne Michigan & 1,763,822 &     1,763 &     24.80 &     18.69 &     24.75 &     12.95 &      2.14 &     12.92 \\ \hline 
Westchester NewYork &   975,321 &        25 &     10.84 &      6.59 &  &    168.62 &     52.87 &  \\ \hline 
Will Illinois &   687,727 &       638 &     13.23 &     10.48 &     13.26 &      7.03 &      5.00 &      6.96 \\ \hline 
Worcester Massachusetts &   818,249 &       383 &     12.28 &     10.26 &     12.34 &     -0.89 &      8.09 &     -0.89 \\ \hline 
\hline
\end{tabular}
}
}
\end{center}
\vspace{-20pt}
\caption*{\scriptsize \textit{Notes}: Columns 6-8 (\textit{Disparity}) correspond to the coefficients on the interaction between a county fixed effect and the ``Fraction Black'' variable from the voter-level regression of wait time on the full set of county fixed effects and the interaction of those fixed effects with ``Fraction Black'', omitting the constant and  clustering standard errors at the polling place level. Column 7 provides  empirical-Bayes-adjusted estimates of these congressional-district-level disparities to account for measurement error. Similarly, Column 5 provides empirical-Bayes-adjusted estimates of the unadjusted congressional-district-level means shown in Column 3. Column 2 displays the population of each listed county; we just show the 100 largest counties (by population in the 2017 American Community Survey's five-year estimates).}
\end{table}

\end{document}